\pdfoutput=1
\documentclass[cernpreprint, atlasdraft=false, texlive=2023, UKenglish, coverpage=false, texmf, orcidlogo]{atlasdoc}
 
\usepackage{atlaspackage}
\usepackage[backref=false]{atlasbiblatex}
 
\usepackage[BSM=true]{atlasphysics}
 
\usepackage{adjustbox}
 
\graphicspath{{logos/}{figures/}}
 
\usepackage{ANA-SUSY-2023-03-PAPER-defs}

\AtlasTitle{
Search for electroweak production of supersymmetric particles in final states with two $\tau$-leptons in $\sqrt{s} = 13\,\ensuremath{\text{Te\kern -0.1em V}}$ $pp$ collisions with the ATLAS detector
}

\AtlasAbstract{
Three searches for the direct production of $\tau$-sleptons or charginos and neutralinos in final states with at least two hadronically decaying $\tau$-leptons are presented.
For chargino and neutralino production, decays via intermediate $\tau$-sleptons or $W$ and $h$ bosons are considered.
The analysis uses a dataset of $pp$ collisions corresponding to an integrated luminosity of $139\,$fb$^{-1}$, recorded with the ATLAS detector at the Large Hadron Collider at a centre-of-mass energy of $13\,\ensuremath{\text{Te\kern -0.1em V}}$.
No significant deviation from the expected Standard Model background is observed and supersymmetric particle mass limits at 95\% confidence level are obtained in simplified models.
For direct production of $\tilde{\chi}^+_1\tilde{\chi}^-_1$, chargino masses are excluded up to $970\,\ensuremath{\text{Ge\kern -0.1em V}}$, while $\tilde{\chi}^{\pm}_1$ and $\tilde{\chi}^0_2$ masses up to $1160\,\ensuremath{\text{Ge\kern -0.1em V}}$ ($330\,\ensuremath{\text{Ge\kern -0.1em V}}$) are excluded for $\tilde{\chi}^{\pm}_1\tilde{\chi}^0_2$/$\tilde{\chi}^+_1\tilde{\chi}^-_1$\ production with subsequent decays via $\tau$-sleptons ($W$ and $h$ bosons).
Masses of $\tau$-sleptons up to $500\,\ensuremath{\text{Ge\kern -0.1em V}}$ are excluded for mass degenerate $\tilde{\tau}_{L,R}$ scenarios and up to $425\,\ensuremath{\text{Ge\kern -0.1em V}}$ for $\tilde{\tau}_L$-only scenarios. Sensitivity to $\tilde{\tau}_R$-only scenarios from the ATLAS experiment is presented here for the first time, with $\tilde{\tau}_R$ masses excluded up to $350\,\ensuremath{\text{Ge\kern -0.1em V}}$.
 
}

\AtlasRefCode{SUSY-2023-03}

\AtlasJournal{JHEP}

\PreprintIdNumber{CERN-EP-2023-295}
 
\AtlasJournalRef{JHEP 05 (2024) 150}
\AtlasDOI{10.1007/JHEP05(2024)150}

\hypersetup{pdftitle={ATLAS document},pdfauthor={The ATLAS Collaboration}}
 
\newcommand{\pTmiss}{\mathbf{p}_\mathrm{T}^\mathrm{miss}}
 
\newcommand{\pTvec}{\mathbf{p}_\mathrm{T}}
\newcommand{\qTvec}{\mathbf{q}_\mathrm{T}}

\begin{document}
 
\maketitle

\clearpage
 
\section{Introduction}
\label{sec:intro}
 
Supersymmetry (SUSY)~\cite{Golfand:1971iw,Volkov:1973ix,Wess:1974tw,Wess:1974jb,Ferrara:1974pu,Salam:1974ig,Martin:1997ns}
postulates the existence of a bosonic (fermionic) partner for each fermionic (bosonic) particle of the Standard Model (SM), whose spin differs by one half unit from each corresponding SM particle.
In models that conserve $R$-parity~\cite{Farrar:1978xj}, the lightest supersymmetric particle (LSP) is stable and can be an excellent dark-matter candidate~\cite{Goldberg:1983nd,Ellis:1983ew}.
 
In the simplified SUSY models considered in this article, the sector of SUSY particles with only electroweak interactions contains charginos ($\susy{\chi}^{\pm}_{i}$, $i = 1, 2$, in order of increasing masses), neutralinos
($\susy{\chi}^{0}_{j}$, $j = 1, 2, 3, 4$, in order of increasing masses), charged sleptons (\slepton), and sneutrinos ($\susy{\nu}$).
Charginos and neutralinos are the mass eigenstates formed from linear superpositions of the superpartners of the
Higgs bosons and electroweak gauge bosons.
The charged sleptons are the superpartners of the charged leptons and are denoted by subscripts reflecting the chirality of the SM partner fields, referred to as \sleptonL or \sleptonR, respectively.
The slepton mass eigenstates are a mixture of \sleptonL and \sleptonR, and are labeled as $\susy{\ell}_{k}$ ($k = 1, 2$ in order of increasing mass).
In this work, the scalar superpartners of the left-handed $\tau$-lepton ($\stau_\mathrm{L}$) and right-handed $\tau$-lepton ($\stau_\mathrm{R}$) are assumed to be mass degenerate unless explicitly stated, and are referred to as ``staus.''
All SUSY scenarios considered in this publication conserve $R$-parity and the LSP is the $\ninoone$.
 
Although they are experimentally challenging, final states with $\tau$-leptons are of particular interest in SUSY searches.  Light sleptons could play a role in the co-annihilation of neutralinos in the early universe, and models with light stau decays into light neutralinos can shed light on the nature of dark matter~\cite{Vasquez:2011}. Furthermore, should SUSY or any other physics beyond the Standard Model (BSM) involving leptons be discovered, independent studies of all three lepton flavors are necessary to investigate the coupling structure of the new physics, especially concerning lepton universality.
 
\begin{figure}[!htb]
\centering
\subfloat[][]{\includegraphics[width=.3\textwidth]{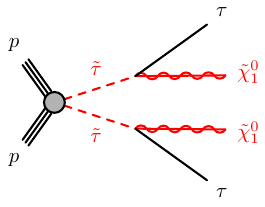}\label{fig:feynmanslepton_DS}}\hspace{0.5cm}
\subfloat[][]{\includegraphics[width=.3\textwidth]{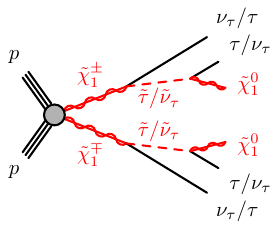}\label{fig:feynmanslepton_CCStau}}\\
\subfloat[][]{\includegraphics[width=.3\textwidth]{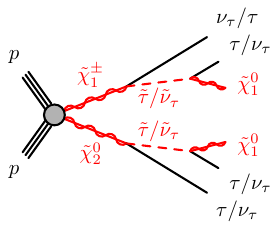}\label{fig:feynmanslepton_CNStau}}\hspace{0.5cm}
\subfloat[][]{\includegraphics[width=.3\textwidth]{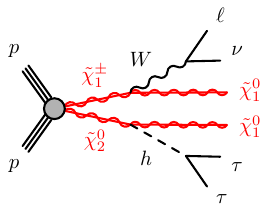}\label{fig:feynmanslepton_Wh}}
\caption{Representative diagrams of SUSY scenarios, which are being searched for in this publication. Direct stau production is shown in (a), while (b) and (c) show the processes considered for the Intermediate stau channel. The process for the Intermediate $Wh$ channel is shown in (d).  In all cases, the subsequent decays contain a final state with at least two $\tau$-leptons.  In the case of \ConepmNtwo production (c), the final state can contain more than two $\tau$-leptons.}
\label{fig:feynmanslepton_summary}
\end{figure}

The first SUSY scenario considered is the direct production of stau pairs, either with mass degenerate or non-degenerate $\stau_\mathrm{L,R}$, which decay directly into a $\tau$-lepton and the LSP 100\% of the time.
The second scenario, referred to as the ``Intermediate stau'' channel, includes the production of mass-degenerate neutralinos and charginos, \ConepmNtwo and \ConepConem, which decay into the lightest neutralino only through intermediate staus or $\tau$-sneutrinos with equal branching fraction. The stau and sneutrino masses are assumed to be halfway between the  $\ninotwo/\chinoonepm$ and $\ninoone$ masses. The search for \ConepmNtwo production is separated into final states with two same-sign (SS) or opposite-sign (OS) $\tau$-lepton pairs.
The third SUSY scenario, referred to as the ``Intermediate \textit{Wh}'' channel, is the direct production of mass-degenerate neutralinos and charginos, \ConepmNtwo,  which decay via the lightest neutral Higgs boson ($h$), consistent with the SM Higgs boson with a mass of $125\,\GeV$, a $W$ boson and two neutralinos. In this case, the final state under consideration contains two hadronically decaying $\tau$-leptons from the Higgs boson decay and one charged light lepton ($e$, $\mu$) from the $W$ boson decay. The light lepton may be the result of a $W\rightarrow\tau\nu$ decay where the $\tau$-lepton decays leptonically. For all three searches presented in this publication, the final state includes two hadronically decaying $\tau$-leptons, low jet activity and large missing transverse momentum, $\pTmiss$, from the neutralinos and neutrinos.
Representative diagrams of the targeted signal processes can be found in Figure~\ref{fig:feynmanslepton_summary}.
 
Previous results from the ATLAS experiment have set exclusion limits at 95\% confidence level on these SUSY models with the Run 1 and partial Run 2
data samples~\cite{SUSY-2013-14,SUSY-2014-05,SUSY-2016-23,SUSY-2018-04}. The CMS experiment also sets comparable exclusion limits~\cite{CMS-SUS-13-006,CMS-SUS-21-001,CMS-SUS-19-012}.
This search extends the current ATLAS reach to higher chargino masses and smaller mass differences between $\ninotwo$/$\chinoonepm/\stau$ and $\ninoone$ using the increased statistics of the full Run 2 data sample. It also achieves sensitivity to $\stau_R$ pair production scenarios for the first time from ATLAS, by introducing the use of machine learning algorithms.

The publication is structured as follows: the ATLAS detector is briefly introduced in Section~\ref{sec:detector} followed by a description of the data and simulated samples used in Section~\ref{sec:sample} and the reconstruction of events in Section~\ref{sec:EvenRec}; the general analysis strategy is outlined in Section~\ref{sec:strategy}, followed by the details of the Direct stau, Intermediate stau, and Intermediate $Wh$ searches in
Sections~\ref{sec:BDTstau}, \ref{sec:stau}, and \ref{sec:wh}, respectively; the systematic uncertainties are discussed in Section~\ref{sec:systematics} and the results are presented in Section~\ref{sec:Results}, before conclusions are drawn in Section~\ref{sec:conclusion}.

\section{ATLAS detector}
\label{sec:detector}
\newcommand{\AtlasCoordFootnote}{
ATLAS uses a right-handed coordinate system with its origin at the nominal interaction point (IP)
in the center of the detector and the \(z\)-axis along the beam pipe.
The \(x\)-axis points from the IP to the center of the LHC ring,
and the \(y\)-axis points upwards.
Cylindrical coordinates \((r,\phi)\) are used in the transverse plane,
\(\phi\) being the azimuthal angle around the \(z\)-axis.
The pseudorapidity is defined in terms of the polar angle \(\theta\) as \(\eta = -\ln \tan(\theta/2)\).
Angular distance is measured in units of \(\Delta R \equiv \sqrt{(\Delta\eta)^{2} + (\Delta\phi)^{2}}\).}

The ATLAS experiment~\cite{PERF-2007-01} at the Large Hadron Collider (LHC) is a multipurpose particle detector
with a forward--backward symmetric cylindrical geometry and a near \(4\pi\) coverage in
solid angle.\footnote{\AtlasCoordFootnote}
It consists of an inner tracking detector (ID) surrounded by a thin superconducting solenoid
providing a \SI{2}{\tesla} axial magnetic field, electromagnetic and hadron calorimeters, and a muon spectrometer.
The ID covers the pseudorapidity range \(|\eta| < 2.5\).
It consists of silicon pixel, silicon microstrip, and transition radiation tracking detectors.
Lead/liquid-argon (LAr) sampling calorimeters provide electromagnetic (EM) energy measurements
with high granularity.
A steel/scintillator-tile hadron calorimeter covers the central pseudorapidity range (\(|\eta| < 1.7\)).
The endcap and forward regions are instrumented with LAr calorimeters
for both the EM and hadronic energy measurements up to \(|\eta| = 4.9\).
The muon spectrometer (MS) surrounds the calorimeters and is based on
three large superconducting air-core toroidal magnets with eight coils each.
The field integral provided by the toroids ranges between \num{2.0} and \SI{6.0}{\tesla\metre}
across most of the detector.
The muon spectrometer includes a system of precision tracking chambers and fast detectors for triggering.
A two-level trigger system is used to select events~\cite{TRIG-2019-04}.
The first-level trigger is implemented in hardware and uses a subset of the detector information
to accept events at a rate below \SI{100}{\kHz}.
This is followed by a software-based trigger that
reduces the accepted event rate to \SI{1}{\kHz} on average
depending on the data-taking conditions.
An extensive software suite~\cite{ATL-SOFT-PUB-2021-001} is used in the reconstruction and analysis of real
and simulated data, in detector operations, and in the trigger and data acquisition systems of the experiment.

\section{Data and simulated event samples }
\label{sec:sample}
 
The data sample considered in this publication corresponds to 139\,fb$^{-1}$ of proton--proton ($pp$)
LHC collision data collected between 2015 and 2018 with the ATLAS detector, at a center-of-mass energy of $13\,\TeV$. 
Data quality requirements are imposed to ensure that only events
in which the entire ATLAS detector was functioning well are used~\cite{DAPR-2018-01}.

Simulated events produced with several Monte Carlo (MC) event generators are used to predict yields for background contributions from SM processes and for possible SUSY signals.
To account for pileup, all simulated events are overlaid with multiple $pp$ collisions simulated with the
soft QCD processes of \PYTHIA[8.186]~\cite{Sjostrand:2007gs} using the A3 set of tuned
parameters~\cite{ATL-PHYS-PUB-2016-017} and the \NNPDF[2.3lo] leading-order (LO)
parton distribution functions (PDFs)~\cite{Ball:2012cx}. For all samples showered with \PYTHIA[8], \EVTGEN[1.2.0]~\cite{EvtGen} is used to simulate the decays of bottom and charmed hadrons. The simulated events are weighted such that the pileup conditions match
those of the data and are required to satisfy the trigger selections. The response of the detector to particles is modeled with an ATLAS
detector simulation~\cite{SOFT-2010-01} based on \geant~\cite{Agostinelli:2002hh} for almost all SM background simulation. A few minor electroweak $Z$+jet processes with very small yields for these searches use a fast simulation based on a parameterisation of the performance of the
ATLAS EM and hadronic calorimeters~\cite{ATL-SOFT-PUB-2014-001} and
on \GEANT elsewhere.

Final states with at least two hadronically decaying $\tau$-leptons, low jet activity and a large $\pTmiss$ are included in this analysis.  As a result, SM background processes containing both real and misidentified $\tau$-lepton final state contributions are considered.  These backgrounds are summarised in the following paragraphs.
 
The production of top-quark pairs (\ttbar) and single top quarks in the $Wt$ and $s$-channels is performed with \POWHEGBOX[v2]~\cite{Frixione:2007nw,Nason:2004rx,Frixione:2007vw,Alioli:2010xd}, with the \NNPDF[2.3lo] PDF set at next-to-leading-order (NLO)
in the Matrix element (ME) calculations and the ATLAS underlying-event tune A14~\cite{ATL-PHYS-PUB-2014-021}.
Electroweak $t$-channel single-top-quark events are simulated using the \POWHEGBOX[v2] event generator.
The parton shower (PS), fragmentation, and the underlying event are simulated using \PYTHIA[8.186] with the \NNPDF[2.3lo] PDF set and a corresponding set of A14 tuned parameters.
The top-quark mass is set to $172.5\,\GeV$.
The \ttbar\ sample is normalized to the cross-section prediction at next-to-next-to-leading-order (NNLO)
in QCD including the resummation of next-to-next-to-leading-logarithmic (NNLL) soft-gluon terms calculated using
\TOPpp[2.0]~\cite{Beneke:2011mq,Cacciari:2011hy,Baernreuther:2012ws,Czakon:2012zr,Czakon:2012pz,Czakon:2013goa,Czakon:2011xx}.
The cross-section for single-top-quark is computed
for the $Wt$-channel at NLO in QCD with NNLL soft gluon
corrections~\cite{Aliev:2010zk,Kant:2014oha}, and to NLO in QCD for the $t$- and $s$-channels.
Top-quark pair production with an additional $W$ or $Z$ boson is calculated using \AMCatNLO[2.2.2]~\cite{Alwall:2014hca} at NLO in the ME calculations, while fragmentation and hadronization
are simulated with \PYTHIA[8.186].
The underlying-event tune A14 is used with the
\NNPDF[2.3lo] PDF set, and the cross-sections are normalized using NLO predictions~\cite{Lazopoulos:2008de,Campbell:2012dh}.
 
Events with $Z/\gamma^*\to\ell\ell$ $(\ell=e,\mu,\tau)$ and $W\to\ell\nu$ produced in association with jets (including jets initiated by heavy flavor quarks) are simulated with \SHERPA[2.2.1]~\cite{Gleisberg:2008ta,Hoeche:2012yf}.
MEs are calculated for up to two additional partons at NLO and four additional partons at LO, using the Comix~\cite{Gleisberg:2008fv} and \openloops~\cite{Cascioli:2011va,Denner:2016kdg} generators and merged with the \SHERPA PS~\cite{Schumann:2007mg} using the \MENLOPS prescription.
The \NNPDF[3.0nnlo]~\cite{Ball:2014uwa} PDF set is used in conjunction with a dedicated PS tuning developed by the \SHERPA authors.
The $W/Z$+jets events are normalized using their NNLO cross-sections~\cite{Catani:2009sm}.
 
Fully leptonically and semileptonically decaying diboson and tribsoson samples ($VV$ and $VVV$, where $V=W,Z$) are simulated with the
\SHERPA[2.2.1] and \SHERPA[2.2.2] generator at NLO\@. In this setup, multiple
matrix elements are matched and merged with the \SHERPA parton shower
based on Catani--Seymour dipole factorisation using the \MENLOPS
prescription~\cite{Catani:2001cc,Hoeche:2009rj}.
The virtual QCD corrections for matrix elements at NLO accuracy are
provided by the \OPENLOOPS library. Samples are simulated using the
\NNPDF[3.0nnlo] set, along with the dedicated set of
tuned parton-shower parameters developed by the \SHERPA authors.
 
Contributions from Higgs boson events produced by gluon--gluon fusion and vector-boson fusion are modeled using \POWHEGBOX[v2] with the \NNPDF[3.0nnlo] PDF and showered using \PYTHIA[8.186].  Associated production of a Higgs boson with a vector boson and a Higgs boson in association with two top quarks are simulated using \PYTHIA[8.186] and \AMCatNLO, respectively.  All Higgs boson samples are normalized to the cross-sections from Ref.~\cite{deFlorian:2016spz}.

SUSY signal model samples are simulated to allow the
interpretation of the search results in terms of SUSY parameters and use the ATLAS fast detector simulation. Signal samples are simulated using \AMCatNLO[2.2.3] interfaced to \PYTHIA[8.186] with the A14 tune for the PS modeling,
hadronization, and underlying event. The ME calculation is performed at tree level and includes the emission of up to two additional partons.
The PDF set used for the generation is \NNPDF[2.3lo]. The ME--PS merging uses the CKKW-L~\cite{CKKW} prescription, with a matching scale set to one quarter of the mass of the pair of produced particles.
Signal cross-sections are calculated with \resummino v2.0.1 to NLO in the strong
coupling constant, adding the resummation of soft gluon emission at
next-to-leading-logarithm accuracy (NLO+NLL)~\cite{Fuks:2012qx,Fuks:2013vua}.
The nominal cross-section and the uncertainty are taken from an envelope of cross-section predictions
using different PDF sets and factorisation and renormalization scales, as described in Ref.~\cite{Borschensky:2014cia}.

\section{Event reconstruction}
\label{sec:EvenRec}
 
After data quality requirements are applied, events with at least one reconstructed primary vertex~\cite{ATL-PHYS-PUB-2015-026} are selected. A primary vertex is defined to have
at least two associated charged-particle tracks with transverse momentum $\pt>500\,\MeV$ and be consistent with the beam spot envelope.
If there are multiple primary vertices in an event, the one with the largest $\sum\pt^2$ of the associated
tracks is chosen.
 
Jets are reconstructed from particle-flow objects calibrated at the EM scale~\cite{PERF-2015-09} using the anti-$k_t$ algorithm~\cite{Cacciari:2008gp,Cacciari:2011ma}
with a radius parameter of 0.4. Jet energies are corrected for detector inhomogeneities,
the non-compensating response of the calorimeter, and pileup effects~\cite{PERF-2011-03, ATL-PHYS-PUB-2015-015}.
The impact due to pileup is accounted for using a technique based on jet areas that provides an
event-by-event and jet-by-jet correction~\cite{jetarea}.
Jets that are likely to have originated from pileup are not considered in the analysis~\cite{PERF-2014-03}.
Jets are required to have $\pt>20\,\GeV$ and $\vert\eta\vert < 2.8$, and
events containing jets that are likely to have arisen from detector noise or cosmic rays are removed~\cite{ATLAS-CONF-2015-029}.

Jets containing $b$-hadrons (\bjets) are identified using the \emph{DL1r} algorithm~\cite{FTAG-2018-01,FTAG-2019-07}, a multivariate discriminant making use of track impact parameters and
reconstructed secondary vertices.  Candidate \bjets are required to have $|\eta| < 2.5$.  A working point is used that has a $\pt$-independent $b$-tagging efficiency of \percent{77} and light-jet ($c$-jet) rejection factor of 140 (4), based on studies using simulated \ttbar events.
 
Electron candidates are reconstructed by matching clusters in the EM calorimeter
with charged-particle tracks in the ID\@. Electrons are required to have $\pt>25\,\GeV$, $|\eta|<2.47$, and to satisfy the
``loose'' working point according to a likelihood-based identification algorithm detailed in Ref.~\cite{EGAM-2018-01}.
Muon candidates are reconstructed from MS tracks matching ID tracks; they are required to have $\pt > 25\,\GeV$ and $\vert\eta\vert < 2.7$ and satisfy the ``medium'' quality criteria described in Ref.~\cite{MUON-2018-03}.
Events containing a muon candidate with a poorly measured charge-to-momentum ratio of $\frac{\sigma(q/p)}{|q/p|}>0.2$ are rejected.
Events are required not to contain any candidate muon with large transverse ($d_0$) and longitudinal ($z_0$) impact parameter, $|z_0|>1$\,mm or $|d_0|>0.2$\,mm,
to reduce contributions from those originating from cosmic rays.
The efficiencies for electrons and muons to satisfy the reconstruction, identification, and isolation criteria are measured using samples of leptonic $Z$ and $J/\psi$ decays, and corrections are applied to the simulated samples to reproduce the efficiencies observed in data~\cite{EGAM-2018-01,MUON-2018-03}.
 
The reconstruction of hadronically decaying $\tau$-leptons is based on information from tracks in the ID and three-dimensional clusters in the EM and hadronic calorimeters.
The $\tau$-lepton reconstruction algorithm is seeded by jets reconstructed from topological clusters of energy deposits in the calorimeter and uses a looser requirement of $\pt>10\,\GeV$ and $|\eta|<2.5$.
The reconstructed energies of the hadronically decaying $\tau$-lepton candidates are corrected from the local hadron topocluster scale to the $\tau$-lepton energy scale, which is calibrated based on simulation and in situ measurements using $Z \rightarrow \tau \tau$ decays~\cite{ATLAS-CONF-2017-029}.
Hadronically decaying $\tau$-lepton candidates are required to have one or three associated charged-particle tracks (prongs) and the total electric charge of those tracks must be $\pm 1$ times the electron charge.
To improve the discrimination between hadronically decaying $\tau$-leptons and jets, electrons, or muons, multivariate algorithms are used~\cite{ATL-PHYS-PUB-2015-045}.
A recurrent neural network discriminant is used to reject jets that do not originate from a hadronically decaying $\tau$-lepton, with a ``medium'' or ``tight'' working point as described in Ref.\cite{ATL-PHYS-PUB-2019-033}.
The medium (tight) working point has an efficiency of \percent{75} (\percent{60}) for 1-prong $\tau$-leptons and \percent{60} (\percent{45}) for 3-prong $\tau$-leptons, based on $\gamma^*\rightarrow \tau\tau$ MC simulation.
The background rejection factor for the medium (tight) working point is 35 (70) for 1-prong $\tau$-leptons and 240 (700) for 3-prong $\tau$-leptons, based on dijet MC simulation.
A boosted decision tree is used to discriminate 1-prong $\tau$-leptons against electrons. This discriminant is built using information from the EM calorimeter and the tracking detector.  This requirement has about \percent{95} efficiency, and a rejection factor from 10 to 50 depending on the $\eta$ range.
The $\tau$-leptons are required to have $\pt > 20\,\GeV$ and $\vert\eta\vert < 2.47$, and must lie outside the transition region between the barrel and endcap calorimeters ($1.37 < |\eta| < 1.52$).
 
The MC simulation is corrected for differences in the efficiencies of the $\tau$-lepton identification at both trigger and reconstruction level between data and simulation.
For hadronically decaying $\tau$-leptons originating from prompt gauge boson decays, the corrections are calculated with a \textit{tag-and-probe} method in a sample of $Z \rightarrow \tau \tau$ events where one $\tau$-lepton decays hadronically and the other leptonically into a muon and two neutrinos~\cite{PERF-2013-06}.
 
The measured $\pTmiss$, and its magnitude $\met$, is defined as the negative vectorial sum of the $\mathbf{\pT}$ of all identified jets, $\tau$-leptons, electrons, photons, muons, and an additional soft term. The soft term is constructed from all tracks
that are associated with the primary vertex but not with any identified particle or jet~\cite{ATL-PHYS-PUB-2015-023,ATL-PHYS-PUB-2015-027}.
 
To avoid the possible double counting of reconstructed objects, an overlap removal procedure is used, following these steps.
The $\tau$-leptons that are close to electron or muon candidates ($\Delta R < 0.2$, where $\Delta R = \sqrt{(\Delta y)^2 + (\Delta\phi)^2}~$) are
removed, as are electrons that share a track with a muon.  For electrons close to a jet ($\Delta R < 0.4$), the electron is removed, except when $\Delta R < 0.2$ and the jet is not $b$-tagged, in which case the jet is removed.
Any remaining jet close to a muon or $\tau$-lepton ($\Delta R < 0.4$) is removed.

\section{General analysis strategy and event variables}
\label{sec:strategy}
 
Events for all scenarios are required to have at least two hadronically decaying $\tau$-leptons.  Different signal regions (SRs) are defined to target the specific SUSY scenario, using kinematic variables that provide a good signal-to-background separation, described in this section.   For $\tau$-leptons, kinematic variables are calculated from the visible decay products.  The event selections and background estimates are described for Direct stau production SRs in Section~\ref{sec:BDTstau}, for the Intermediate stau channel SRs in Section~\ref{sec:stau}, and for the Intermediate $Wh$ channel SRs in Section~\ref{sec:wh}.
 
The main SM backgrounds are estimated by normalizing MC simulation samples to data in dedicated control regions (CRs); backgrounds resulting from non-prompt and misidentified leptons are derived from data, while sub-dominant backgrounds are estimated by using MC simulation only.  To validate the modelling of the SM backgrounds, the yields and shapes of key kinematic variables are compared with data in dedicated validation regions (VRs). SRs are designed for the best expected sensitivity to the simplified SUSY signal models studied. Where appropriate, looser, or merged SRs are used to enhance discovery prospects or to set model-independent limits.

The observed number of events in the CRs and SRs are used in a combined profile likelihood fit to determine the expected SM background yields.
The statistical interpretation of the results is performed using the profile likelihood method implemented in the
HistFitter framework~\cite{Baak:2014wma}. 
Systematic uncertainties are included as nuisance parameters in the likelihood fits and are assumed to follow a Gaussian distribution with a width determined from the size of the uncertainty.
Correlations of systematic uncertainties between control and signal regions, and between background processes are taken into account with common nuisance
parameters. The fit parameters are determined by maximising the product of the Poisson probability functions and the constraints for
the nuisance parameters.
 
To constrain the SM backgrounds normalized to data in CRs, a \textit{background-only} fit is used. This uses the observed yields in the CRs with the expected SM contributions (other than multi-jet) in the CRs and SRs, and the corresponding background transfer factors described in Section~\ref{sec:BG_QCD}.
The free parameters in the fit are the normalizations of the SM processes.
To estimate the SM background in the SRs (or VRs), a combined fit is performed that includes the observed data in the CRs and SRs (or VRs) with the SM background estimates, and accounts for correlations between CRs and SRs (or VRs), as well as between background processes.
The presence of BSM signal in any CR, SR, or VR is assumed to be zero in the combined and background-only fits.
 
To assess the possibility of the presence of any BSM events in each SR, a \textit{model-independent} fit is used. This uses the observed yields in both the CRs and SRs with the SM background estimates to test whether any BSM events could be present in the SR\@.
The presence of BSM signal in any CR is assumed to be zero in the model-independent fit.
The significance of a possible excess of observed events over the SM prediction is quantified by the one-sided probability, $p (\text{signal} = 0)$ denoted by $p_0$, of the background alone to fluctuate to the observed number of events or higher using the asymptotic formula described in Ref.~\cite{statforumlimitsl}.
The upper limit on the visible cross-section of BSM events in an SR, $\sigma_{\mathrm{vis}}^{95}$, is also calculated, which includes the acceptance and efficiency effect of any BSM signal possibly present.
 
Finally, to assess the compatibility of the signal scenarios with the data observation, a \textit{model-dependent} fit is used, which accounts for the SUSY signal in all CRs and SRs scaled by a floating signal normalization factor.
The background normalization factors are also determined simultaneously in the fit.
A SUSY scenario is rejected if the upper limit at \percent{95} confidence level (CL) of the signal normalization factor obtained in this fit is smaller than the predicted cross-section of the scenario ~\cite{CLs:2002}.

The following variables are used to discriminate SUSY signals from the SM background:\label{sec:variables}
 
\begin{itemize}
\item $\mt$, the transverse mass values obtained from a reconstructed lepton with the
$\pTmiss$, defined by
\[
\mt(\pTvec,\pTmiss) = \sqrt{2(|\pTvec||\pTmiss|-\pTvec\cdot\pTmiss)},
\]
where $\pTvec$ is the transverse momenta of the lepton. The transverse mass calculated using light leptons only ($e$ or $\mu$) is denoted by $m_{{\mathrm T},\ell}$.
 
\item $m_{{\mathrm T}\mathrm{sum}}$, the
sum of the  transverse mass values of the two highest-$\pt$ $\tau$-lepton candidates with the $\pTmiss$.
For Intermediate $Wh$ scenarios, $m_{{\mathrm T},\ell}$ is also added to $m_{{\mathrm T}\mathrm{sum}}$.
 
\item $\mttwo$, the ``stransverse mass,'' which has a kinematic endpoint for events  where two massive particles are pair produced, then each particle decays into a detected object (the lepton) and an undetected object (the neutralino) \cite{Lester:1999tx,Barr:2003rg}.
It is defined as:
\[
\mttwo = \min_{\qTvec}\left[\max\left(m_{{\mathrm T}}(\mathbf{p}_{\mathrm{T}1},\qTvec),m_{{\mathrm T}}(\mathbf{p}_{\mathrm{T}2},\pTmiss-\qTvec)\right)\right],
\]
where $\mathbf{p}_{\mathrm{T}1}$ and $\mathbf{p}_{\mathrm{T}2}$ are the transverse momenta of the two leptons and $\qTvec$ is the transverse vector chosen to minimise the larger of the two transverse masses.
In this case, $\qTvec$ or $\pTmiss-\qTvec$ replaces $\pTmiss$ in the calculation of $\mt$.
In events with more than two $\tau$-lepton candidates, the pair that maximises $\mttwo$ is used.
Similarly, in the Intermediate $Wh$ analysis, the pairing of $e\tau$, $\mu\tau$, or $\tau\tau$ that maximises $\mttwo$ is used.
For $\ttbar$ and $WW$ events, the $\mttwo$ distribution has a kinematic endpoint at the $W$ boson mass, but the distribution for SUSY scenarios extends significantly beyond this endpoint if the mass difference between the produced SUSY particle and the $\ninoone$ is large.
The $\ninoone$ is assumed to be massless in the calculation of $\mttwo$.
 
\item $\meff$, the effective mass, is defined as the scalar sum of the $\met$ and the $\pt$ of the two highest-$\pt$ $\tau$-leptons. The effective mass for SUSY processes is typically higher than for SM processes.
\item $\Delta R(\tau_{1},\tau_{2}) =\sqrt{(\Delta \eta(\tau_{1},\tau_{2})^{2}+\Delta\phi(\tau_{1},\tau_{2})^{2}}$, the angular distance between the two highest-$\pt$ $\tau$-leptons. An upper requirement on this variable is used to discriminate against back-to-back objects in SM events.
\item $m(\tau_{1},\tau_{2})$: the invariant mass of the two highest-$\pt$ $\tau$-leptons. A similar variable is also used for the invariant mass of a $\tau$-lepton and a muon, $m(\tau,\mu)$.
\item  $|\Delta\phi(\tau_1,\tau_{2})|$: the absolute value of the difference of azimuthal angle around the \(z\)-axis between the two highest-$\pt$ $\tau$-lepton candidates. A similar variable is also used for the $\pTmiss$ and $\tau$-leptons, e.g.,  $|\Delta\phi(\tau,\pTmiss)|$.
\item  $|\Delta\eta(\tau_1,\tau_{2})|$: the absolute value of the difference of pseudorapidity between the two highest-$\pt$ $\tau$-leptons. 
 
\end{itemize}

\section{Direct stau production}
\label{sec:BDTstau}
The Direct stau analysis targets the production of $\stauL\stauL$ and $\stauR\stauR$, with the stau decaying into a $\tau$-lepton and a $\ninoone$, as shown in Figure~\ref{fig:feynmanslepton_DS}. This analysis aims to improve upon previous results, particularly for moderate mass splittings between the stau and $\ninoone$, and for $\stau_R\stau_R$ production. Multiple Boosted Decision Trees (BDTs) are trained for sensitivity to different areas of the $\stau-\ninoone$ phase space. The event selection and BDT training are described in Section~\ref{sec:stauEvSel}, followed by the background estimate in Section~\ref{sec:stauBgEst}.

\subsection{Event selection}
\label{sec:stauEvSel}
 
Events are selected using an asymmetric di-$\tau$ trigger, where the first $\tau$-lepton must satisfy $\pt> 95\,\GeV$ and the second $\tau$-lepton must satisfy $\pt > 60\, (75)\,\GeV$ for 2015--2017 (2018) data-taking.
The $\tau$-lepton $\pt$ thresholds applied are to ensure the trigger efficiencies are in the plateau region.

Four BDTs are trained using the LightGBM~\cite{LightGBM} package on four groupings of signal scenarios chosen for their different $\stau$ masses and $\Delta m(\stau,\ninoone)$.
Events used in training have at least two OS medium $\tau$-leptons, no electron, muon, or $b$-tagged jet candidates and satisfy loose selections on kinematic variables, as shown in Table~\ref{tab:stauSel}.
Three BDT SRs (SR-BDT1, SR-BDT2, SR-BDT3) have two bins in their BDT score, while the fourth BDT has only one bin (SR-BDT4) due to the low statistics available.
The binned regions are used for setting exclusion limits in simplified models, while inclusive regions that merge each of the two bins in the binned regions are used to enhance discovery prospects and set model-independent limits.
The single-bin region, SR-BDT4, is unchanged when used for setting exclusion limits and model-independent limits.
The events selected by the four BDT SRs significantly overlap, thus the SR with the best expected sensitivity is chosen for interpretations.
 
The BDT training uses inputs including $\met$, the $\pt$, and $\mt$ of the two $\tau$-leptons, $\Delta \phi(\tau_1,\pTmiss)$, $\Delta \phi(\tau_2,\pTmiss)$, $\Delta \eta(\tau_1, \tau_2)$, $m(\tau_1,\tau_2)$, $\meff$, and $m_{{\mathrm T}\mathrm{sum}}$.
Cross-validation is used with three equal folds for a training, validation, and test set.
To improve the reliability of the background modelling during training, input variables are binned  such that the statistical error on the MC simulated background is less than 30\% per bin.
Furthermore, the maximum depth of the individual decision trees is restricted to one so that each tree makes only one selection on a variable.
To avoid over-fitting, early-stopping is used, where trees cease to be added during training if performance on the relevant validation set does not improve after ten trees.
The fitting procedure is further constrained to avoid over-fitting by using regularisation techniques, where BDT2 and BDT4 use a higher amount of regularisation than BDT1 and BDT3, and consequently, BDT2 and BDT4 occupy a smaller fraction of the total BDT score range compared with the others.
 
\begin{table}[!h]
\centering
\caption{Summary of the selection requirements for the stau pair production SRs.
\label{tab:stauSel}}
\footnotesize
\begin{tabular}{l| c|c|c|c|c|c|c}
\toprule
& \multicolumn{7}{c}{BDT Training Preselection} \\
\midrule
$N$ medium $\tau$ & \multicolumn{7}{c}{$\geq$ 2 }\\
Charge combination & \multicolumn{7}{c}{OS }\\
Trigger & \multicolumn{7}{c}{asymm. di-$\tau$} \\
$N$ $e/\mu$ & \multicolumn{7}{c}{$=0$ }\\
$N$ $b$-jets & \multicolumn{7}{c}{$=0$}       \\
$\met$ [$\GeV$] & \multicolumn{7}{c}{$ > 20$ }         \\
$\mttwo$ [$\GeV$] & \multicolumn{7}{c}{$ > 30$ }         \\
$m(\tau_1,\tau_2)$ [$\GeV$] & \multicolumn{7}{c}{ $> 120$ }         \\
$\Delta R(\tau_1, \tau_2)$ & \multicolumn{7}{c}{$ < 4$ }         \\
\midrule  \midrule
& \multicolumn{2}{c|}{SR-BDT1} & \multicolumn{2}{c|}{SR-BDT2} & \multicolumn{2}{c|}{SR-BDT3} & SR-BDT4 \\
& Bin 1 & Bin 2 & Bin 1 & Bin 2 & Bin 1 & Bin 2 & \\
\midrule  \midrule
Target & \multicolumn{2}{c|}{Low $m(\stau)$} & \multicolumn{2}{c|}{Mid $m(\stau)$} & \multicolumn{2}{c|}{Mid $m(\stau)$} & High $m(\stau)$ \\
scenario & \multicolumn{2}{c|}{Small $\Delta m(\stau,\ninoone)$} &  \multicolumn{2}{c|}{Large $\Delta m(\stau,\ninoone)$} &  \multicolumn{2}{c|}{Small $\Delta m(\stau,\ninoone)$} &   \\
\midrule
$N$ medium $\tau$ & \multicolumn{7}{c}{$=2$}\\
BDT1 score & $\in(0.73,0.78)$ & $>0.78$ & -- & -- & -- & -- & -- \\
BDT2 score & -- & -- & $\in(0.78,0.82)$ & $>0.82$ & -- & -- & -- \\
BDT3 score & -- & -- & -- & -- & $\in(0.79,0.86)$ & $>0.86$ & -- \\
BDT4 score & -- & -- & -- & -- & -- & -- & $>0.64$ \\
\bottomrule
\end{tabular}
\end{table}
 
\subsection{Background estimate}
\label{sec:stauBgEst}
 
The main backgrounds contributing to the Direct stau SRs are from
multi-jet production with misidentified $\tau$-leptons, $W$ and $Z$ boson production in association with
jets, multi-boson production, and events containing at least one top quark, referred to as \textit{Top} background.
Top quark background events originate mostly from \ttbar~production in
association with additional jets or an additional  $W$ or $Z$ boson.
The dominant SM backgrounds, $W$+jets, $Z$+jets, and top quark processes, are estimated by using MC simulation normalized to data in dedicated control regions.
The multi-jet background yield is estimated from data using the so-called ABCD method, described below, and MC simulation is used for all other minor SM backgrounds.
The normalization factors obtained from control regions are applied for backgrounds in the BDT training, as is the data-driven multi-jet estimate.

\subsubsection{Multi-jet background estimate}
\label{sec:BG_QCD}
Background events contain a combination of `real' $\tau$-leptons, defined as
correctly identified $\tau$-leptons, or `misidentified' $\tau$-leptons, which can originate from a
misidentified light-flavor quark or gluon jet, an electron, or a muon.
Selected events from multi-jet production contain mostly misidentified $\tau$-leptons originating from misidentified jets.
 
\begin{figure}[!h]
\centering
\includegraphics[width=0.7\textwidth]{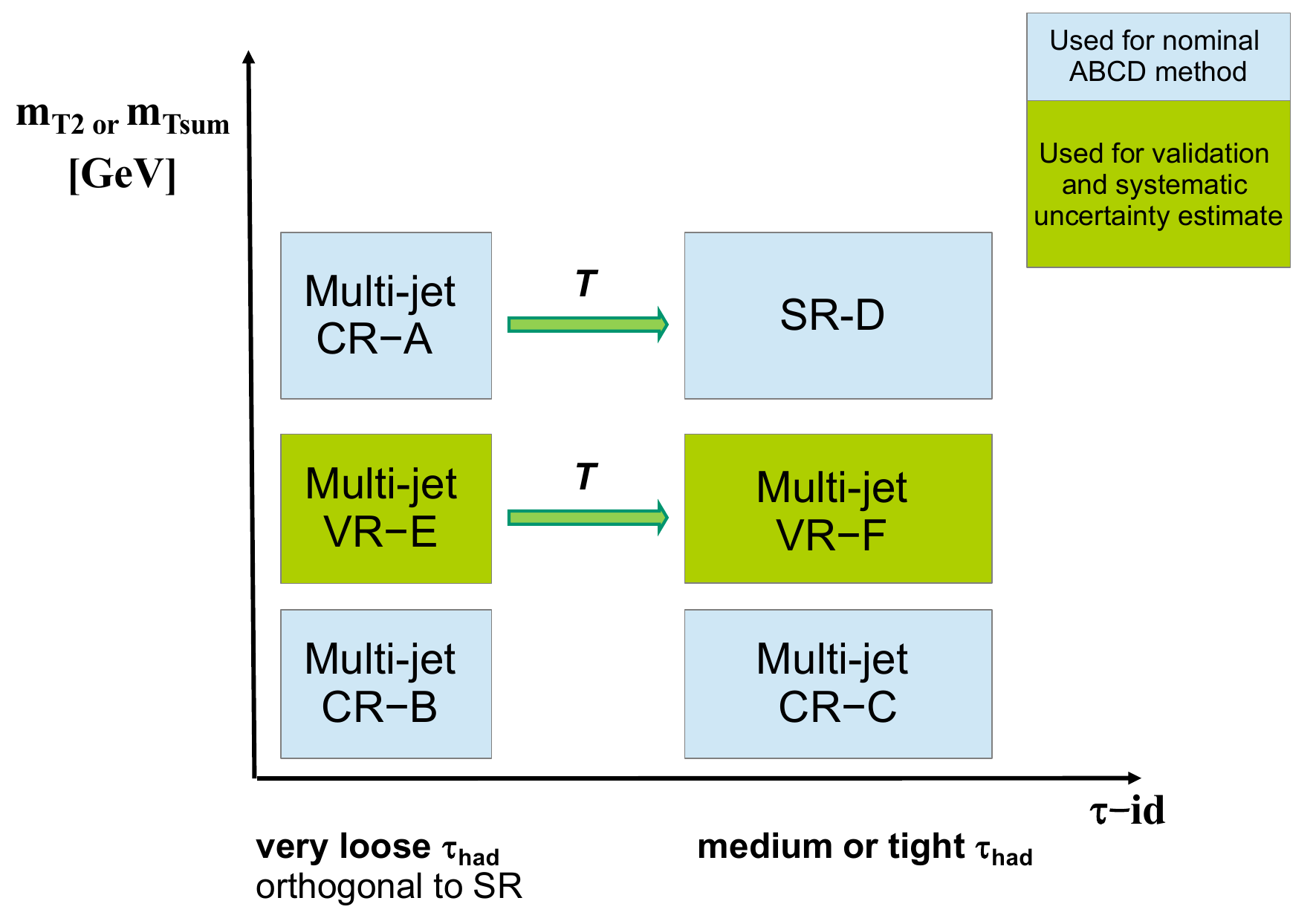}
\caption{ Illustration of the ABCD method for the multi-jet background determination.
The control regions A, B, and C and signal region D are drawn as light blue boxes.
Shown in green and labeled as VR are the regions E and F,
which are used to validate the ABCD method and to estimate the
systematic uncertainty.
}
\label{fig:abcdDS}
\end{figure}

The multi-jet contribution is estimated by using a data-driven ABCD method, which is common to both the Direct stau channel and the Intermediate stau channel described in Section~\ref{sec:stau}.
Orthogonal regions are defined using two approximately uncorrelated discriminating variables, as shown schematically in Figure~\ref{fig:abcdDS}.
The number of multi-jet events in region D, $N_{\mathrm D}$, can
be calculated from that in region A, $N_{\mathrm A}$,
multiplied by the transfer factor $\mathrm{T}=N_{\mathrm C}/N_{\mathrm B}$, where region D corresponds to one of the SRs and regions A--C are CRs.
Intermediate VRs are defined to verify the reliability of the transfer factor obtained from the ABCD estimate and to estimate the systematic uncertainty from the residual correlation between the discriminating variables.
Non-multi-jet SM contributions are subtracted from data in each region using MC simulation, and in the case of the Direct stau channel, the $W$+jets, $Z$+jets, and top quark backgrounds are subtracted after their normalization to data in their dedicated CRs.
 
The discriminating variables used here are the $\tau$-lepton identification (``very loose'' $\tau$-leptons as described in Ref.\cite{ATL-PHYS-PUB-2019-033}, that fail to meet either the medium criterion or OS requirement) and $\mttwo$ ($10<\mttwo<20\,\GeV$ for regions B and C, $20<\mttwo<30\,\GeV$ for regions E and F, $\mttwo>30\,\GeV$ for A and D).
The $\tau$-lepton $\pt$ thresholds to ensure high triggering efficiency are not applied, effectively lowering the $\tau$-lepton $\pt$ thresholds to the online thresholds of $80\,\GeV$ and $50\,\GeV$ ($60\,\GeV$) for 2015--2017 (2018) data-taking. This increases the statistics available and reduces the statistical uncertainty without affecting the transfer factor within statistical uncertainties.

\begin{figure}[htpb!] 
\centering
\subfloat[][]{\includegraphics[width=0.49\textwidth]{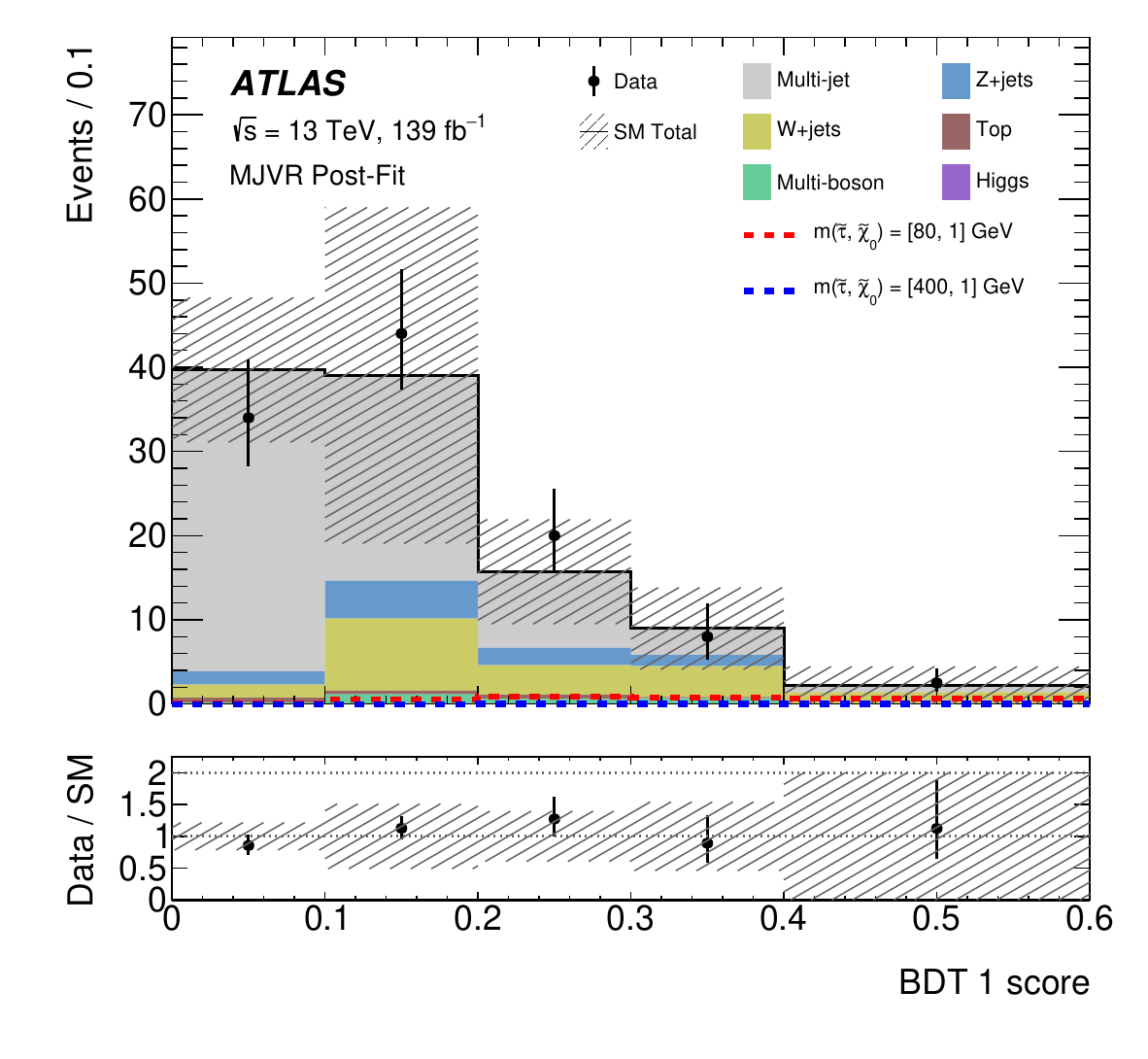}}
\subfloat[][]{\includegraphics[width=0.49\textwidth]{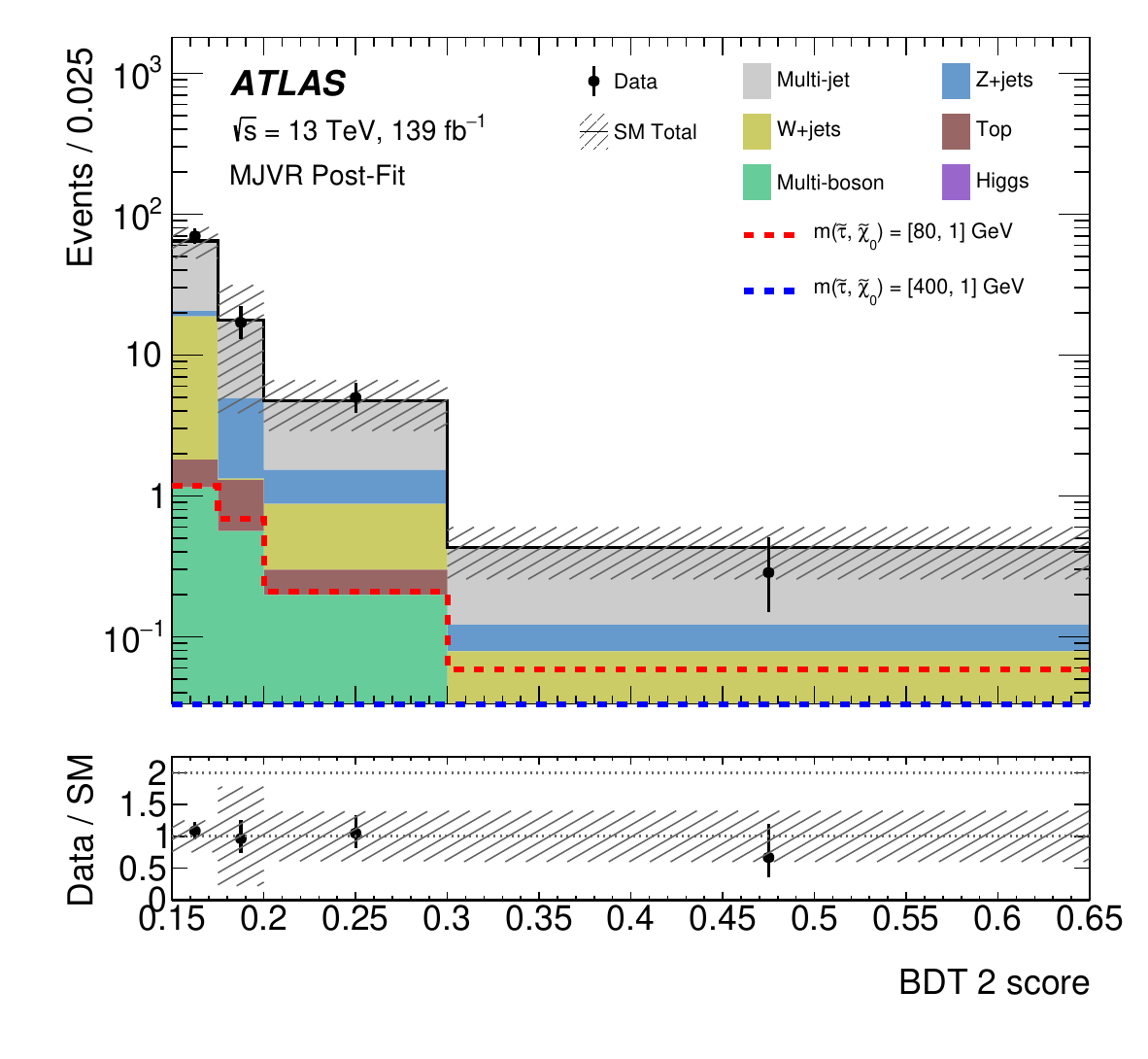}}\\
\subfloat[][]{\includegraphics[width=0.49\textwidth]{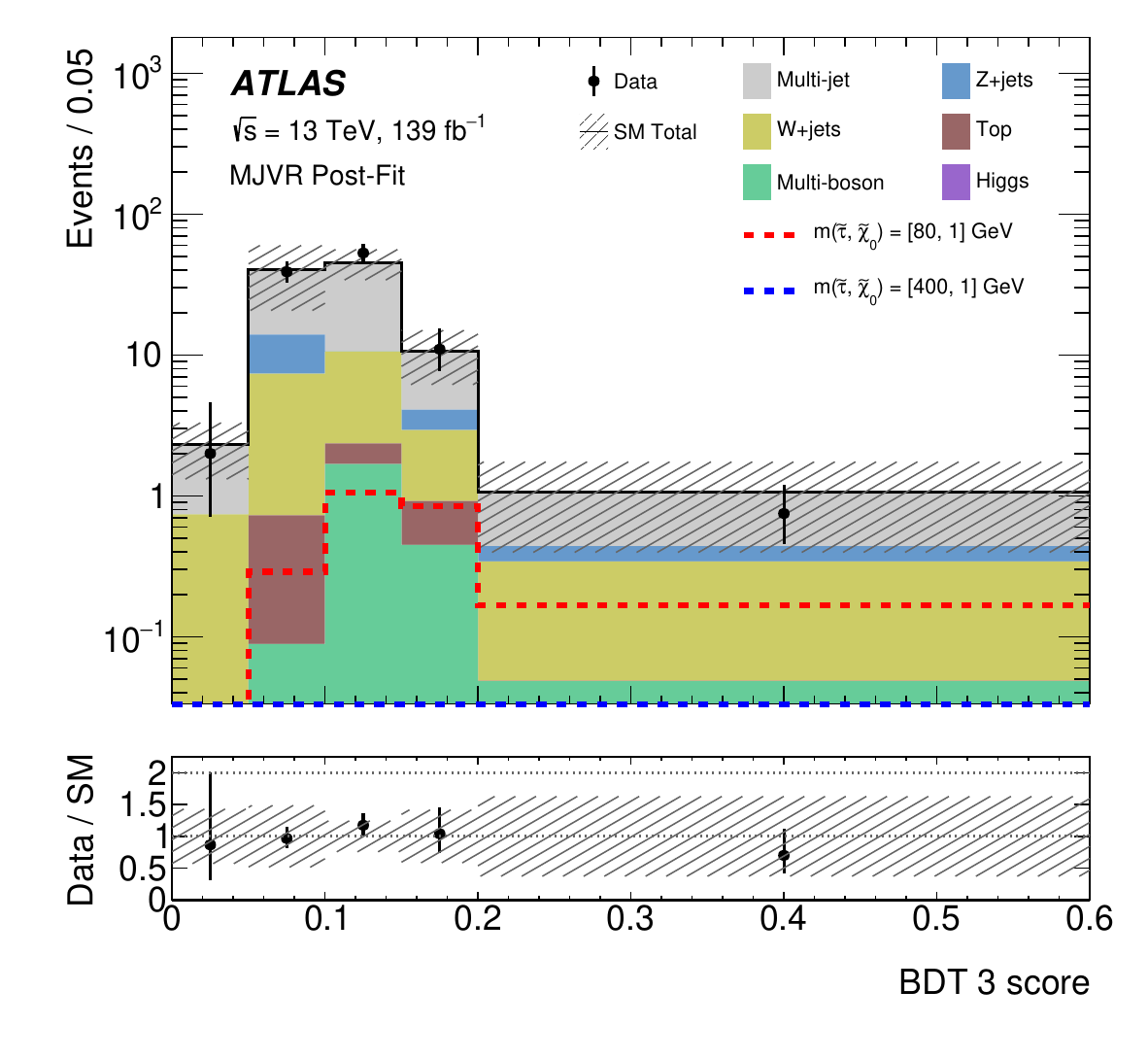}}
\subfloat[][]{\includegraphics[width=0.49\textwidth]{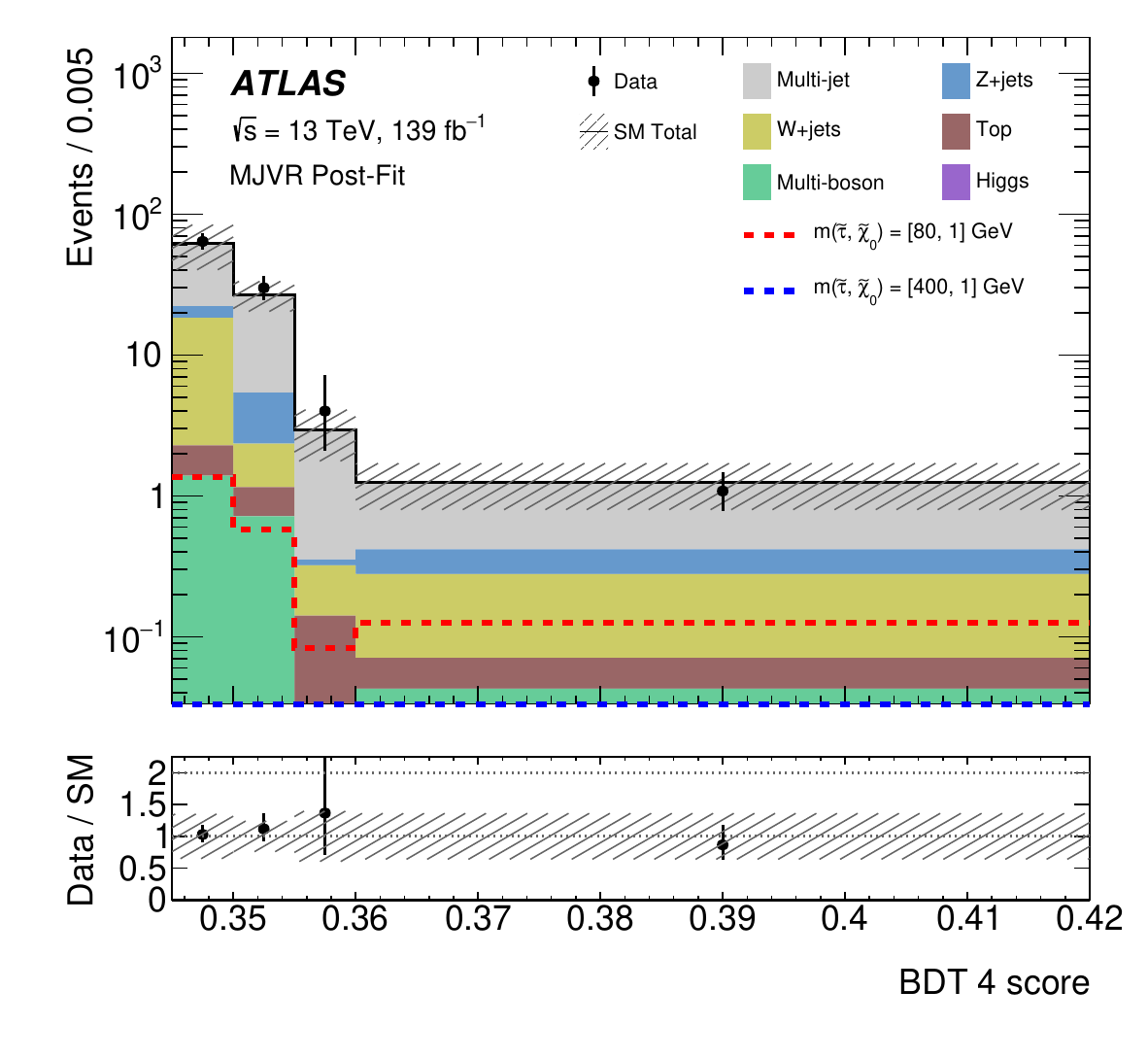}}\\
\caption{The BDT score distributions for the Direct stau multi-jet background in MJVR after the background-only fit, showing the scores for (a) BDT1, (b) BDT2, (c) BDT3, and (d) BDT4.
A reduced range of BDT2 in (b) and BDT4 score in (d) are shown as they populate only a narrow range of the full score allowed.
Representative SUSY scenarios are overlaid for illustration.
The hatched bands represent the combined statistical and systematic uncertainties of the total SM background.
The lower panels show the ratio of data to the total SM background estimate.
}
\label{fig:DS-MJVR}
\end{figure}

Agreement between data and the estimated SM background is found in VR-F, within statistical uncertainties.
A validation region (MJVR), with higher multi-jet purity and with the $\tau$-lepton $\pt$ thresholds applied (see Table~\ref{tab:AllDScrvr_defDS}), is also checked.
Good agreement is seen, as shown in Figure~\ref{fig:DS-MJVR}, and a discussion of the systematic uncertainties of the total SM background can be seen in Section~\ref{sec:systematics}.
The correlation between the $\tau$-lepton identification and the kinematic variables is verified by studying the variation of the
transfer factor as a function of the kinematic variables used and is found to be negligible.

\begin{table}[!h]
\centering
\caption{The definition of the control and validation regions for the Direct stau channel.
}
\label{tab:AllDScrvr_defDS}
\footnotesize
\begin{tabular}{l | c|c| c|c }
\toprule
Process & \multicolumn{2}{c|}{$W$+jets} &  \multicolumn{2}{c}{Top} \\
\midrule
Region & $W$CR                 & $W$VR & $T$CR                 & $T$VR \\
\midrule     \midrule
Charge combination & \multicolumn{4}{c}{OS}\\
\midrule
$N$ medium $\tau$ &\multicolumn{2}{c|}{$=1$} & $=1$ & $=2$ \\
$N$ tight $\tau$ &\multicolumn{2}{c|}{$=0$} & -- & -- \\
$N$ $e$/$\mu$ &\multicolumn{2}{c|}{$=1$ $\mu$} & $=1$ $\mu$ & $=0$\\
$N$ $b$-jets & \multicolumn{2}{c|}{$=0$ } & $\geq 1$ & $\geq 1$ \\
Trigger & \multicolumn{2}{c|}{single $\mu$} & single $\mu$ & asymm. di-$\tau$ \\
$\pt(\tau_1)$ [$\GeV$] & \multicolumn{2}{c|}{-} & -- & $>95$ \\
$\pt(\tau_2)$ [$\GeV$] & \multicolumn{2}{c|}{-} & -- & $>65$ \\
max$[\pt(\tau),\pt(\mu)]$ [$\GeV$] &  \multicolumn{2}{c|}{$>95$} & $>95$ & -- \\
min$[\pt(\tau),\pt(\mu)]$ [$\GeV$] &  \multicolumn{2}{c|}{$>60$} & $>60$ & -- \\
$\met$ [$\GeV$] & \multicolumn{2}{c|}{$> 20$ } & $ > 20$ & $ > 20$ \\
$m_{{\mathrm T}, \mu}$ [$\GeV$] & \multicolumn{2}{c|}{$\in(50,150]$ } & $\in(50,150]$ & -- \\
$\mttwo$ [$\GeV$] & \multicolumn{2}{c|}{$> 30$ } & $> 30$ & $>30$ \\
$m(\tau, \mu)$ [$\GeV$] & \multicolumn{2}{c|}{$ > 120$ } & $> 120$ & --- \\
$m(\tau_1, \tau_2)$ [$\GeV$] &\multicolumn{2}{c|}{ -- } & -- & $> 120$ \\
\midrule
BDT score & All $<0.5$ & Any $>0.5$ & \multicolumn{2}{c}{-} \\
\bottomrule
\end{tabular}
\\
\vspace{0.15cm}
 
\begin{tabular}{l | c|c |c|c|c }
\toprule
Process & \multicolumn{2}{c|}{$Z$+jets} & Multi-boson & Multi-jet & Inclusive \\
\midrule
Region &  $Z$CR                 & $Z$VR  & MBVR    & MJVR             & InclVR \\
\midrule      \midrule
Charge combination & \multicolumn{5}{c}{OS}\\
\midrule
$N$ medium $\tau$ & \multicolumn{2}{c|}{$=2$} & $=2$ & $=2$ & $=2$\\
$N$ $e$/$\mu$ & \multicolumn{2}{c|}{$=0$} & $=0$& $=0$& $=0$ \\
$N$ $b$-jets & \multicolumn{2}{c|}{$=0$ } & $=0$& $=0$ & $=0$ \\
Trigger & \multicolumn{2}{c|}{asymm. di-$\tau$} & asymm. di-$\tau$ & asymm. di-$\tau$ & asymm. di-$\tau$\\
$\met$ [$\GeV$] & \multicolumn{2}{c|}{-} & $>20$ & $<50$ & $>20$\\
$\mttwo$ [$\GeV$] & \multicolumn{2}{c|}{$> 30$ } & $> 30$ & $> 30$ & $> 30$\\
$m(\tau_1, \tau_2)$ [$\GeV$] & \multicolumn{2}{c|}{$ < 120$ } & $<120$& $>120$  & $>120$ \\
$\Delta R(\tau_1, \tau_2)$ & \multicolumn{2}{c|}{$ < 4$ } & $<4$ & $>3$ & $<4$\\
\midrule
BDT1 score & $\leq 0.10$ & $>0.10$ & -- & \multicolumn{2}{c}{$\leq 0.60$}\\
BDT2 score & -- & -- & -- & \multicolumn{2}{c}{$\leq 0.70$}\\
BDT3 score & -- & -- & -- & \multicolumn{2}{c}{$\leq 0.70$}\\
BDT4 score & $\leq 0.60$ & $\leq 0.60$ & $> 0.61$ & \multicolumn{2}{c}{$\leq 0.60$}\\
\bottomrule
\end{tabular}
\end{table}
 
The signal contamination in a certain region is defined as the ratio of the number of signal events to the sum of the number of signal events and SM background processes.
A significant signal contamination in SM CRs can affect the model-dependent fit and reduce the sensitivity to SUSY scenarios.
Thus the CRs (and VRs) are designed to have negligible signal contamination for the SUSY scenarios considered in this paper, particularly for the SUSY particle mass ranges not excluded by previous searches.

\subsubsection{$W$+jets background estimate}
\label{sec:BG_W}
 
The production of $W$+jets events with at least one misidentified $\tau$-lepton is an important background, composing up to 50\% of the expected SM background in the Direct stau SRs.
To correct the misidentified $\tau$-lepton MC modelling and reduce the theoretical uncertainty from the $W$+jets background, dedicated control regions pure in $W$ bosons decaying into $\mu\nu$ ($W$CR) are used to normalize the $W$+jets MC estimate to data.
 
Events are required to satisfy a single-muon trigger with increasing $\pT$ thresholds of $21-27\,\GeV$ throughout data-taking to ensure the trigger efficiencies are in the plateau region~\cite{TRIG-2018-01}.
Events must contain exactly one isolated $\mu$ matching the trigger signature and one $\tau$-lepton with the same charge selection as the signal regions.
The $\tau$-lepton is required to satisfy the medium ID criterion, but fail to meet the tight criterion, to allow future searches targeting direct stau production using one leptonically decaying $\tau$-lepton and one hadronically decaying $\tau$-lepton.
The contribution from events with top quarks is suppressed by rejecting events containing $b$-tagged jets. The contributions from $Z$+jets, top-quark and multi-boson production are reduced with $\mt$ requirements.
 
Thresholds on $\met$ and lepton $\pt$ are applied to select events with kinematics similar to the SR definition.
Events in the $W$CR are selected by requiring all four BDT scores to be low, while the modelling is checked using a $W$VR with any one of the four BDT scores being high.
The definitions of the $W$CR and $W$VR are given in Table~\ref{tab:AllDScrvr_defDS}.
The purity of the selection in $W$+jets events is around $83$\% in the control and validation regions. The signal contamination in the $W$CR and $W$VR is negligible due to the requirement of an isolated muon in these regions.

\subsubsection{$Z$+jets background estimate}
\label{sec:BG_ZDS}
The production of $Z$+jets where the $Z$ boson decays into two $\tau$-leptons is an irreducible background in the Direct stau SRs, composing $7-50$\% of the total SM background.
The $Z$+jets background is estimated by using MC simulation normalized to data in a control region, $Z$CR, which is similar to the SRs, but reverses the selection on $m(\tau_1, \tau_2)$.
Events with a low BDT4 score are selected to avoid overlap with the validation region for multi-bosons (Section~\ref{sec:BG_irreDS}). A selection on the score for BDT1 is used to create an orthogonal validation region $Z$VR to check the modelling of the normalized $Z$+jets background. The purity of the selection in $Z$+jets events is around $85-90$\% in the control and validation regions.
The definitions of the $Z$CR and $Z$VR are shown in Table~\ref{tab:AllDScrvr_defDS}.
 
\subsubsection{Top quark background estimate}
\label{sec:BG_topDS}
Events containing one or more top quarks also contribute significantly to the total background in the Direct stau SRs at $7-30$\%.
The top quark background is estimated by using MC simulation normalized to data in a control region, $T$CR, which,
like the $W$CR, contains one isolated muon and one medium $\tau$-lepton, but additionally requires a $b$-tagged jet.
A validation region, $T$VR, is used to check the modelling of the top quark background in events closer to the signal regions with two medium $\tau$-leptons and a $b$-tagged jet.
The purity of the selection in top quark events is around $80-92$\% in the control and validation regions.
The definitions of the $T$CR and $T$VR are shown in Table~\ref{tab:AllDScrvr_defDS}.

\subsubsection{Multiboson background estimate}
\label{sec:BG_irreDS}
\begin{figure}[h!]  
\centering
\subfloat[][]{\includegraphics[width=0.49\textwidth]{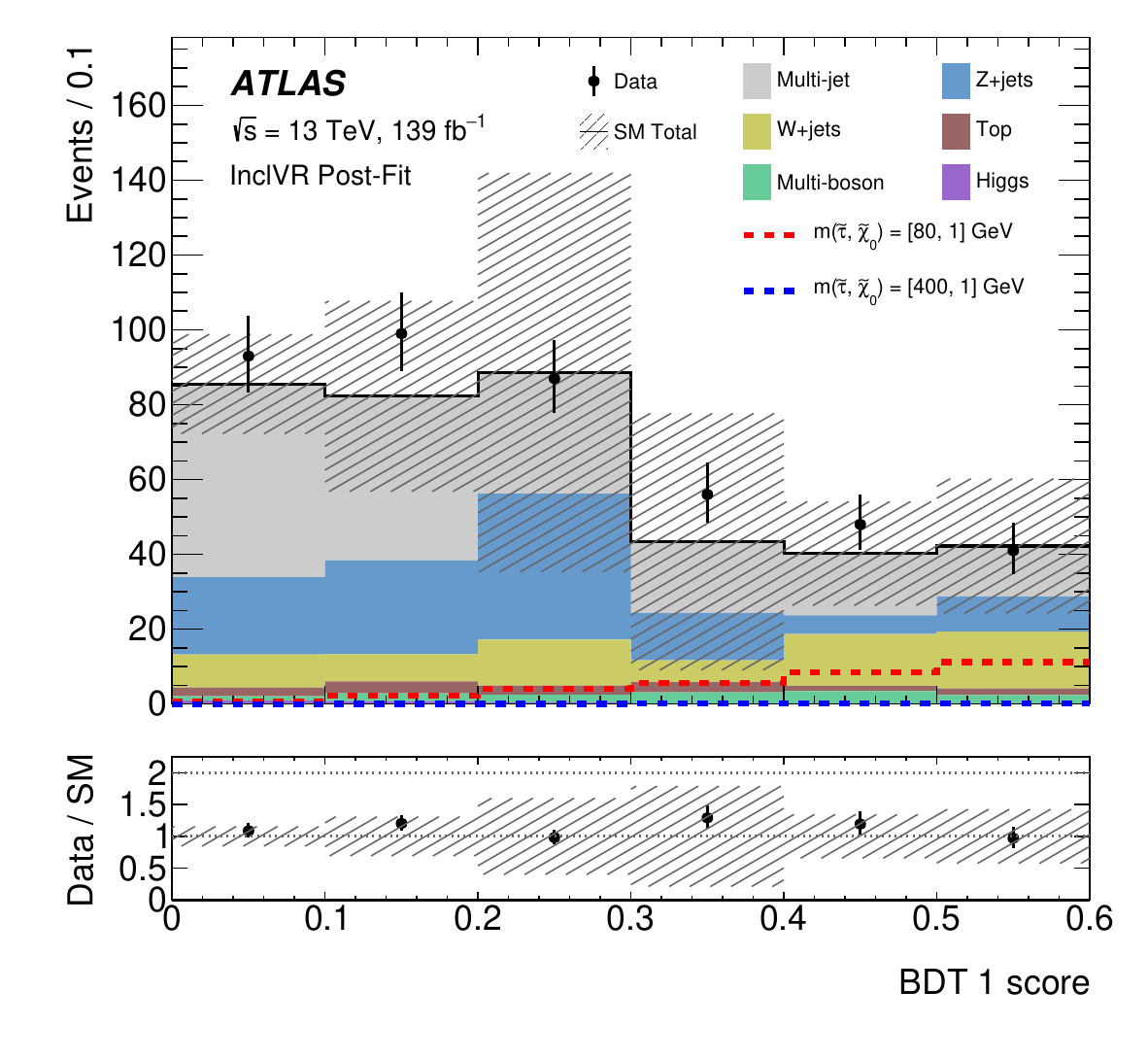}}
\subfloat[][]{\includegraphics[width=0.49\textwidth]{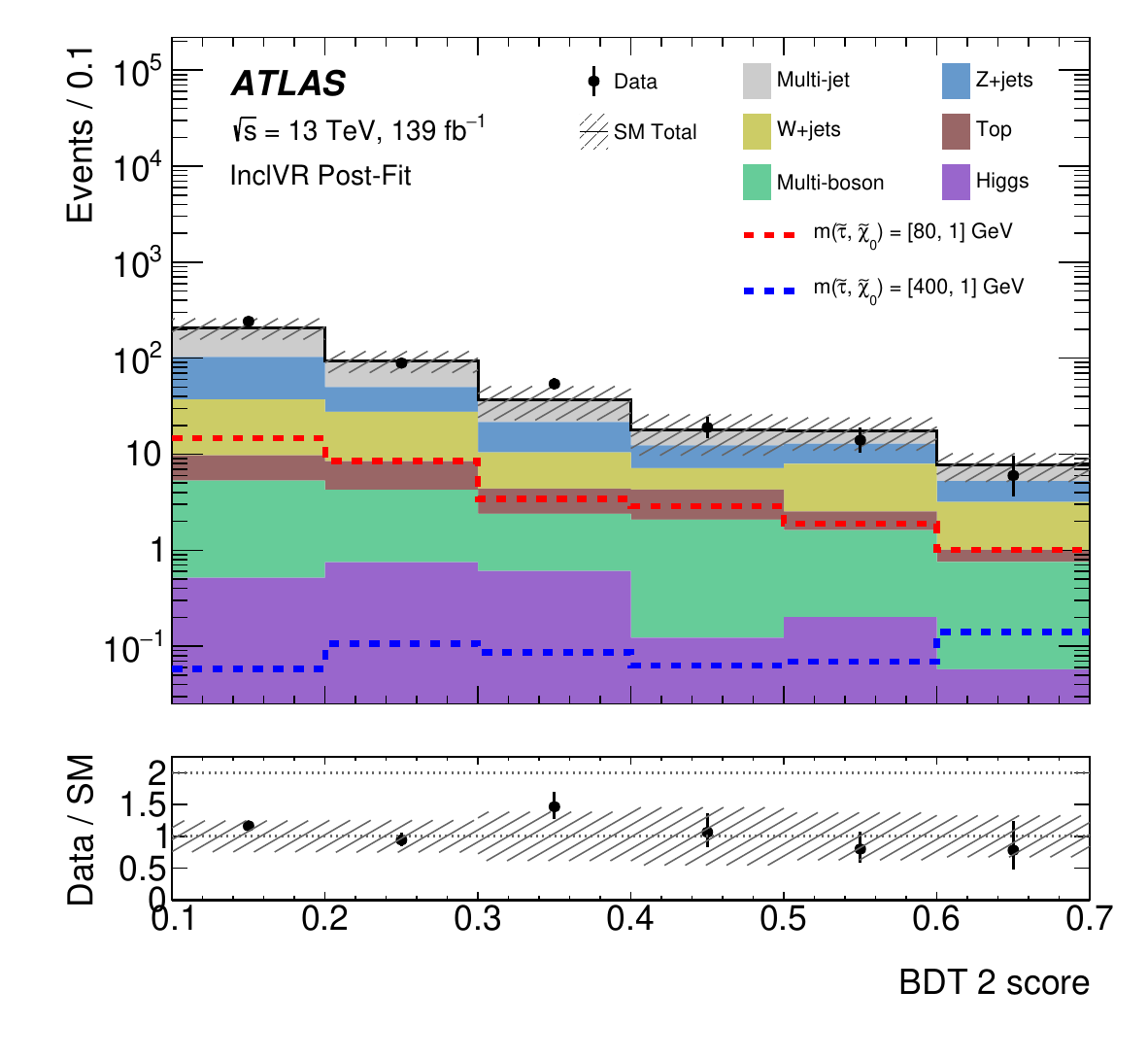}}\\
\subfloat[][]{\includegraphics[width=0.49\textwidth]{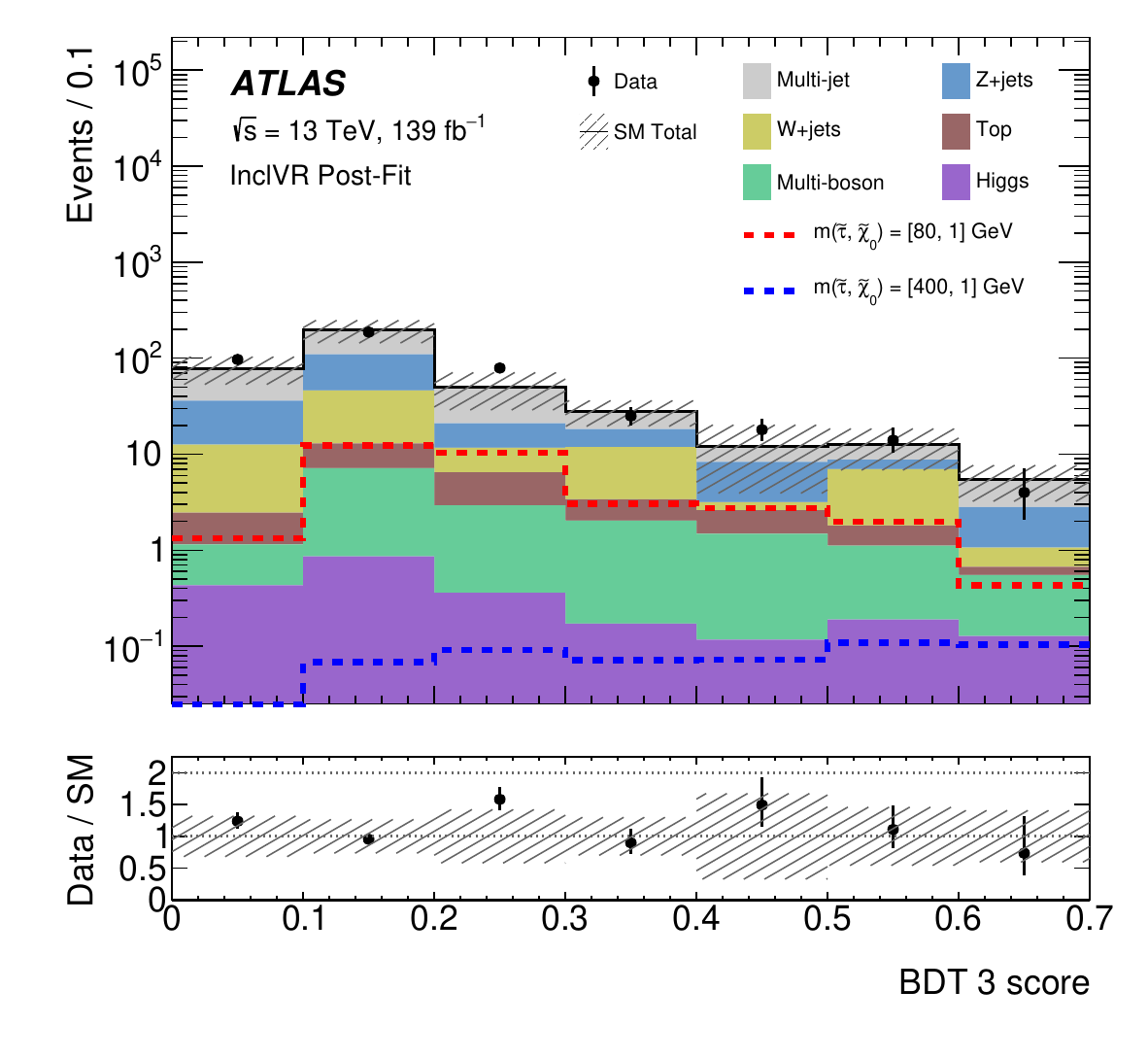}}
\subfloat[][]{\includegraphics[width=0.49\textwidth]{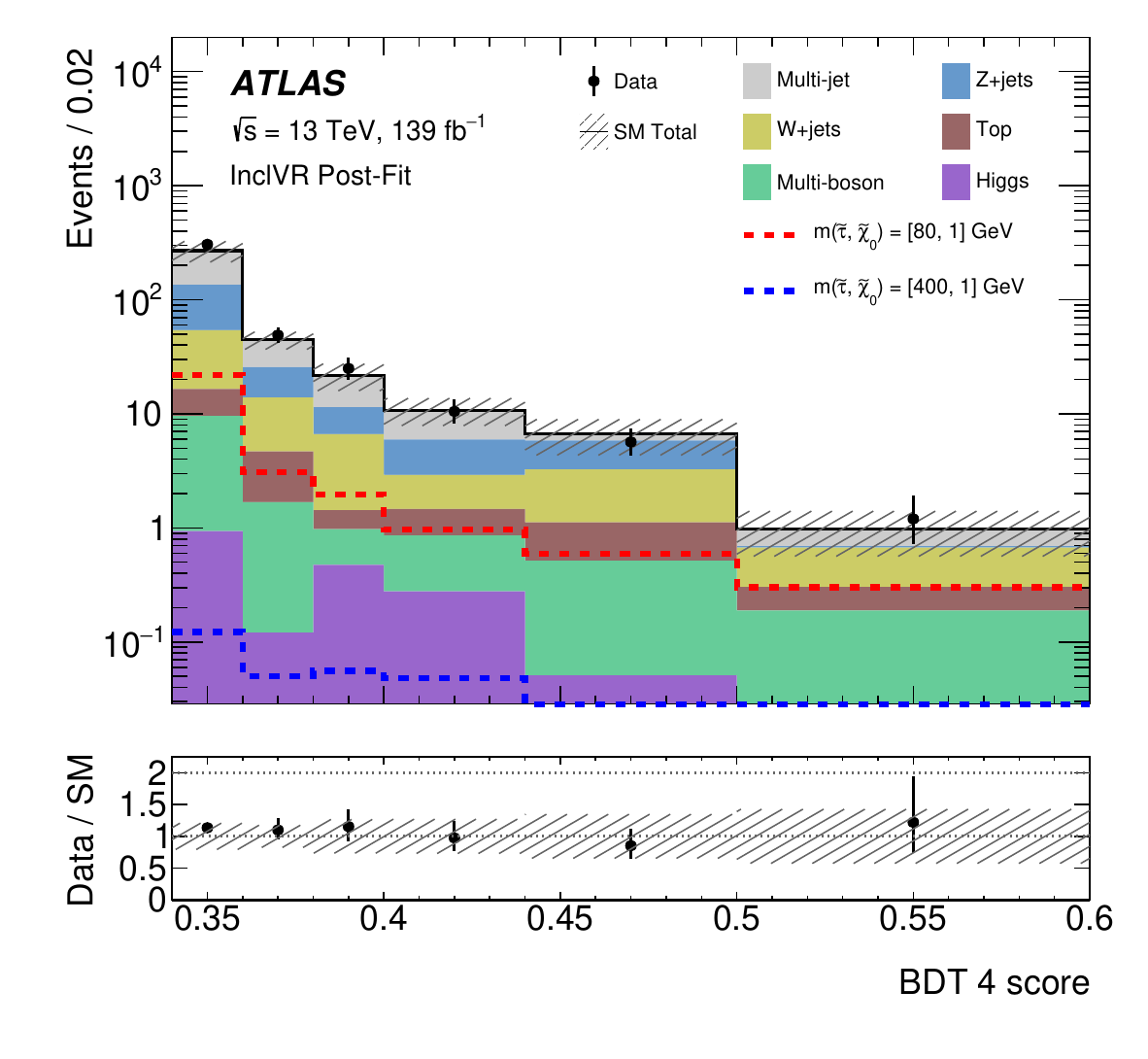}}\\
\caption{The BDT score distributions for the Direct stau channel in InclVR after the background-only fit, showing the scores for (a) BDT1, (b) BDT2, (c) BDT3, and (d) BDT4.
Representative SUSY scenarios are overlaid for illustration.
The hatched bands represent the combined statistical and systematic uncertainties of the total SM background. Large uncertainties are seen in some bins due to a few highly weighted $Z$+jets MC simulation events migrating across bins when considering systematic variations.
The lower panels show the ratio of data to the total SM background estimate.
}
\label{fig:vrgen_DS}
\end{figure}

Events with more than one SM $W/Z$ boson are the dominant background in SR-BDT2 and SR-BDT4, with very small contributions from processes involving Higgs bosons in all SRs.
These processes produce two real $\tau$-leptons and $\met$, and this irreducible background is taken from MC simulation. The modelling is checked using two validation regions.
The first targets multi-boson production (MBVR) using a similar selection to SR-BDT4 but with the selection on $m(\tau_1, \tau_2)$ reversed. The second is a more inclusive validation region (InclVR) that checks the overall background estimate strategy and the modelling of the BDT score distributions. The definitions of these regions are shown in Table~\ref{tab:AllDScrvr_defDS}.
The background estimate is seen to model the data well in these two regions and the agreement in the BDT score distributions in InclVR is shown in Figure~\ref{fig:vrgen_DS}.

\begin{figure}[h!] 
\centering
\includegraphics[trim = 0mm 12mm 0mm 0mm, clip, width=0.95\textwidth]{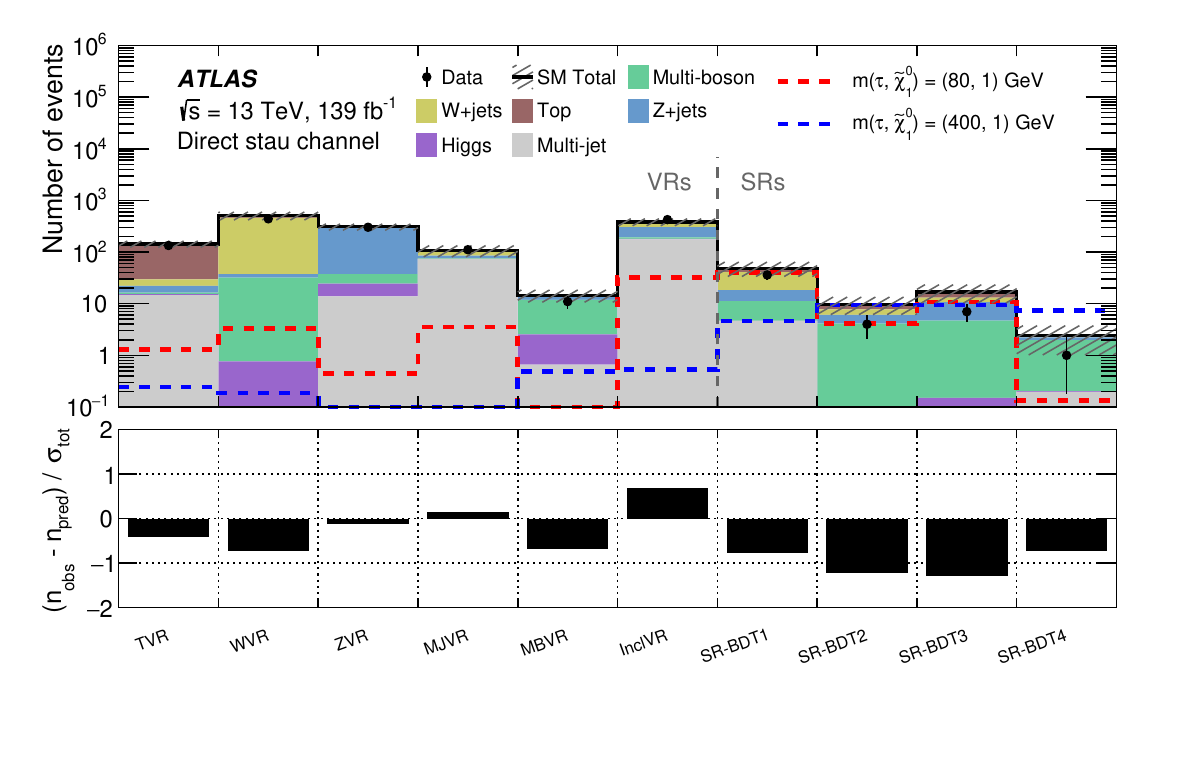}
\caption{ Comparison of the observed and expected event yields in all VRs and SRs after the background-only fit for the Direct stau channel.
Representative SUSY scenarios are overlaid for illustration.
The hatched band represents the combined statistical and systematic uncertainties of the total SM background.
The lower panel shows the significance of any difference between the data and total SM background estimate yields.
}
\label{fig:DSyields}
\end{figure}

Overall, the background estimate strategy is seen to model the data well in all validation regions used for the Direct stau channel. The yields for the estimated SM background and data are shown in Figure~\ref{fig:DSyields} after the background-only fit to data is applied, as described in Section~\ref{sec:fitStrategy}.
 
\section{Intermediate stau channel}
\label{sec:stau}
\label{sec:intro1}
The Intermediate stau channel targets two different production mechanisms,
mass-degenerate \ConepmNtwo (C1N2) and \ConepConem (C1C1), with decays into the
lightest neutralino only through intermediate stau and
$\tau$-sneutrinos, as shown in Figure~\ref{fig:feynmanslepton_CCStau} and Figure~\ref{fig:feynmanslepton_CNStau}.  The C1N2 analysis is then sub-divided into final
states where the two highest-$\pt$ $\tau$-lepton candidates have opposite sign (OS) charge (C1N2OS) or have
same-sign (SS) charge (C1N2SS).  The SRs are separated into low-mass (LM) and high-mass (HM) regions to target
respectively low or high \chinoonepm/\ninotwo mass regions.
The high-mass regions target SUSY scenarios with masses above that probed by the partial Run 2 result from the ATLAS Collaboration~\cite{SUSY-2016-23}, and the low-mass regions target SUSY scenarios with smaller mass splittings.
The event selections of C1C1, C1N2OS and C1N2SS analyses are described in Section~\ref{sec:variables1}, and the background estimates are described in Section~\ref{sec:bgesti1}.
 
\subsection{Event selection}
\label{sec:variables1}
The SRs are optimised by varying the kinematic selection criteria resulting in
six SRs defined to cover LM and HM regions for C1C1, C1N2OS and C1N2SS channels.
Events used in this channel must satisfy the same di-$\tau$ trigger described in Section~\ref{sec:stauEvSel}, or a combined di-$\tau$+$\met$ trigger.
To ensure the trigger efficiencies are in the plateau region, events that satisfy the di-$\tau$ + \met trigger for 2015--2017 (2018) data sample must have a $\tau$-lepton with $\pt > 50\, (75)\,\GeV$, a second $\tau$-lepton with $\pt > 40\,\GeV$ (both matching the trigger signature), and the reconstructed $\met > 150\,\GeV$.
 
For the C1N2OS (C1N2SS) channels, events are required to have at least two medium $\tau$-leptons with OS (SS), while for the C1C1 channel, events are required to have exactly two medium $\tau$-leptons with OS\@.
In the LM signal regions, SR-C1C1-LM and SR-C1N2OS-LM, at least one of the $\tau$-leptons must satisfy a tighter identification criteria, referred to as a tight $\tau$-lepton~\cite{ATL-PHYS-PUB-2019-033}, to suppress quark or gluon jets misidentified as $\tau$-leptons in the lower $\met$ region.
Additionally, $\met >60\,\GeV$ is required in these two SRs to further suppress background from misidentified $\tau$-leptons.
 
To discriminate the SUSY signal events from SM background processes, additional requirements are applied, as outlined in Table~\ref{tab:SR1-def}.
Background processes producing $\tau$-lepton pairs from low-mass resonances, or from $Z$ or Higgs bosons, are suppressed using high $m(\tau_1,\tau_2)$ selections, while top quark processes are suppressed with a $b$-jet veto.
Contributions from $\ttbar$ and $WW$ events are reduced using lower bounds on $\mttwo$ and $m_{\mathrm{T}\mathrm{sum}}$.

\begin{table}[!h]
\centering
\caption{Summary of the selection requirements for the $\chinoonep\chinoonem$/$\chinoonepm\ninotwo$ pair
production SRs for the Intermediate stau channel.
\label{tab:SR1-def}}
\footnotesize
 
\begin{tabular}{l | c|c|c | c|c|c}
\toprule
~~~~~~~~~~~~~~~~~~~~~~~~~~SR- & C1C1-LM & C1N2OS-LM & C1N2SS-LM & C1C1-HM &  C1N2OS-HM &  C1N2SS-HM \\
\midrule      \midrule
Trigger & \multicolumn{3}{c|}{asymm. di-$\tau$} &  \multicolumn{3}{c}{di-$\tau$ + $\met$} \\
$\met$ [$\GeV$] & \multicolumn{3}{c|}{$ < 150$ }  & \multicolumn{3}{c}{$ > 150$ }         \\
\midrule
$N$ medium $\tau$ & $=2$ & $\geq 2$ & $\geq 2$ & $=2$  & $\geq 2$ & $\geq 2$ \\
$N$ tight $\tau$ & $\geq 1$ & $\geq 1$ & -- & -- & -- & -- \\
Charge combination & OS & OS & SS & OS & OS & SS \\
$N$ $b$-jets & $=0$& $=0$& $=0$& $=0$& $=0$& $=0$       \\
$N$ jets  & -- &  $<3$&  $<3$ & -- & -- & --\\
$|\Delta\phi(\tau_1,\tau_{2})|$ & $> 1.6$    & -- &  $> 1.5$ & -- & -- & --  \\
$m (\tau_{1},\tau_{2})$ [$\GeV$] & $> 120$ & $>120$  & --  & $> 120$ & $>120$  & -- \\
$\met$ [$\GeV$] & $>60$ & $>60$ & -- & -- & -- & -- \\
$m_{\mathrm{T}\mathrm{sum}}$ [$\GeV$] & -- & -- & $> 200$ & $>400$ & $>400$ & $>450$      \\
$\mttwo$ [$\GeV$] & $> 80$ & $> 70$ & $>80$ & $>85$ & $>85$ & $>80$  \\
 
\bottomrule
 
\end{tabular}
\end{table}

\subsection{Background estimate}
\label{sec:bgesti1}
The dominant backgrounds in the Intermediate stau SRs are multi-jet production with misidentified $\tau$-leptons, $W$+jets and $Z$+jets, multi-boson production, and top quark backgrounds.
The multi-jet background accounts for \percent{28--48} ($<$\percent{17}) of the total SM yield in high-mass SRs (low-mass SRs) for all three scenarios, while $W$+jets accounts for \percent{4--16} of the total background.
The multi-jet and $W$+jets background estimates are presented in Section~\ref{sec:BG_QCDXX} and Section~\ref{sec:BG_WXX}, respectively, and use the same methods described in Sections~\ref{sec:BG_QCD} and \ref{sec:BG_W}.
A dedicated CR is used to estimate the top quark backgrounds for the SS final state, where misidentified $\tau$-lepton contributions are a dominant source, as described in Section~\ref{sec:BG_top}.
 
Multi-boson production contributes mainly through events containing real $\tau$-leptons resulting from
$WW$ and $ZZ$ decaying into a $\tau\tau\nu\nu$ final state in C1C1
and C1N2OS scenarios, while in the C1N2SS scenario, the main process is $WZ$ decaying into a $\tau\tau\tau\nu$ final state.
The contribution from real $\tau$-leptons exceeds 85--90\% in $Z$+jets and diboson production.  Additionally, the real $\tau$-lepton contribution exceeds 80\% in backgrounds containing
top quarks in OS final states.  The estimation of the Top, multiboson, and $Z$+jets backgrounds is described in Section~\ref{sec:BG_irre}.

\subsubsection{Multi-jet background estimate}
\label{sec:BG_QCDXX}
The same approach for the multi-jet background estimate described in Section~\ref{sec:BG_QCD} is used for the Intermediate stau channels, with the following changes.
The sign of the electric charge of the two $\tau$-leptons (OS or SS),
and \mttwo\ and m$_{{\mathrm T}\text{sum}}$ are used to define the control regions and validation regions, as shown in Table~\ref{tab:abcddef}.
In the C1C1 and C1N2OS scenarios, the \mttwo\ variable is used to distinguish the regions of the ABCD method, while $m_{{\mathrm T}\text{sum}}$ is used for the C1N2SS channel.
In all validation regions and both sets of CR-B and CR-C, the events are required to satisfy a di-$\tau$ trigger, instead of the di-$\tau$ + \met\ trigger in the Intermediate stau channel to increase the statistics from the lower \met\ requirements.  The offline $\met > 150\,\GeV$ requirement is also removed.
The di-$\tau$ trigger requires the identification of two hadronically decaying $\tau$-leptons with transverse momenta exceeding the same set of thresholds for the di-$\tau$ + \met trigger, such that no bias on the $\tau$-leptons is introduced.

\begin{table}[!h]
\centering
\caption{The definition of the ABCD regions for all channels in the
Intermediate stau scenarios. Only those requirements that differ
between the CRs/VRs and the SRs are listed.
}
\label{tab:abcddef}
\footnotesize
\begin{tabular}{l|l||c|c|c}
\toprule
Channel    & variable & CR-B / CR-C & VR-E / VR-F  & CR-A / SR \\
\midrule     \midrule
C1C1-LM  &$\mttwo$ [$\GeV$]  &$\in (15, 35]$ &  $\in (35, 80]$ & $> 80$  \\
&$\met$ [$\GeV$]  & $\in (10, 150]$  &  $\in (10, 150]$ & $\in (60, 150]$ \\[+0.5mm]
\midrule
C1C1-HM  &$\mttwo$ [$\GeV$]  & $\in (35, 60]$  &  $\in (60, 85]$ & $> 85$  \\
&$m_{{\mathrm T}\text{sum}}$ [$\GeV$] & $\in (100, 300]$ &  $\in (200, 400]$ & $> 400$ \\
&$\met$ [$\GeV$] & $> 50$ &  $> 50$ & $> 150$ \\[+0.5mm]
\midrule
C1N2OS-LM &$\mttwo$ [$\GeV$] & $\in (15, 35]$  & $\in (35, 70]$ & $> 70$ \\
&$\met$ [$\GeV$]  & $\in (10, 150]$  &  $\in (10, 150]$ & $\in (60, 150]$ \\[+0.5mm]
\midrule
C1N2OS-HM &$\mttwo$ [$\GeV$]  & $\in (35, 60]$  &  $\in (60, 85]$ & $> 85$  \\ [+0.5mm]
&$m_{{\mathrm T}\text{sum}}$ [$\GeV$] & $\in (150, 300]$ & $\in (200, 400]$ & $> 400$ \\
&$\met$ [$\GeV$] & $> 50$ &  $> 50$ & $> 150$ \\[+0.5mm]
\midrule
C1N2SS-LM  &$m_{{\mathrm T}\text{sum}}$ [$\GeV$] & $< 100$ &  $\in (100, 200]$  & $> 200$ \\
&$|\Delta\phi(\tau_1,\tau_{2})|$ & $<$ 1.5  & $<$ 1.5  & $> 1.5$ \\[+0.5mm]
\midrule
C1N2SS-HM &$m_{{\mathrm T}\text{sum}}$ [$\GeV$] & $\in (100, 200]$  &  $\in (200, 450]$  & $> 450$ \\
&$\met$ [$\GeV$] & $> 50$ &  $> 50$ & $> 150$ \\
\bottomrule
\end{tabular}
\end{table}
 
\begin{figure}[!h]
\centering
\subfloat[][]{\includegraphics[width=0.495\textwidth]{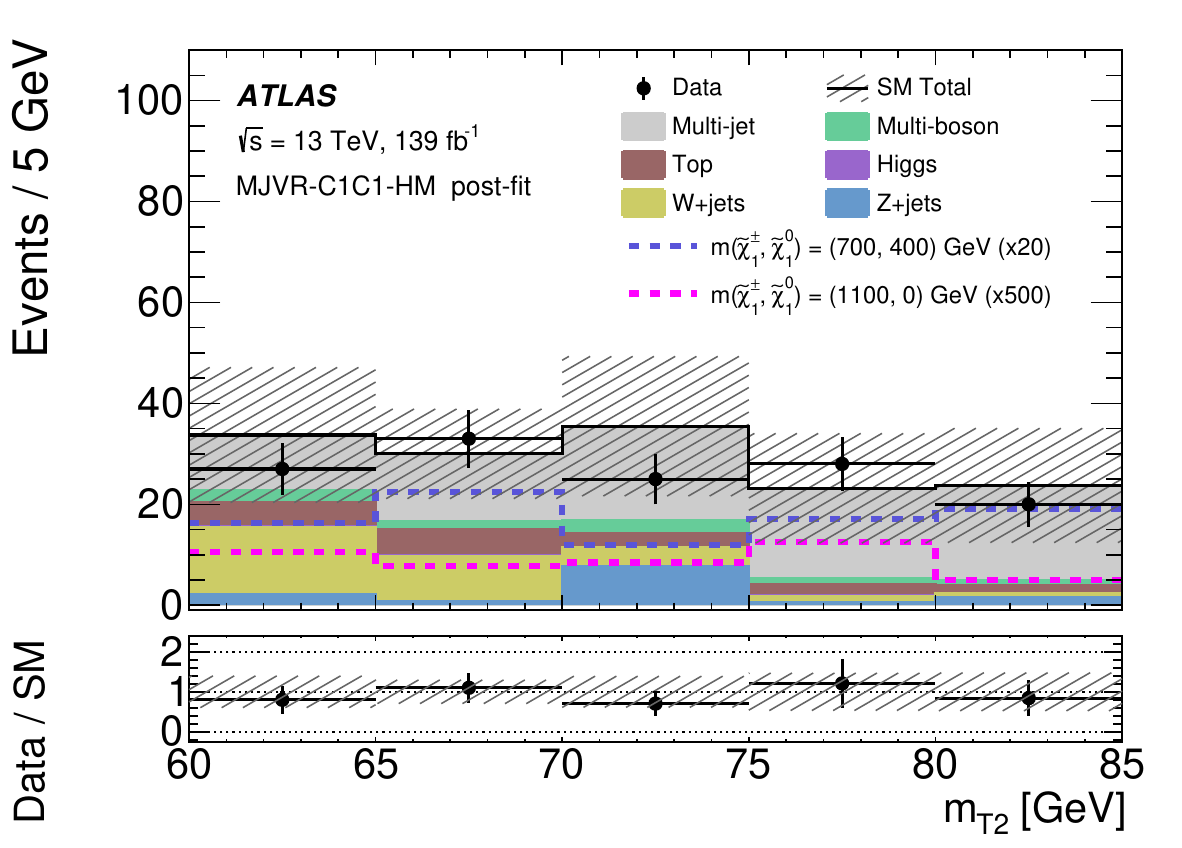}}  
\subfloat[][]{\includegraphics[width=0.495\textwidth]{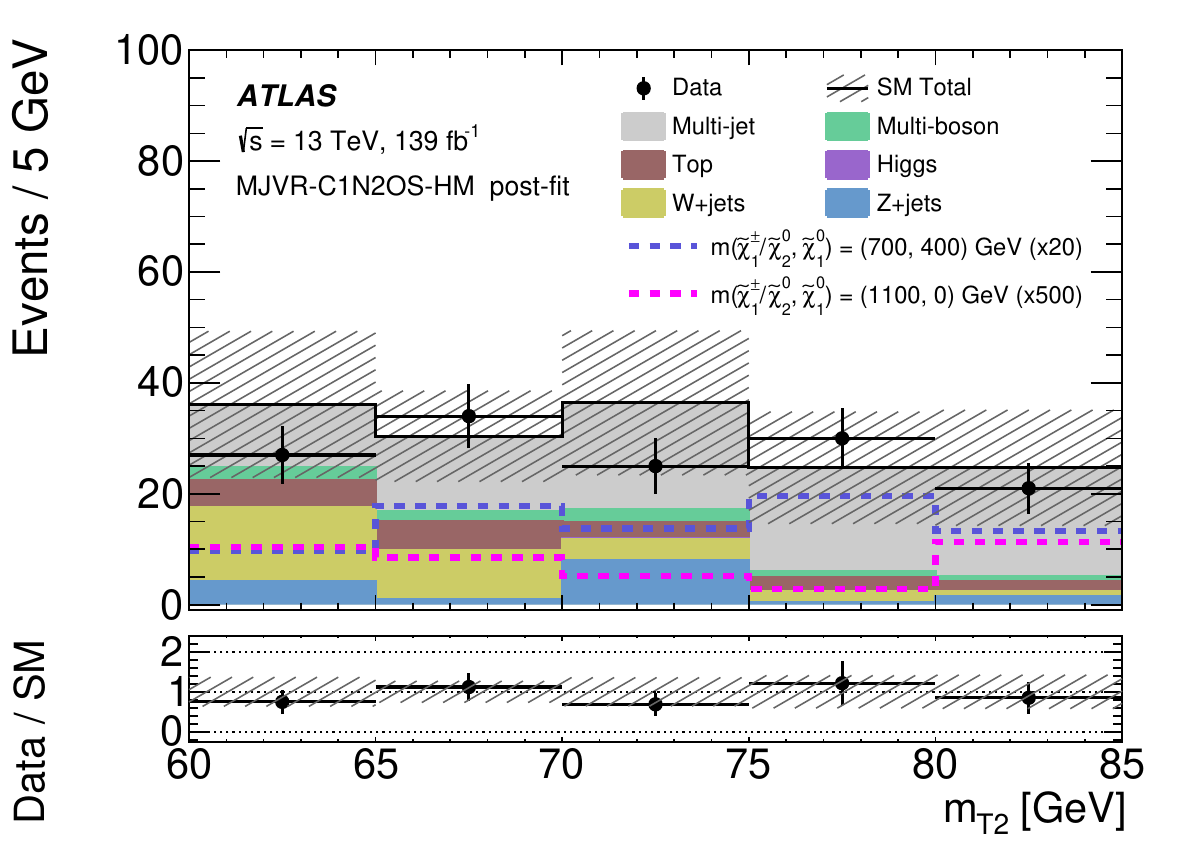}}\\ 
\subfloat[][]{\includegraphics[width=0.495\textwidth]{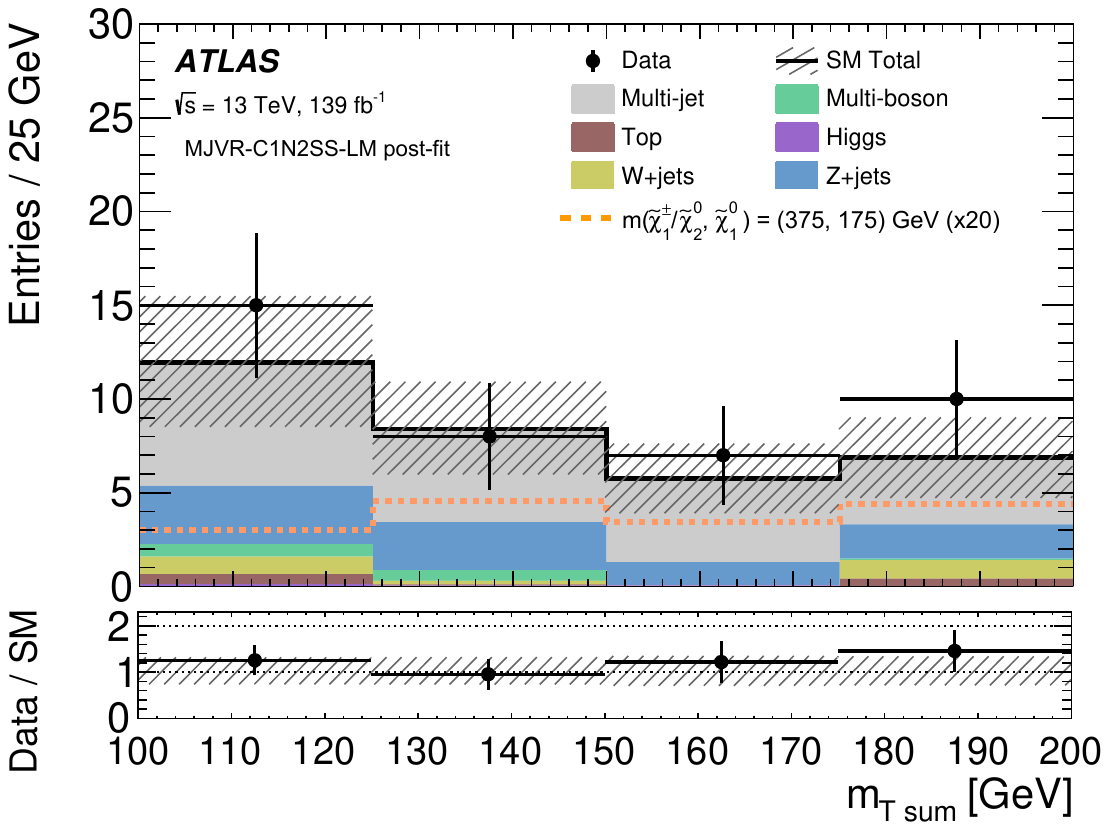}} 
\caption{The kinematic distributions of \mttwo\ and $m_{\mathrm{T}\mathrm{sum}}$ in the
multi-jet background VR-F after the background-only fit for (a) SR-C1C1-HM (MJVR-C1C1-HM), (b)
SR-C1N2OS-HM (MJVR-C1N2OS-HM)
and (c) SR-C1N2SS-LM (MJVR-C1N2SS-LM), respectively.
Representative SUSY scenarios are overlaid for illustration and scaled appropriately for improved visibility.
The hatched bands represent the combined statistical and systematic uncertainties of the total SM background.
The lower panels show the ratio of data to the total SM background estimate.
}
\label{fig:ABCD-vali}
\end{figure}

The estimated SM background is found to agree with the data for the \mttwo\ and $m_{\mathrm{T}\mathrm{sum}}$ distributions in the validation regions, as shown in Figure~\ref{fig:ABCD-vali}.
Several signal reference points targeting sensitivity to LM and HM SRs are also shown to highlight the small signal contamination in the VRs.

\subsubsection{$W$+jets background estimate}
\label{sec:BG_WXX}
 
The $W$+jets background is an important background for the Intermediate stau channel. A similar approach as in Section~\ref{sec:BG_W} is used and definitions for the $W$+jet CRs and VRs ($W$CR and $W$VR) for OS/SS selections are shown in Table~\ref{tab:IntStaucrvr_def}.
Top quark background events in the OS final state are labeled as ``top-tagged'' if they satisfy a dedicated selection; these events are vetoed in the $W$CR to suppress top quark background processes.
Events are top-tagged using the contransverse mass variable~\cite{Tovey:2008}, $m_{\mathrm{CT}}$, to identify events that are kinematically compatible with $\ttbar$ pair production.
Furthermore, top-tagged events must have at least two jets and the scalar sum of the $\pt$ of at least one combination of two jets and the two leptons in the event must exceed $100\,\GeV$.
The orthogonality of the $W$CRs and $W$VRs is ensured using $\mttwo(\tau,\mu)$ selections.
The purity of the selection in $W$+jets events is \percent{73--85} in all $W$ control and validation regions. The signal contamination in the $W$CR and $W$VR is negligible due to the requirement of a muon in the event.
 
The multi-jet contribution in the $W$CRs and $W$VRs is small and is estimated using the lack of correlation of the charge of the faked muon and $\tau$-lepton in multi-jet production, resulting in equal SS and OS lepton contributions.
This is not the case for other processes like $W$+jets production, which is dominated by $gu/gd$-initiated processes that often give rise to a jet originating from a quark, the charge of which is anti-correlated with the $W$-boson charge.
The contribution from multi-jet production in the $W$CRs and $W$VRs is estimated using data in regions selected with the same requirements as the CR/VR but requiring the electric charge of the two leptons to be different, i.e.\ SS for OS CR/VR, and OS for SS CR/VR.
Event yields from SM processes other than multi-jet production are subtracted from the data in these regions using MC simulation, leaving only the multi-jet estimate, which is assumed to be equal to that in the CR/VR.
Based on studies with simulated samples, a conservative systematic uncertainty of \percent{100} is assigned to the estimate of the multi-jet event yield in the $W$+jets CRs and VRs.

\begin{table}[!h]
\centering
\caption{The definition of the control and validation regions for the Intermediate stau channel.
}
\label{tab:IntStaucrvr_def}
\scriptsize
\begin{adjustbox}{width=1\textwidth}
\begin{tabular}{l| c|c|c|c |c|c |c|c}
\toprule
Process & \multicolumn{4}{c|}{$W$+jets} & \multicolumn{4}{c}{Top} \\\midrule
Region & $W$CR-OS   & $W$VR-OS &$W$CR-SS    & $W$VR-SS & $T$CR-SS-HM & $T$VR-SS-HM & $T$VR-OS-LM & $T$VR-OS-HM  \\
\midrule      \midrule
Charge combination & \multicolumn{2}{c|}{OS} & \multicolumn{2}{c|}{SS} & \multicolumn{2}{c|}{SS} & \multicolumn{2}{c}{OS} \\
Trigger & \multicolumn{4}{c|}{single $\mu$} & \multicolumn{2}{c|}{di-$\tau$ + $\met$} & asymm. di-$\tau$ & di-$\tau$ + $\met$ \\
$N$ medium $\tau$ & \multicolumn{4}{c|}{$=1$}  & \multicolumn{2}{c|}{$< 2$} & \multicolumn{2}{c}{$\geq 2$}  \\
$N$ tight $\tau$ & \multicolumn{4}{c|}{-}  & \multicolumn{2}{c|}{-} & \multicolumn{2}{c}{$\geq 1$} \\
$N$ loose $\tau$ & \multicolumn{4}{c|}{-}  & \multicolumn{2}{c|}{$\geq 1$} & \multicolumn{2}{c}{-} \\
$N$ ``very loose'' $\tau$ & \multicolumn{4}{c|}{-}  & \multicolumn{2}{c|}{$\geq 2$} &\multicolumn{2}{c}{-} \\
$N$ $e/\mu$ &\multicolumn{4}{c|}{$=1$ $\mu$} &\multicolumn{2}{c|}{-}  &\multicolumn{2}{c}{-} \\
$N$ $b$-jets & \multicolumn{4}{c|}{$=0$ } & \multicolumn{2}{c|}{$\geq 1$} & $\geq 1$  & $=0$\\
$\pt(\tau)$ [$\GeV$] & \multicolumn{4}{c|}{$>50$} & \multicolumn{2}{c|}{-} &\multicolumn{2}{c}{-} \\
$\pt(\mu)$ [$\GeV$] & \multicolumn{4}{c|}{$>40$} & \multicolumn{2}{c|}{-} &\multicolumn{2}{c}{-} \\
Top tagged &   \multicolumn{2}{c|}{veto}  & \multicolumn{2}{c|}{-} & \multicolumn{2}{c|}{-} &\multicolumn{2}{c}{-} \\
$m_{{\mathrm T}, \mu}$ [$\GeV$] & \multicolumn{2}{c|}{$<140$}  & \multicolumn{2}{c|}{$\in(50,150]$} & \multicolumn{2}{c|}{-} &\multicolumn{2}{c}{-} \\
$m_{{\mathrm T}, \mu}+m_{{\mathrm T}, \tau}$ [$\GeV$] & \multicolumn{2}{c|}{-}  & \multicolumn{2}{c|}{$>80$} & \multicolumn{2}{c|}{-} &\multicolumn{2}{c}{-} \\
$\met$ [$\GeV$] & \multicolumn{2}{c|}{$>60$ } & \multicolumn{2}{c|}{$>50$ } & \multicolumn{2}{c|}{$>150$ } & $\in (20,150]$ & $>150$ \\
 
$m(\tau_1, \tau_2)$ [$\GeV$] & -- & -- & -- & -- & -- &- & $>120$ & $>120$ \\
$\mttwo$ [$\GeV$] & $\in(40,70]$ & $>70$ & $<60$ & $>60$ & -- & -- & $>40$ & $>30$ \\
$m_{\mathrm{T}\mathrm{sum}}$ [$\GeV$] & -- & -- & -- & -- & $<400$ & $>400$ & $>150$ & $>150$ \\
$|\Delta \phi(\tau_1,\tau_2)|$ & -- & -- & -- & --  & -- & -- & $>1.0$ & $>1.0$ \\
\bottomrule
\end{tabular}
 
\end{adjustbox}
 
\begin{tabular}{l| c|c|c|c|c }
\toprule
Process & \multicolumn{2}{c|}{$Z$+jets} & \multicolumn{3}{c}{Multi-bosons} \\\midrule
Region & $Z$VR-OS-LM & $Z$VR-OS-HM & $MB$VR-OS-LM & $MB$VR-OS-HM & $MB$VR-SS \\
\midrule      \midrule
Charge combination & \multicolumn{2}{c|}{OS} & \multicolumn{2}{c|}{OS} & SS \\
Trigger & asymm. di-$\tau$ & di-$\tau$ + $\met$ & asymm. di-$\tau$ & di-$\tau$ + $\met$ & single $\mu$ \\
$N$ medium $\tau$ & \multicolumn{4}{c|}{$\geq 2$} & $=1$ \\
$N$ tight $\tau$ & \multicolumn{4}{c|}{$\geq 1$} & -- \\
$N$ $\mu$ & \multicolumn{4}{c|}{-} & $=2$ \\
$N$ $b$-jets & \multicolumn{4}{c|}{$=0$ } & $=0$ \\
$\met$ [$\GeV$] & $\in (40,150]$ & $>150$ & $\in (70,150]$ & $>150$ & $>100$ \\
$m(\tau_1, \tau_2)$ [$\GeV$] & $<70$ & $<60$ & $<80$ & $<90$  & -- \\
$\Delta R(\tau_1,\tau_2)$ & $<1.0$ & $<1.0$ & $<1.2$ & $<1.2$ & -- \\
$|\Delta \phi(\tau_1,\tau_2)|$ & -- & -- & $<1.0$ & $<1.0$ & -- \\
$|\Delta \phi(\tau_1,\pTmiss)|$ & -- & -- & -- & -- & $\leq 1.75$ \\
$m_{\mathrm{T}\mathrm{sum}}$ [$\GeV$] & -- & -- & $>180$ & $>180$ & -- \\
$\mttwo$ [$\GeV$] & $<60$ & $<60$ & $>60$ & $>60$ & -- \\
 
\bottomrule
\end{tabular}
\end{table}

\subsubsection{Top quark background estimate for C1N2SS}
\label{sec:BG_top}
 
The top quark background in SR-C1N2SS-LM is very small ($<$1\%) and mostly composed of two real $\tau$-leptons, and hence is estimated from MC simulation.
However, the top quark background is a dominant contribution in SR-C1N2SS-HM and mostly consists of one or two misidentified $\tau$-leptons from \ttbar~and $Wt$ production.
To estimate this background in SR-C1N2SS-HM, a data-driven approach is used to normalize
the top quark background MC simulation to data using a dedicated top-enriched CR ($T$CR-SS-HM)
and validated in a top-enriched VR ($T$VR-SS-HM), described in Table~\ref{tab:IntStaucrvr_def}.
The $\tau$-lepton identification working point was loosened to increase the statistics in these regions, while high $\met$ is required to suppress the contribution from multi-jet processes.
The top quark background purity in the top quark CR and VR for SR-C1N2SS-HM ($T$CR-SS-HM and $T$VR-SS-HM) is high and exceeds 83\% in both regions.

\subsubsection{Top, multiboson, and $Z$+jets background estimates}
\label{sec:BG_irre}
 
Additional irreducible SM backgrounds are estimated from MC simulation and checked in dedicated VRs.
The top quark background contributions in OS final states are small and amount to about \percent{7--14} of the total background in all SRs.
Two regions enriched in top-quark events are defined to validate the top quark modelling in OS events, $T$VR-OS-LM and $T$VR-OS-HM for the low and high-mass SR kinematics, respectively, and are defined in Table~\ref{tab:IntStaucrvr_def}.
 
The $Z$+jets contribution is \percent{16--21} of the total background in all OS SRs and mainly arises from $Z\rightarrow \tau\tau$ decays.
The multi-boson background accounts for \percent{25--50} of the total SM contribution in the SRs and mainly arises from $WW\rightarrow \tau\nu \tau\nu$  and $ZZ \rightarrow \tau\tau \nu\nu$ events. Over \percent{96} of $\tau$-leptons from $Z$+jets and multi-bosons events in the SRs are real.
To validate the MC modelling and normalization of the $Z$+jets and multi-boson
processes, four dedicated VRs are defined:  $Z$VR-OS-LM ($MB$VR-OS-LM) is defined to validate $Z$+jets (multi-boson) MC
modelling in the low-mass SRs, while $Z$VR-OS-HM ($MB$VR-OS-HM) is defined to validate
the MC modelling in the high-mass SRs, as shown in Table~\ref{tab:IntStaucrvr_def}.
 
The purity of the selection in \ttbar\ and $Z$+jets events is in the range of \percent{81--99} in the respective validation regions, and the purity of the selection in multi-boson events is \percent{41--68} in the OS $MB$VRs.
The signal contamination in the VRs is small due to the $b$-jet requirement in the top-quark VRs and the $m (\tau_{1},\tau_{2})$ upper thresholds in the $Z$+jets and multi-boson VRs.
 
In the SS final state, the multi-boson background is estimated from MC simulation and validated in a dedicated VR, $MB$VR-SS, also defined in Table~\ref{tab:IntStaucrvr_def}. The selection is enriched in multi-boson production, resulting in a multi-boson purity of 73\%.

\begin{figure}[h!]
\centering
\includegraphics[trim = 0mm 25mm 0mm 0mm, clip, width=0.95\textwidth]{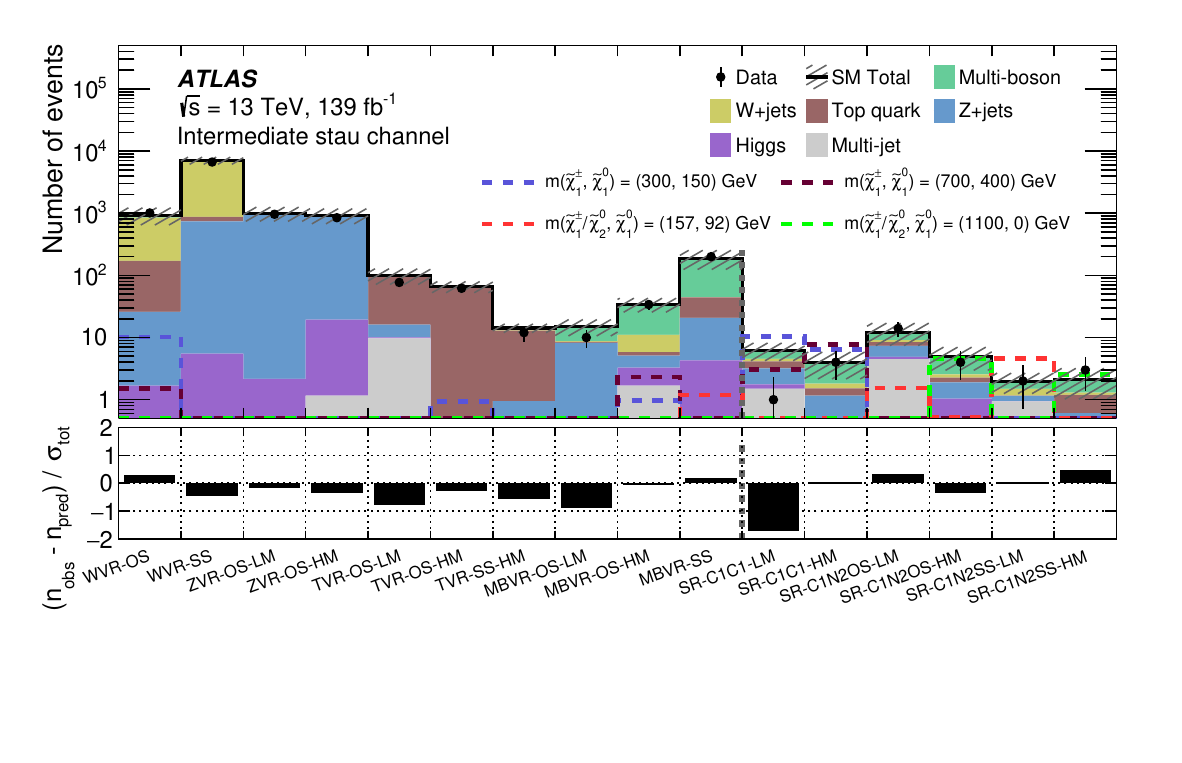} 
\caption{Comparison of the observed and expected event yields in all VRs and SRs after the background-only fit for the signal regions targeting chargino/neutralino production and decay via intermediate staus.
Representative SUSY scenarios are overlaid for illustration.
The hatched band represents the combined statistical and systematic uncertainties of the total SM background.
The lower panel shows the significance of any difference between the data and total SM background estimate yields.
}
\label{fig:bkg-vr-stau}
\end{figure}

The agreement between data and the SM prediction is shown in all
validation and signal regions for the Intermediate stau channels in Figure~\ref{fig:bkg-vr-stau} after the background-only fit to data is applied, as described in Section~\ref{sec:fitStrategy}.
 
\section{Intermediate $Wh$ channel}
\label{sec:wh}
The search for the production of mass-degenerate \ConepmNtwo\ decaying via an intermediate $W$ and $h$ boson, as shown in Figure~\ref{fig:feynmanslepton_Wh}, is described in this section.
The analysis targets the leptonic decay of the $W$ boson to an electron or muon, and the decay of the Higgs boson to two OS $\tau$-leptons.
The event selection of the Intermediate $Wh$ channel is described in Section~\ref{sec:variables2} and the background estimates are described in Section~\ref{sec:bgesti2}.
 
\subsection{Event selection}
\label{sec:variables2}
Events for the Intermediate $Wh$ channel must have at least two hadronically decaying $\tau$-leptons with OS and exactly one light lepton ($e$ or $\mu$).
The selected light lepton must satisfy the signal electron or muon requirements described in Section~\ref{sec:EvenRec} and also satisfy the single electron or single muon trigger with a threshold of $\pt>27\,\GeV$ to ensure the triggering efficiency is in the plateau region.
Two SRs are defined to cover low-mass (SR-Wh-LM) and high-mass (SR-Wh-HM) \ConepmNtwo\ production, with an overlapping selection, as shown in Table~\ref{tab:SR2-def}.
To set limits in a SUSY scenario, the SR with the best expected sensitivity is used.
The invariant mass of the two $\tau$-leptons, $m(\tau_{1},\tau_{2})$, is required to be compatible with the Higgs boson mass, while the $b$-jet veto and $|\Delta R(\tau_1,\tau_2)| $ requirements suppress the top quark and fake $\tau$-lepton backgrounds, respectively.
High $\mttwo$, $m_{{\mathrm T}, \ell}$ and $m_{\mathrm{T}\mathrm{sum}}$ requirements provide good discrimination between SUSY signal events and SM background.
 
\begin{table}[h!]
\centering
\caption{Summary of selection requirements for the SRs of
$\chinoonepm\ninotwo$ pair production decaying into an intermediate $Wh$ for low-mass and high-mass regions. The two SRs are not orthogonal.
\label{tab:SR2-def}}
\footnotesize
\begin{tabular}{l |c|c}
\toprule
& SR-Wh-LM & SR-Wh-HM \\
\midrule     \midrule
Trigger & \multicolumn{2}{c}{Single $e$ or $\mu$} \\
$N$ medium $\tau$ & \multicolumn{2}{c}{$\geq 2$ } \\
$N$ $e/\mu$ & \multicolumn{2}{c}{$=1$} \\
Charge combination & \multicolumn{2}{c}{OS} \\
$N$ $b$-jets & \multicolumn{2}{c}{$=0$} \\
$|\Delta \phi(\tau_{1},\tau_{2})|$ & \multicolumn{2}{c}{ $< 3$ } \\
\midrule
$\Delta\text{R}(\tau_{1},\tau_{2})$ & --  &  $ < 2.2$ \\
$m (\tau_{1},\tau_{2})$ [$\GeV$] &  $\in [90, 130]$ & $\in [80,160]$ \\
$\mttwo$ [$\GeV$] & $> 100$     & $> 80$ \\
$ m_{{\mathrm T}, \ell}$ [$\GeV$] & -- &  $ > 80$    \\
$ m_{\mathrm{T}\mathrm{sum}}$ [$\GeV$]  &  --  &  $ > 450$   \\
\bottomrule
\end{tabular}
\end{table}

\subsection{Background estimate}
\label{sec:bgesti2}
The SM backgrounds for a final state with two hadronically decaying $\tau$-leptons from a Higgs boson decay and one light lepton from a $W$-boson decay can be separated into two groups.
The first group is composed of events with a real light lepton and two real $\tau$-leptons, and is dominated by multi-boson processes that are estimated by using MC simulation, as described in Section~\ref{sec:wh:bkgestBoson}.
The second group includes processes with one or two $\tau$-leptons from misidentified jets.
Events with a single misidentified $\tau$-lepton are dominated by events with a top quark, which is estimated by using MC normalized to data in a dedicated CR and described in Section~\ref{sec:wh:bkgestTop}.
Events with two misidentified $\tau$-leptons are mostly from $W$+jets events and are estimated by using a data-driven technique -- the fake factor method -- as described in Section~\ref{sec:wh-ff}.
All other SM backgrounds are estimated directly from MC simulation.
Overall, the dominant background contributions in both SRs are from top quark and multi-boson processes, and account for 89\%--90\% of the total background.

\subsubsection{Misidentified $\tau$-lepton background estimate}
\label{sec:wh-ff}
 
The $W$+jets process dominates the background with two misidentified $\tau$-leptons, with less than $7 \%$ contribution from other processes. The fake factor method estimates all processes with two misidentified $\tau$-leptons using a control data sample (FFCR) with $\tau$-leptons that fail to meet the nominal $\tau$-lepton identification requirement. This estimate is obtained as the product of the number of FFCR events and the fake factor (FF), which relates the number of events with looser $\tau$-lepton candidates to the number where $\tau$-leptons meet the nominal identification criteria.
 
To compute the FF, an anti-$\tau$-lepton ($\bar{T}$) is defined that satisfies a looser set of identification criteria, but fails to meet the medium identification criteria of $\tau$-leptons ($T$).
The FF value is the ratio of $T$ events to $\bar{T}$ events.
To estimate the two misidentified $\tau$-lepton contributions, three control regions are defined using the identification criteria applied to the two highest-$\pt$ $\tau$-leptons: $\bar{T}\bar{T}$, $T\bar{T}$, and $TT$.
The estimate of the contribution from processes with two misidentified $\tau$-leptons ($N_\mathrm{fakes}$) can be written as the product of the two fake factors (FF$_{\tau i}^\mathrm{CR}$) and the number of events from the $\bar{T}\bar{T}$ region ($N_\mathrm{\bar{T}\bar{T},~fake~bkg}$):
\begin{equation}
N_\mathrm{fakes}  =  N_\mathrm{\bar{T}\bar{T},~fakes}  \times \mathrm{FF_{\tau1}^{CR}} \times \mathrm{FF_{\tau2}^{CR}},
\end{equation}
where the individual fake factors for the two $\tau$-leptons are written as:
\begin{equation}
~~~~~~~~\mathrm{FF_{\tau1}^{CR}} = \dfrac{N_\mathrm{data,T\bar{T}} - N_\mathrm{MC,T\bar{T}}^{\geq \mathrm{1~truth}~\tau}}{N_\mathrm{data,\bar{T}\bar{T}} - N_\mathrm{MC,\bar{T}\bar{T}}^{\geq \mathrm{1~truth}~\tau}} ~~~~\mathrm{and}
\end{equation}
\begin{equation}
\mathrm{FF_{\tau2}^{CR}} = \dfrac{N_\mathrm{data,TT} - N_\mathrm{MC,TT}^{\geq \mathrm{1~truth}~\tau}}{N_\mathrm{data,T\bar{T}} - N_\mathrm{MC,T\bar{T}}^{\geq \mathrm{1~truth}~\tau}} ~.
\end{equation}
The contamination from events with at least one real $\tau$-lepton
($N_\mathrm{MC}^{\geq \mathrm{1~truth~tau}}$) is estimated from MC
simulation and subtracted when calculating the ratio.

The fake factors are calculated in a $W$-enriched control region (FFCR-Wh), which maximises available statistics by loosening kinematic selection requirements and remains orthogonal to the SRs by inverting the OS requirement, as summarised in Table~\ref{tab:wh-CRVRs}.
 
\begin{table}[!hbp]
\centering
\caption{The definition of the fake factor control and validation regions, FFCR-Wh and FFVR-Wh, respectively.
\label{tab:wh-CRVRs}}
\footnotesize
\begin{tabular}{l |c|c}
\toprule
& FFCR-Wh & FFVR-Wh \\
\midrule     \midrule
$N$ medium $\tau$ & -- & $\geq 2$ \\
$N$ ``very loose'' $\tau$ & $\geq 2$ & -- \\
Charge combination & SS & OS \\
\midrule
$N$ $e/\mu$  &   \multicolumn{2}{c}{$=1$}\\
$N$ $b$-jets &  \multicolumn{2}{c}{$=0$} \\
$|\Delta \phi(\tau_{1},\tau_{2})|$ & \multicolumn{2}{c}{$ < 3$ }\\
\midrule
$m(\tau_{1},\tau_{2})$ [$\GeV$] & $ > 20$ & $ \in[40,160]$  \\
$\mttwo$ [$\GeV$] & $ > 20$ & $> 30$  \\
\bottomrule
\end{tabular}
\end{table}

\begin{figure}[htpb!]
\centering
\subfloat[][]{\includegraphics[width=0.495\textwidth]{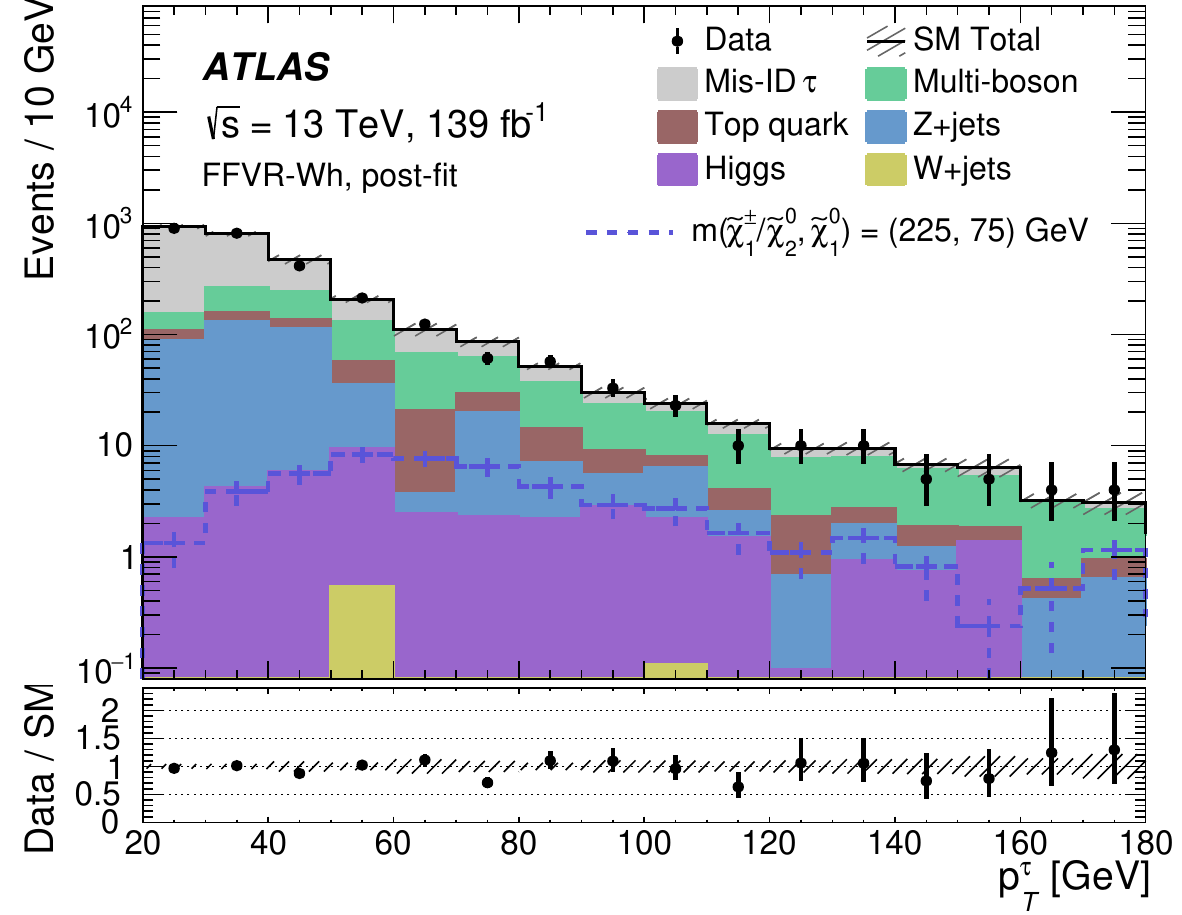}} 
\subfloat[][]{\includegraphics[width=0.495\textwidth]{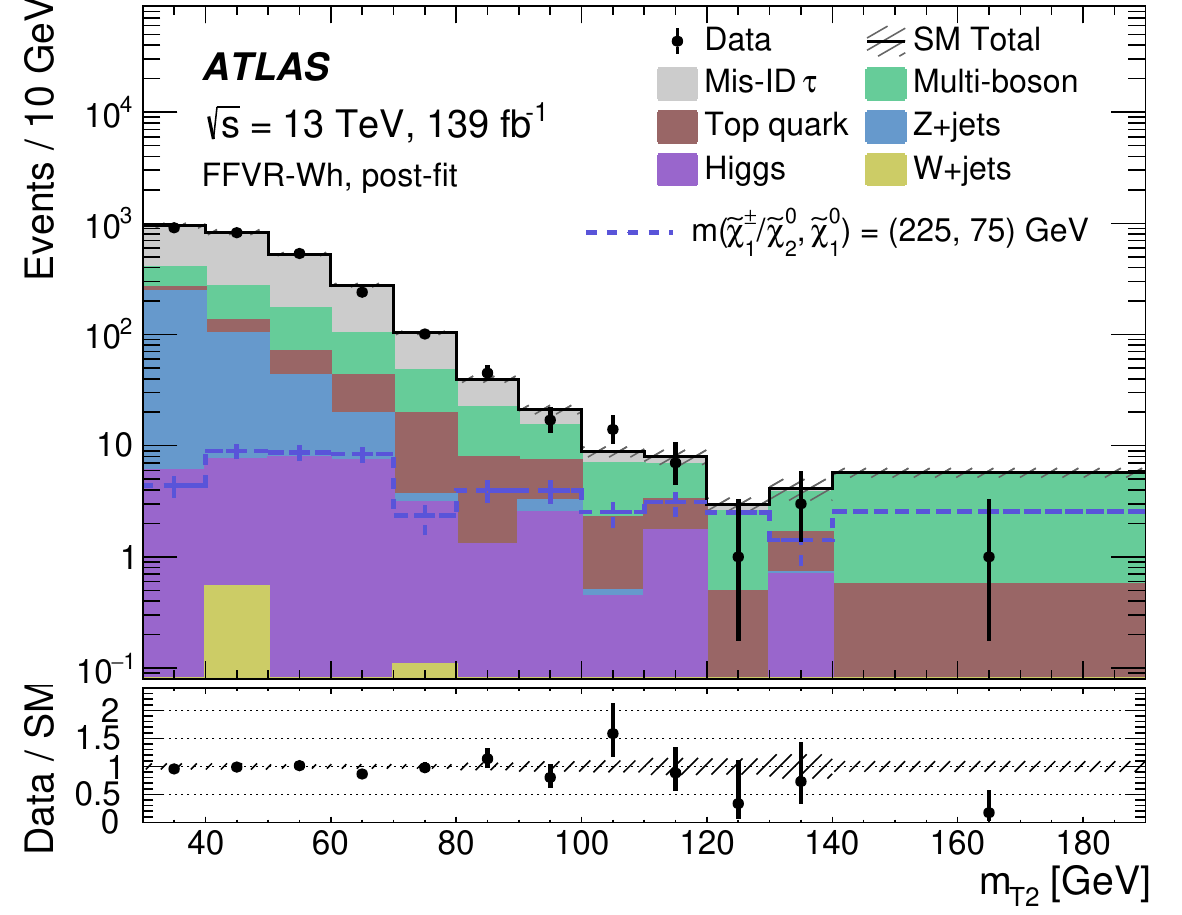}} 
\caption{The distributions of (a) the highest-$\pt$ $\tau$-lepton transverse momentum
$p_{\text{T}}^{\tau}$  and (b) \mttwo\ variables in  FFVR-Wh after the background-only fit.
The misidentified contribution with at least two misidentified
$\tau$-leptons (Mis-ID $\tau$) is estimated from data using the fake factor method.
Representative SUSY scenarios are overlaid for illustration.
The hatched bands represent the combined statistical and systematic uncertainties of the total SM background.
The lower panels show the ratio of data to the total SM background estimate.
\label{fig:Wh-CRVR} }
\end{figure}

The fake factor dependencies on the parameters of the $\tau$-leptons, such
as the number of tracks, $\pt$ and $|\eta|$, are studied and are found to be minimal except for the number of tracks (1-prong or 3-prong $\tau$-leptons).
The fake factors are measured in bins of $\pt$ and $|\eta|$, separately for 1-prong and 3-prong $\tau$-leptons, with the binning  optimised based on the available statistics.
The fake factor for the two highest-$\pt$ $\tau$-leptons is similar at around 0.4 (0.1) for the 1-prong (3-prong) $\tau$-leptons.
The fake factor estimate is validated in a misidentified $\tau$-lepton dominated VR (FFVR-Wh) with a selection similar to the SR-Wh-LM with loosened selections on $m(\tau_{1},\tau_{2}) $ and $\mttwo$, as shown in Table~\ref{tab:wh-CRVRs}.
The kinematic distributions in FFVR-Wh are shown in Figure~\ref{fig:Wh-CRVR} and good agreement between data and SM prediction is observed.

\subsubsection{Top quark background estimate}
\label{sec:wh:bkgestTop}
 
The top quark background is small in SR-Wh-HM and is accounted for in the estimate with the FF method since it is dominated by two misidentified $\tau$-leptons.
However, in SR-Wh-LM the top quark background is more important and is comprised mainly of \ttbar~events with one $W$-boson decaying into
an electron or muon, and the other into a $\tau$-lepton -- the second $\tau$-lepton typically originates from a misidentified jet.
The top quark background is estimated from MC and normalized to data using a top quark enriched control region ($T$CR-Wh) and validated in a top-quark VR ($T$VR-Wh), as defined in Table~\ref{tab:wh-TopMBCRVRs}.
MC simulation shows the top quark background composition is similar across the CR, VR and SR\@.
The selection for the control and validation region are similar to the SR selection, but with one or two $b$-jets required to be orthogonal to the SR\@.
To increase the statistics, the OS requirement is removed and the $m(\tau_{1},\tau_{2})$ requirement is also loosened, and to improve the purity of the top quark background, high $m_{\mathrm{T}\mathrm{sum}}$ is required.
The top quark background purity is between 73\%--81\% in the CR and VR.

\begin{table}[!hbp]
\centering
\caption{The definition of the top quark and multi-boson control and validation regions
for the Intermediate $Wh$ channel.
\label{tab:wh-TopMBCRVRs}}
\footnotesize
\begin{tabular}{l |c|c|c }
\toprule
Process & \multicolumn{2}{c|}{Top} & Multi-boson \\\midrule
Region & $T$CR-Wh & $T$VR-Wh & $MB$VR-Wh\\
\midrule     \midrule
$N$ medium $\tau$ & \multicolumn{3}{c}{$\geq 2$}  \\
Charge combination & \multicolumn{3}{c}{OS} \\
$N$ $e$/$\mu$ &\multicolumn{3}{c}{$=1$} \\
Trigger & \multicolumn{3}{c}{single $e/\mu$} \\
$|\Delta \phi(\tau_{1},\tau_{2})|$ &  \multicolumn{3}{c}{$<3$ } \\
\midrule
$N$ $b$-jets & \multicolumn{2}{c|}{$\in[1,2]$ } & $=0$ \\
$p_{T\tau_{2}}$ [$\GeV$] & \multicolumn{2}{c|}{-} & $>30$ \\
$m_{{\mathrm T}, l}$ [$\GeV$] & \multicolumn{2}{c|}{-} & $>70$ \\
$m(\tau_{1},\tau_{2})$ [$\GeV$] & \multicolumn{2}{c|}{$\in(40,160]$ } & $\in(40,70]$ \\
$m_{\mathrm{T}\mathrm{sum}}$ [$\GeV$] &  \multicolumn{2}{c|}{$>250$ } & -- \\
$\mttwo$ [$\GeV$] & $\in(20,80]$ & $>80$ & $<80$ \\
\bottomrule
\end{tabular}
\end{table}

\subsubsection{Multi-boson background estimate}
\label{sec:wh:bkgestBoson}
Multi-boson production is the dominant SM contribution in both the Intermediate $Wh$ SRs.
The main contribution is $WZ$ production with both the bosons decaying leptonically, giving  one light lepton and two real $\tau$-leptons in the SRs. Smaller contributions in SR-Wh-LM stem from $ZZ$ or $WW$ production decaying into one light lepton, one real $\tau$-lepton, and one misidentified $\tau$-lepton.
 
\begin{figure}[h!]
\centering
\includegraphics[trim = 0mm 25mm 0mm 0mm, clip, width=0.9\textwidth]{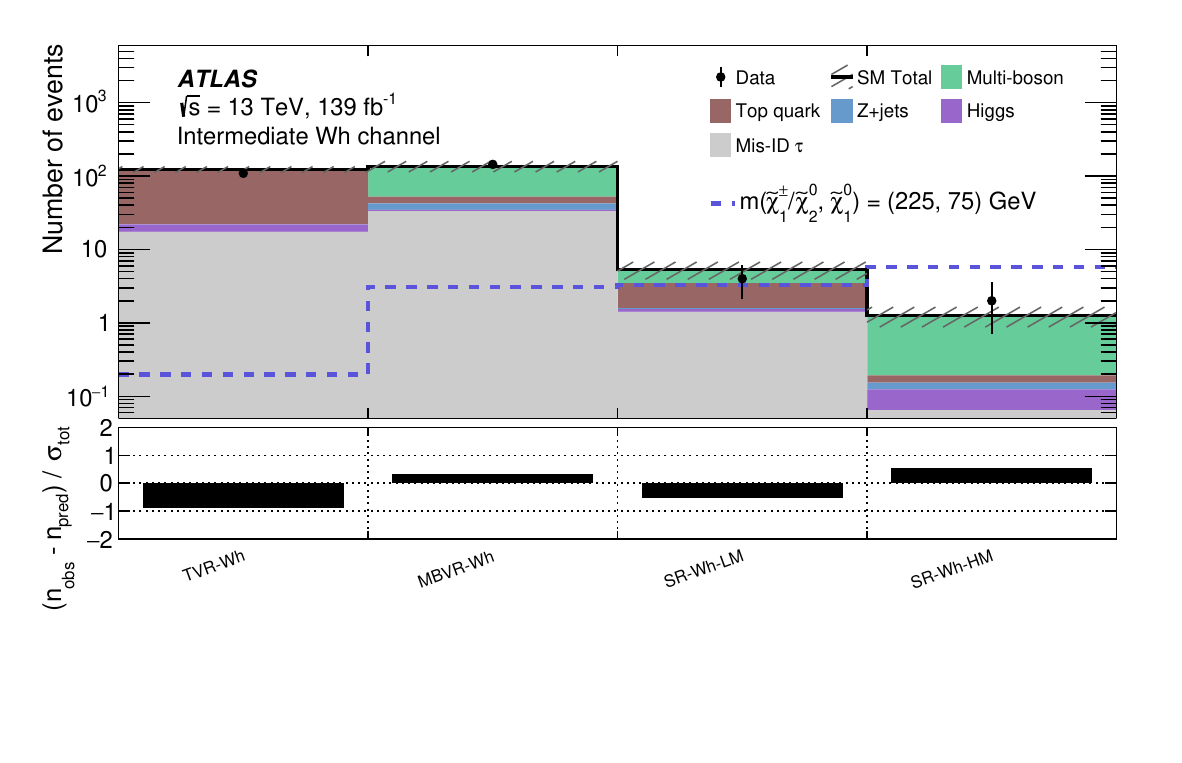} 
\caption{Comparison of the
observed and expected event yields in the $T$VRs and $MB$VRs and two SRs after the
background-only fit.
Representative SUSY scenarios are overlaid for illustration.
The hatched band represents the combined statistical and systematic uncertainties of the total SM background.
The lower panel shows the significance of any difference between the data and total SM background estimate yields.
}
\label{fig:bkg-vr-wh}
\end{figure}

The multi-boson background is estimated from MC simulation and validated in a multi-boson enriched region, $MB$VR-Wh, defined in Table~\ref{tab:wh-TopMBCRVRs}.
The selection on the invariant mass of the $\tau$-leptons is lowered compared with the SRs, and an upper threshold on $\mttwo$ of $80\,\GeV$ is required.
The multi-boson purity is found to be 61\% in $MB$VR-Wh.
The good agreement between data and the SM prediction is shown in the validation and signal regions for the Intermediate $Wh$ channels in Figure~\ref{fig:bkg-vr-wh}.
 
\section{Systematic uncertainties}
\label{sec:systematics}
 
Systematic uncertainties have an impact on the background and signal estimates in the control and signal regions. Uncertainties arising from experimental effects and theoretical sources are considered.
The main sources of experimental systematic uncertainty in the SM
background estimates include $\tau$-lepton and jet energy calibrations and resolution,
$\tau$-lepton identification, systematic effects due to the presence of pileup events, and
uncertainties related to the modelling of \met\ in the simulation. The uncertainties in the energy and momentum scale of each of the objects
entering the \met\ calculation are estimated, as well as the uncertainties in the soft-term resolution and scale~\cite{ATLAS-CONF-2018-023}.
A variation in the pileup reweighting of the MC simulated event samples is included to cover the uncertainty in the ratio of the predicted and
measured inelastic cross-section~\cite{Aaboud:2016mmw}. The uncertainty in the combined 2015--2018 integrated luminosity was measured to be $1.7\%$~\cite{ATLAS-CONF-2019-021}, obtained using the LUCID-2 detector~\cite{LUCID2} for the primary luminosity measurements.
 
Theoretical uncertainties affecting the main irreducible backgrounds $W$+jets, $Z$+jets, top quark processes, and dibosons, are estimated by varying the generator parameters: renormalization and factorisation scales, as well as PDFs, following the PDF4LHC recommendations~\cite{PDF4LHC}. Uncertainties due to the choice of renormalization and factorisation scales are included by varying the scales from their nominal values by a factor of two or one half -- the two scale variations are taken as uncorrelated and the additional coherent up/down variation of the two scales is also considered. Additionally, cross-section uncertainties are assigned to be included in the normalization of the signal and the background processes taken directly from MC simulation.
 
Several sources of uncertainty are considered for the ABCD method used to
determine the multi-jet background estimate for Direct stau channel and the Intermediate stau channel, they include:  the correlation between the
$\tau$-lepton identification and the kinematic variables
\mttwo,
the limited number of events in the CRs, and the subtraction of other SM backgrounds.
The systematic uncertainty in the correlation is estimated by using the transfer factor from VR-E to VR-F instead of that from CR-B to CR-C.
The systematic uncertainty in the non-multi-jet background subtraction in the control regions is estimated by considering the total uncertainties of the MC estimates of the non-multi-jet background in the CRs.
The systematic uncertainty due to the limited number of events in the control regions is estimated by taking the statistical uncertainty of the event yields in these control regions.
The statistical uncertainty from the data yields in every CR is propagated to the uncertainty on the background estimation in the SRs.
 
\begin{figure}[h!]
\centering
\includegraphics[width=0.9\textwidth]{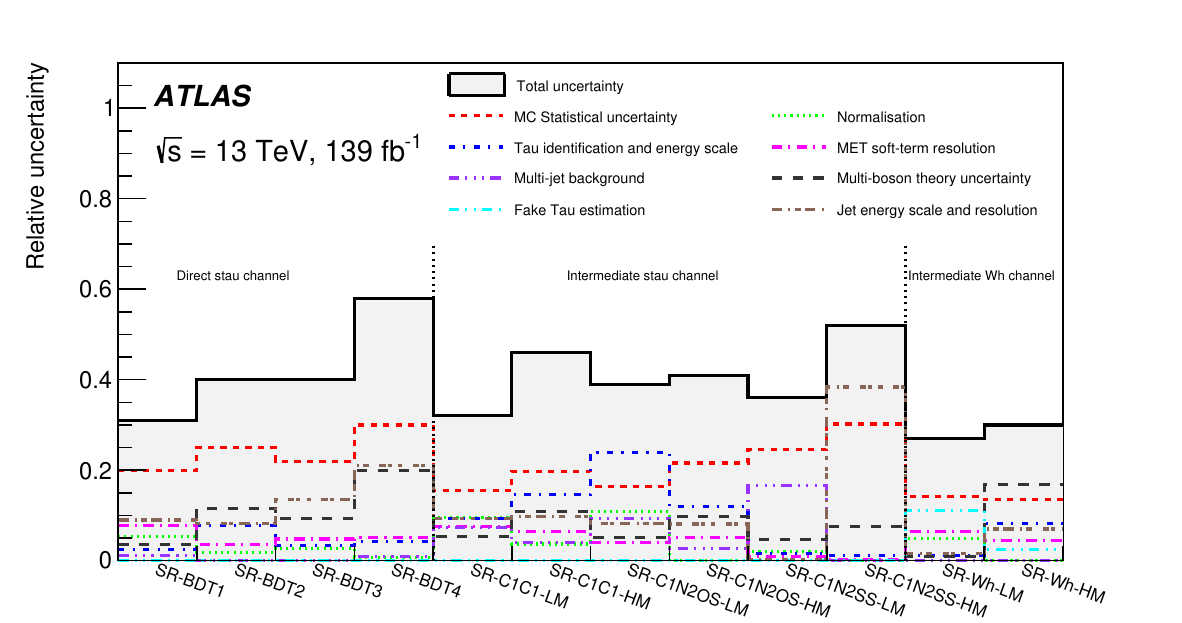}\\  
\caption{Summary of the total uncertainty in the predictions of the background event yields of each SR in the three search channels. The dominant systematic contributions are indicated by individual dashed/dot-dashed/dotted lines.  The total uncertainty in each SR is denoted by the solid black line. For simplicity, the individual uncertainties are added in quadrature for each category, without accounting for correlations.
Theoretical uncertainties in the multiboson MC simulation-based estimates are grouped under the ``Multi-boson theory uncertainty'' category, while the ``Normalisation'' category includes the
statistical uncertainties of the data counts in the CRs and the uncertainty from the fitted normalization
factors.
}
\label{fig:syst-summary}
\end{figure}

The different sources of uncertainty considered for the fake factor method used to
determine the misidentified background with at least two misidentified $\tau$-leptons in the Intermediate $Wh$ channel are as follows. The fake factor values are varied up and down by their
statistical uncertainties and the difference is used as a source of uncertainty. A further 30\%
systematic uncertainty in the subtracted MC processes is used as a conservative
estimate on the systematic uncertainty from MC subtractions in
the FFCR-Wh. The difference between the quark/gluon contributions in the fake factor control regions and signal regions was studied and found to be negligible compared with the uncertainties placed on the overall normalization and statistical contributions.
The systematic uncertainties and their sizes in each SR are summarized in Figure \ref{fig:syst-summary}.
 
The dominant contributions of systematic uncertainties in all scenarios are
mainly from the statistics of the MC samples, the normalization
uncertainties of the multi-jet background, the $\tau$-lepton identification and the energy scale, jet energy scale and resolution.
In the Intermediate stau channel, uncertainties in the multi-jet estimate are also important.
In the Direct stau and Intermediate $Wh$ channels, multi-boson theory uncertainties represent a major contribution to the systematics.

\section{Results}
\label{sec:Results}
\label{sec:fitStrategy}
 
The results of the Direct stau, Intermediate stau, and Intermediate $Wh$ channels are presented in this section as background-only combined fits to the control and signal regions, along with model-dependent and model-independent fits for BSM signal exclusion.
 
\subsection{Direct stau production analysis}
 
\begin{figure}[h!] 
\centering
\subfloat[][]{\includegraphics[width=0.49\textwidth]{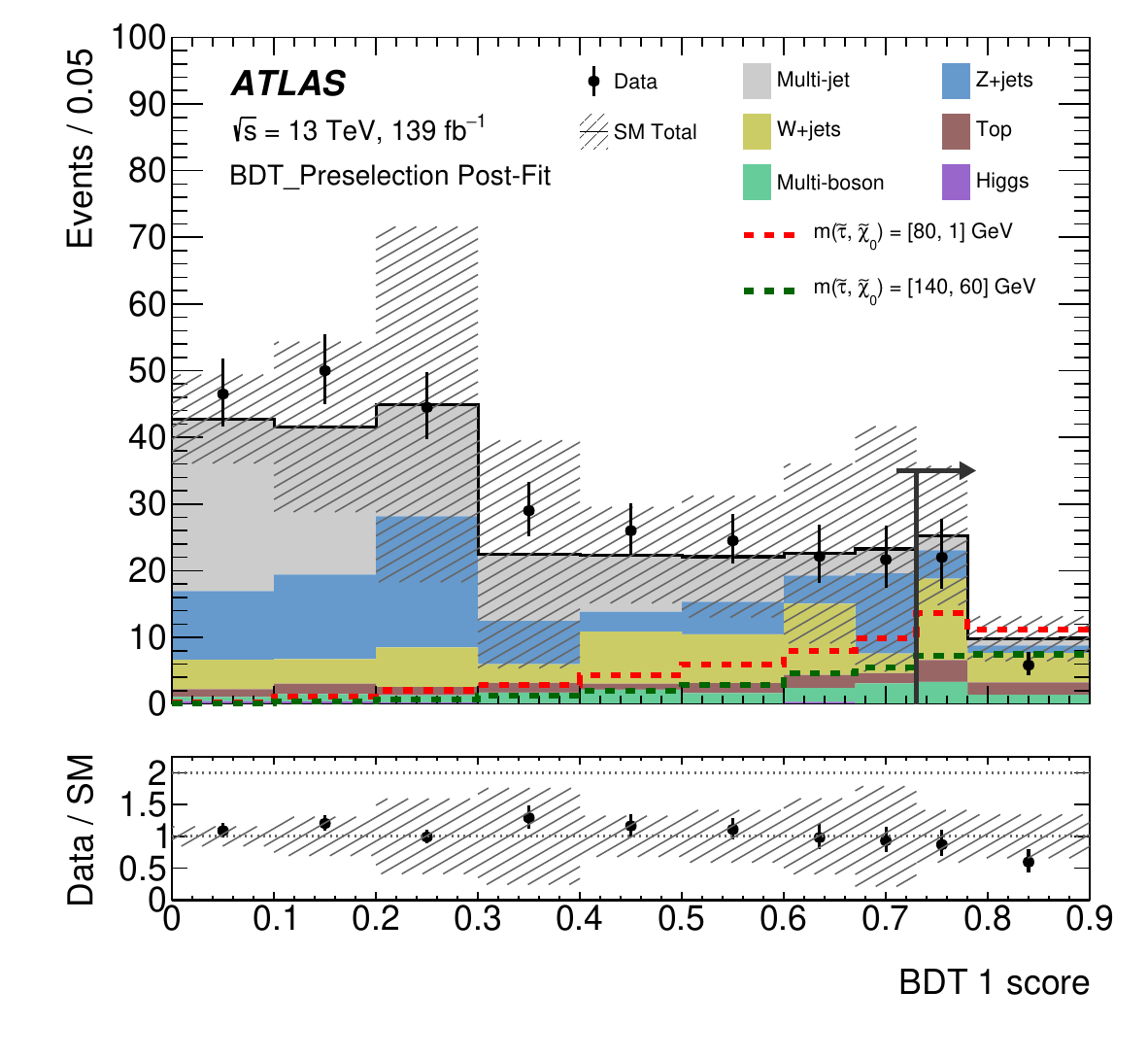}}
\subfloat[][]{\includegraphics[width=0.49\textwidth]{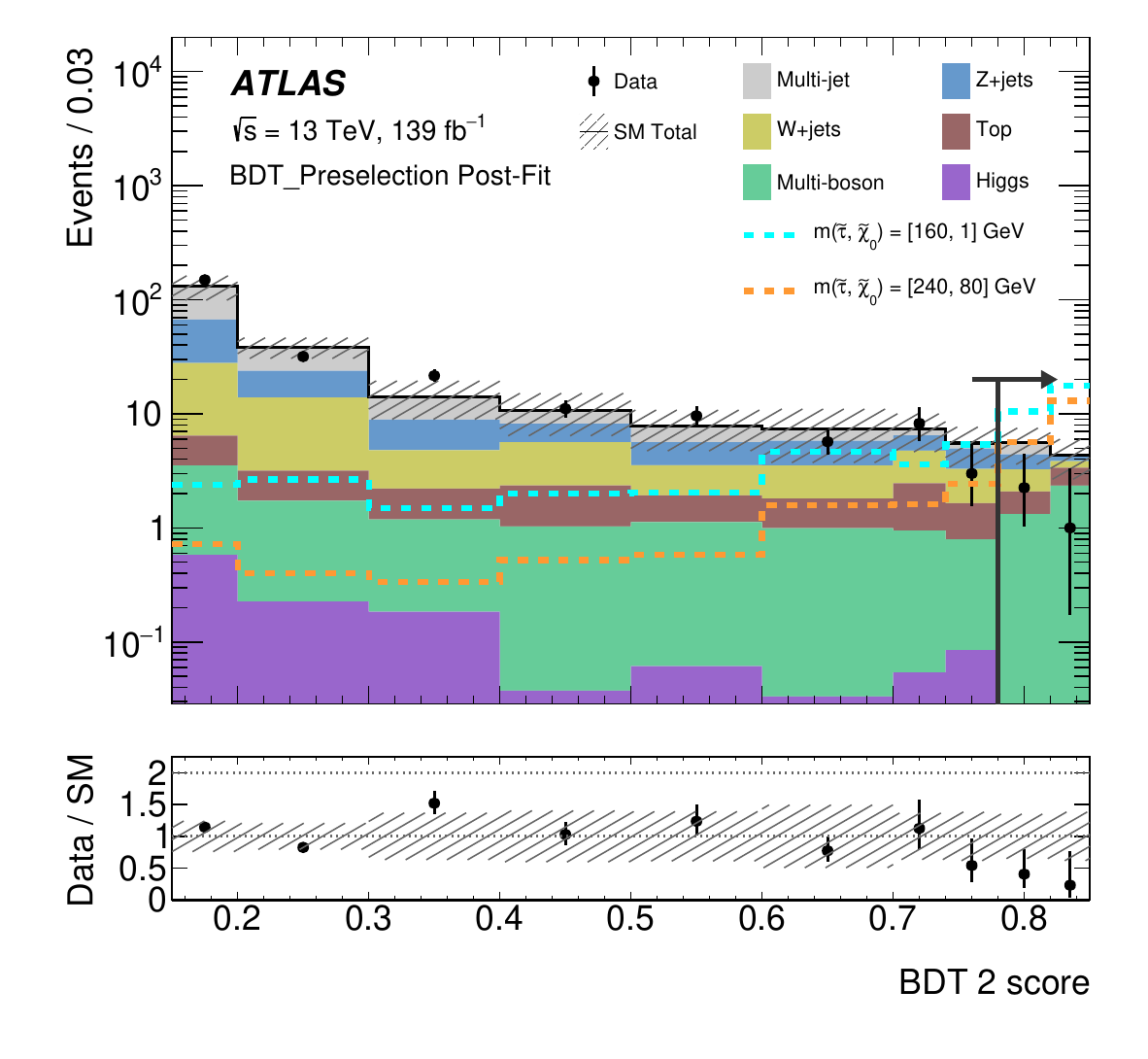}}\\
\subfloat[][]{\includegraphics[width=0.49\textwidth]{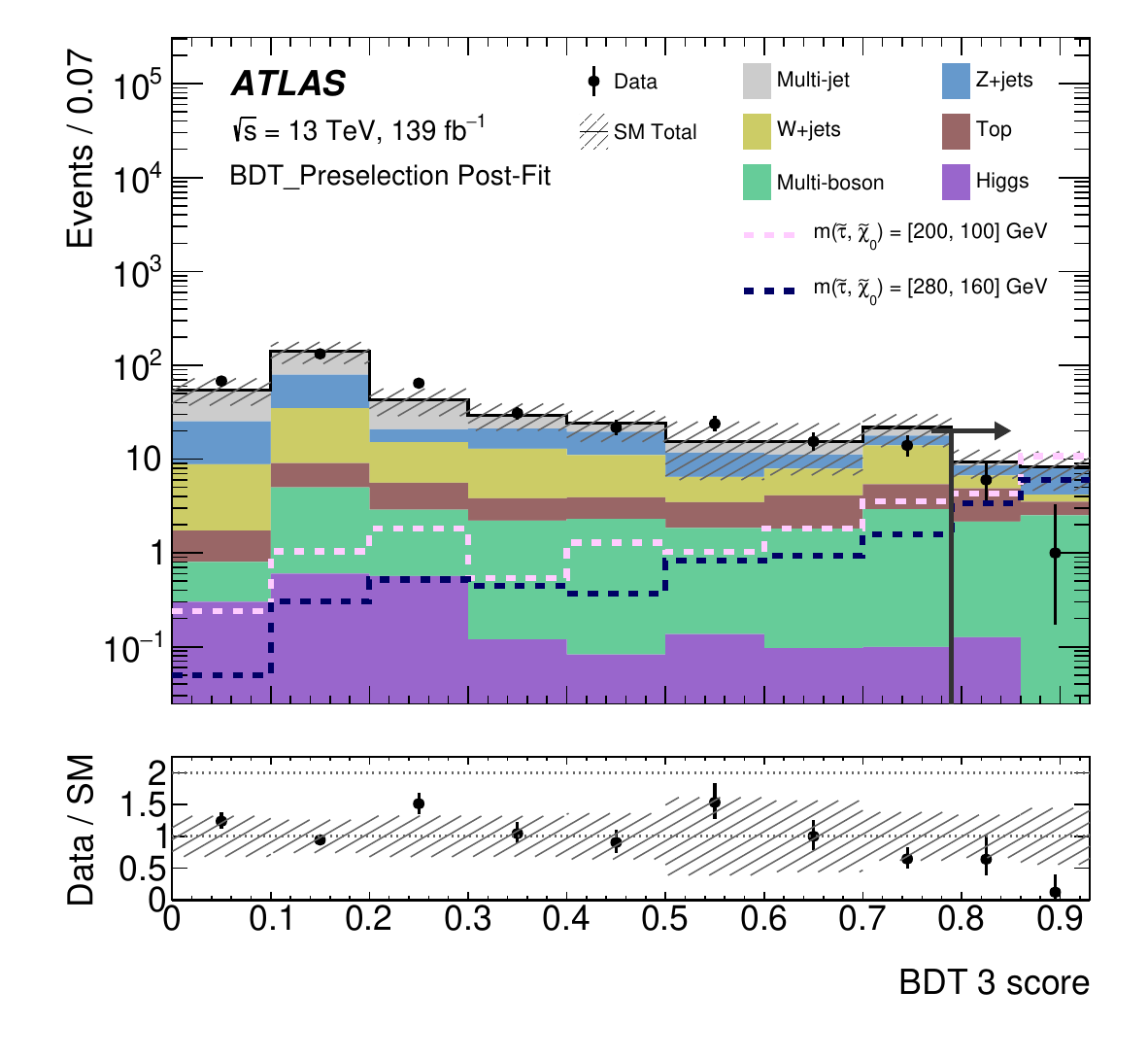}}
\subfloat[][]{\includegraphics[width=0.49\textwidth]{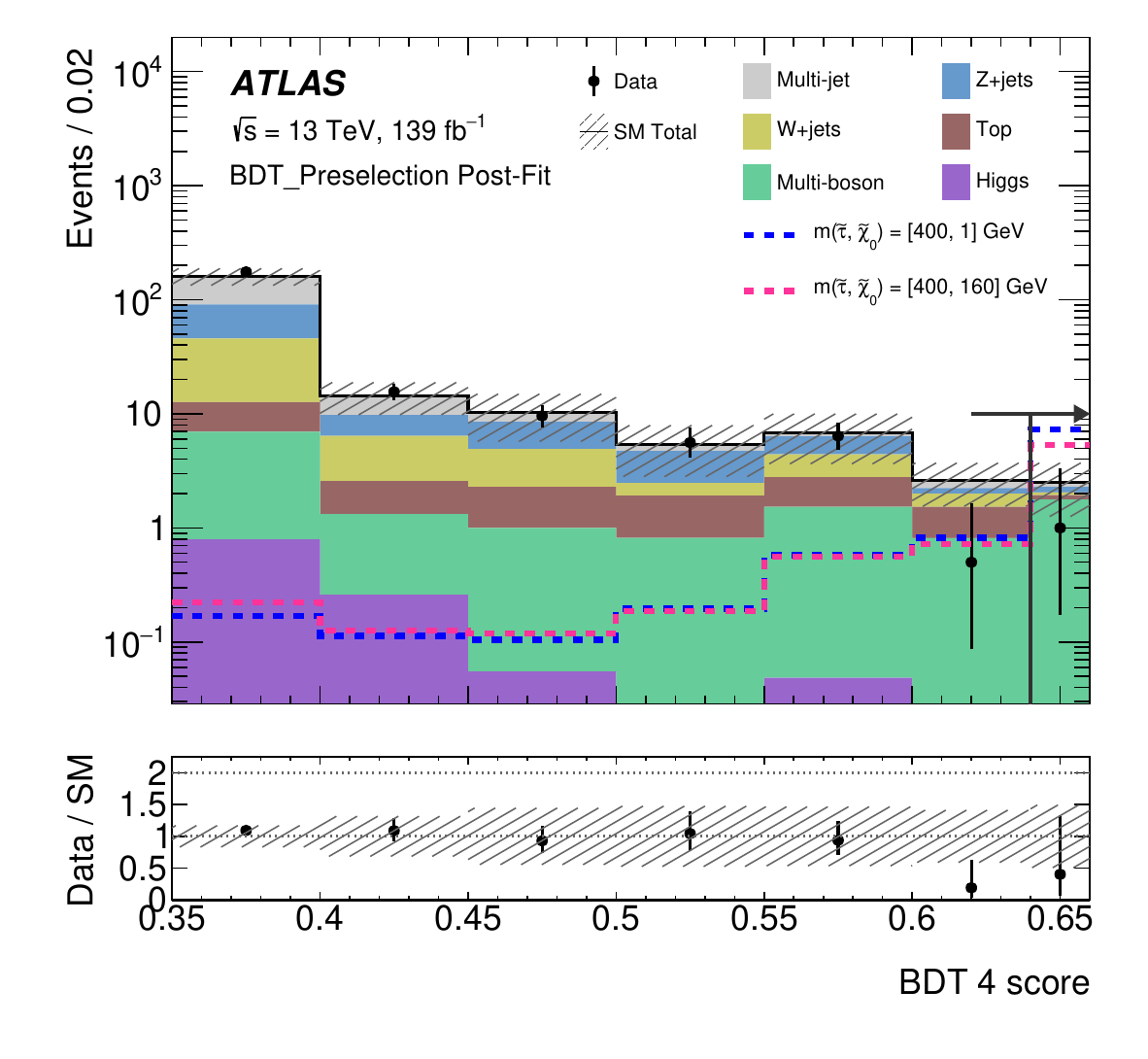}}\\
\caption{The BDT score distributions for the Direct stau channel after the background-only fit, showing the scores for (a) BDT1, (b) BDT2, (c) BDT3, and (d) BDT4, before the selections on the BDT scores are made.
The black arrow depicts the BDT score selection for the SR-BDT\@.
A few example SUSY scenarios targeted by each BDT are overlaid for illustration.
The hashed bands represent the combined statistical and systematic uncertainties of the total SM background.
The lower panels show the ratio of data to the total SM background estimate.
}
\label{fig:SRBDTplots_DS}
\end{figure}

The expected and observed numbers of events in the Direct stau signal regions are shown in Table~\ref{tab:stauSRyields}, where the observations are consistent with the SM expectations. The distributions of the four BDT scores are shown in Figure~\ref{fig:SRBDTplots_DS}.
All the SR-BDT strongly overlap and show a common deficit of significance $0.7-1.3\,\sigma$. For example, the single event in data selected by SR-BDT4 is also selected by the other three SR-BDT, while the four events in SR-BDT2 are also selected by SR-BDT3.
The $W$+jets, $Z$+jets, and top quark backgrounds are normalized to data in their respective control regions, obtaining normalization factors of $0.93\pm0.11$, $0.91\pm0.07$, and $0.85\pm0.05$, respectively, as shown in Table~\ref{tab:fit:NF}.

\begin{table}[h!]
\centering
\caption{
Observed and expected numbers of events after the background-only fit in the signal regions targeting direct stau production.
The uncertainties correspond to the sum of statistical and systematic uncertainties.
The correlation of systematic uncertainties among control regions and
among background processes is taken into account.
The one-sided $p_0$-value, and the observed and expected \percent{95} CL
upper limits on the visible non-SM cross-section ($\sigma_{\mathrm{vis}}^{95}$) from the model-independent fit are given for the unbinned SRs. Values of $p_0>0.5$  are truncated to $p_0=0.5$.
\label{tab:stauSRyields}
}
\footnotesize
\renewcommand\arraystretch{1.3}
\begin{tabular}{l||c|c|c||c|c|c}
\toprule
SM process  & SR-BDT1 & SR-BDT1\_bin0 & SR-BDT1\_bin1 & SR-BDT2 & SR-BDT2\_bin0 & SR-BDT2\_bin1 \\
\midrule
Multi-boson & $6.6 \pm 3.1$ & $3.3 \pm 1.9$ & $3.3 \pm 1.7$ & $4.1 \pm 2.0$ & $1.7 \pm 0.9$ & $2.3 \pm 1.2$ \\
W+jets & $23 \pm 10$ & $12 \pm 7$ & $11 \pm 4$ & $1.6_{-1.6}^{+1.7}$ & $1.6_{-1.6}^{+1.6}$ & $0.09_{-0.09}^{+0.09}$ \\
Top & $7.8 \pm 2.6$ & $3.3 \pm 1.7$ & $4.5 \pm 1.9$ & $2.1 \pm 1.2$ & $1.0 \pm 0.9$ & $1.0 \pm 0.7$ \\
Z+jets & $7 \pm 7$ & $4 \pm 4$ & $2.8 \pm 2.6$ & $2.0 \pm 0.9$ & $1.6 \pm 0.8$ & $0.40 \pm 0.20$ \\
Higgs & $-$ & $0.004_{-0.004}^{+0.006}$ & $-$ & $0.02_{-0.02}^{+0.14}$ & $0.02_{-0.02}^{+0.13}$ & $0.004_{-0.004}^{+0.018}$ \\
Multi-jet & $4.6 \pm 2.8$ & $2.2 \pm 1.6$ & $2.4 \pm 2.3$ & $<1.0$ & $<1.0$ & $<0.16$ \\
\midrule
SM total & $49 \pm 15$ & $25 \pm 10$ & $24 \pm 8$ & $10 \pm 4$ & $5.9 \pm 2.8$ & $3.9 \pm 1.6$ \\
\midrule
Observed & $36$ & $22$ & $14$ & $4$ & $3$ & $1$ \\
\midrule
$m(\stau, \ninoone) =  (80, 1) \,\GeV$ & $40 \pm 9$ & $14 \pm 4$ & $27 \pm 6$ & $4.2 \pm 2.0$ & $3.4 \pm 1.4$ & $0.7 \pm 0.7$ \\
$m(\stau, \ninoone) = (400, 1) \,\GeV$ & $4.6 \pm 0.9$ & $1.2 \pm 0.4$ & $3.4 \pm 0.8$ & $9.5 \pm 1.1$ & $2.1 \pm 0.4$ & $7.4 \pm 0.9$ \\
\midrule
\rule{0pt}{2.6ex}
$p_0$                                 & $0.5$ & -- & -- & $0.5$ & -- & -- \\
Expected $\sigma_\text{vis}^{95}$ [fb]  & $ 0.22$ & -- & -- & $ 0.06$ & -- & -- \\
Observed $\sigma_\text{vis}^{95}$ [fb]  & $ 0.18$ & -- & -- & $ 0.04$ & -- & -- \\
\bottomrule
\end{tabular}

\begin{tabular}{l||c|c|c||c}
\toprule
SM process  & SR-BDT3 & SR-BDT3\_bin0 & SR-BDT3\_bin1 & SR-BDT4 \\
\midrule
Multi-boson & $4.6 \pm 2.2$ & $2.0 \pm 1.2$ & $2.5 \pm 1.3$ & $1.8 \pm 1.0$ \\
W+jets & $2_{-2}^{+4}$ & $2_{-2}^{+4}$ & $0.5 \pm 0.5$ & $<0.7$ \\
Top & $3.7 \pm 1.3$ & $2.7 \pm 1.1$ & $1.0 \pm 0.9$ & $0.16_{-0.16}^{+0.19}$ \\
Z+jets & $6 \pm 4$ & $1.9_{-1.9}^{+2.7}$ & $4.1 \pm 2.6$ & $0.2_{-0.2}^{+0.9}$ \\
Higgs & $0.13_{-0.13}^{+0.15}$ & $0.13_{-0.13}^{+0.14}$ & $0.001_{-0.001}^{+0.012}$ & $<0.015$ \\
Multi-jet & $<1.1$ & $<1.0$ & $<0.26$ & $0.20_{-0.20}^{+0.28}$ \\
\midrule
SM total & $17 \pm 7$ & $9 \pm 6$ & $8.2 \pm 3.3$ & $2.4 \pm 1.4$ \\
\midrule
Observed & $7$ & $6$ & $1$ & $1$ \\
\midrule
$m(\stau, \ninoone) = (80, 1) \,\GeV$ & $11 \pm 4$ & $8.6 \pm 3.0$ & $2.2 \pm 1.4$ & $0.13_{-0.13}^{+0.22}$ \\
$m(\stau, \ninoone) =  (400, 1) \,\GeV$ & $9.4 \pm 1.4$ & $3.7 \pm 0.7$ & $5.7 \pm 1.1$ & $7.3 \pm 0.9$ \\
\midrule
\rule{0pt}{2.6ex}
$p_0$                                 & $0.5$ & -- & -- & $0.5$ \\
Expected $\sigma_\text{vis}^{95}$ [fb]  & $ 0.07$ & -- & -- & $ 0.03$ \\
Observed $\sigma_\text{vis}^{95}$ [fb]  & $ 0.04$ & -- & -- & $ 0.03$ \\
\bottomrule
\end{tabular}
 
\end{table}

\begin{table}[ht]
\centering
\footnotesize
\caption{Normalization factors from the
background-only ﬁt in each scenario for the Direct stau, Intermediate stau, and Intermediate $Wh$ channels.  The normalization factors include corrections to the misidentified $\tau$-lepton efficiency in addition to the cross-section and acceptance effects.}
\begin{tabular}{l|c|c|c|c|c}
\toprule
Channel  & Direct stau & \multicolumn{3}{c|}{Intermediate stau} & Intermediate $Wh$\\
\midrule
Normalization factor  & $\stau\stau$ & C1C1 & C1N2OS  & C1N2SS & C1N2Wh \\
\midrule
$\mu_{W\text{+jets}}$ & $0.93\pm0.11$ & $0.98 \pm 0.12$ & $0.98 \pm 0.12$ & $1.04 \pm 0.09$ & --\\
\midrule
$\mu_{Z\text{+jets}}$ & $0.91\pm0.07$ & -- & -- & -- & --\\
\midrule
$\mu_{\text{Top}}$ & $0.85\pm0.05$ & -- & -- & $0.71 \pm 0.11$ & $1.00\pm0.14$ \\
\midrule
$\mu_{\text{Multi-jet}}$ & -- & $1.0 \pm 0.4$ & $1.00 \pm 0.29$ & $1.00 \pm 0.04$ & -- \\
\bottomrule
\end{tabular}
\label{tab:fit:NF}
\end{table}

\begin{figure}[h!] 
\centering
\subfloat[][]{\includegraphics[width=0.49\textwidth]{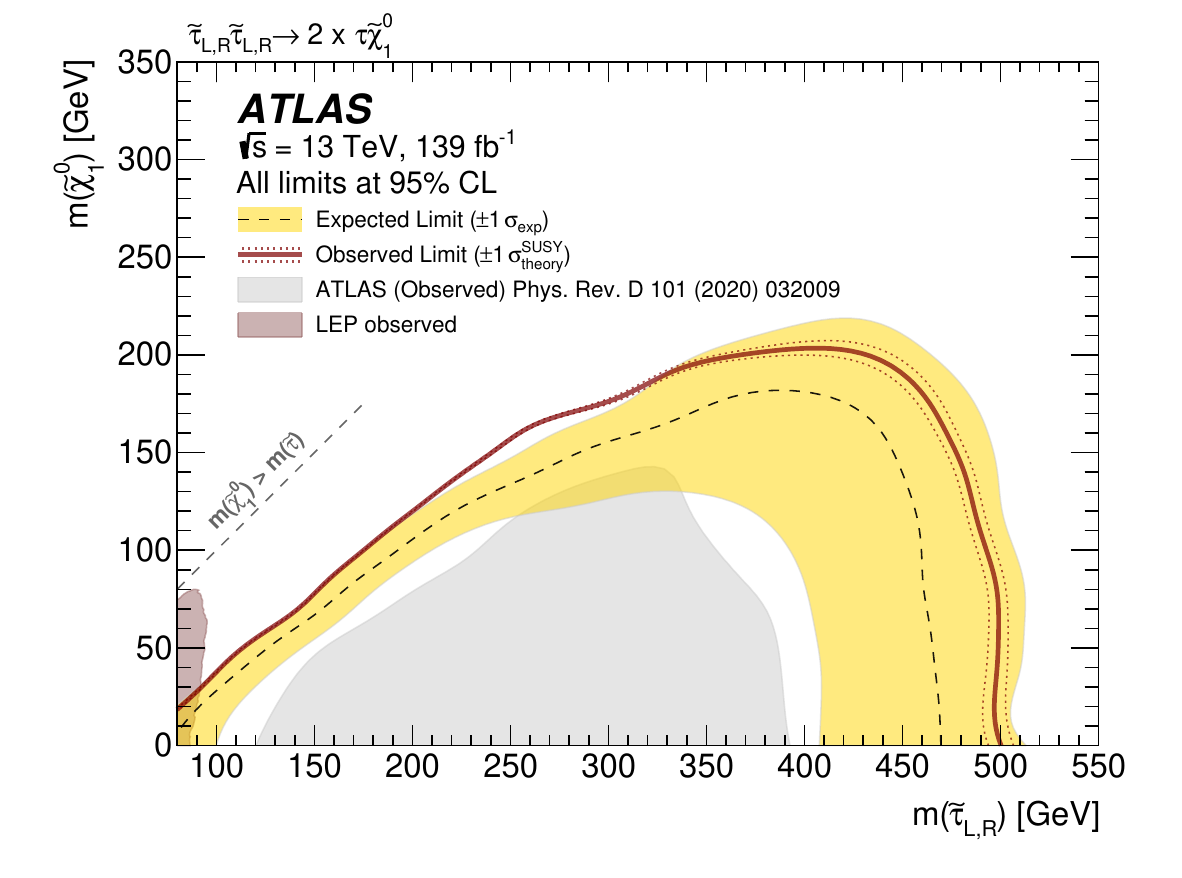}\label{fig:ExclStauLR}} \\
\subfloat[][]{\includegraphics[width=0.49\textwidth]{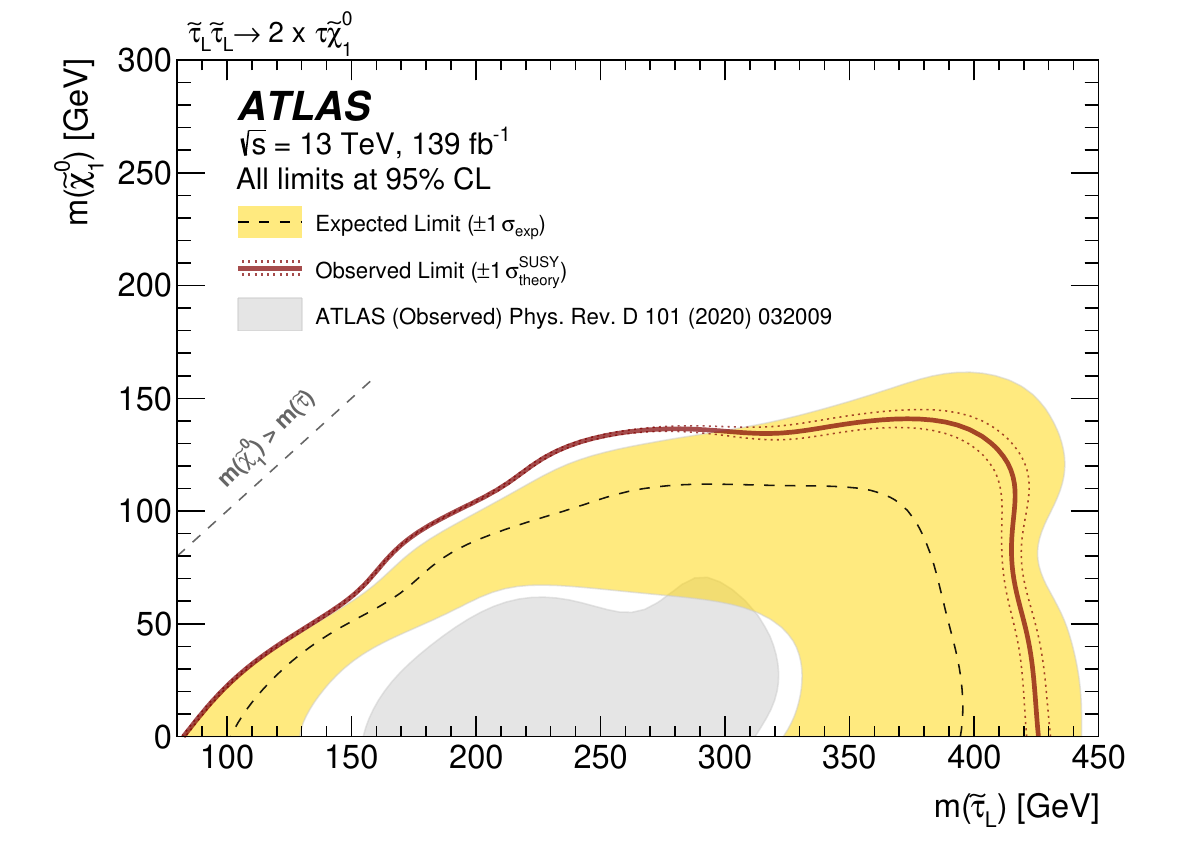}\label{fig:ExclStauL}}
\subfloat[][]{\includegraphics[width=0.49\textwidth]{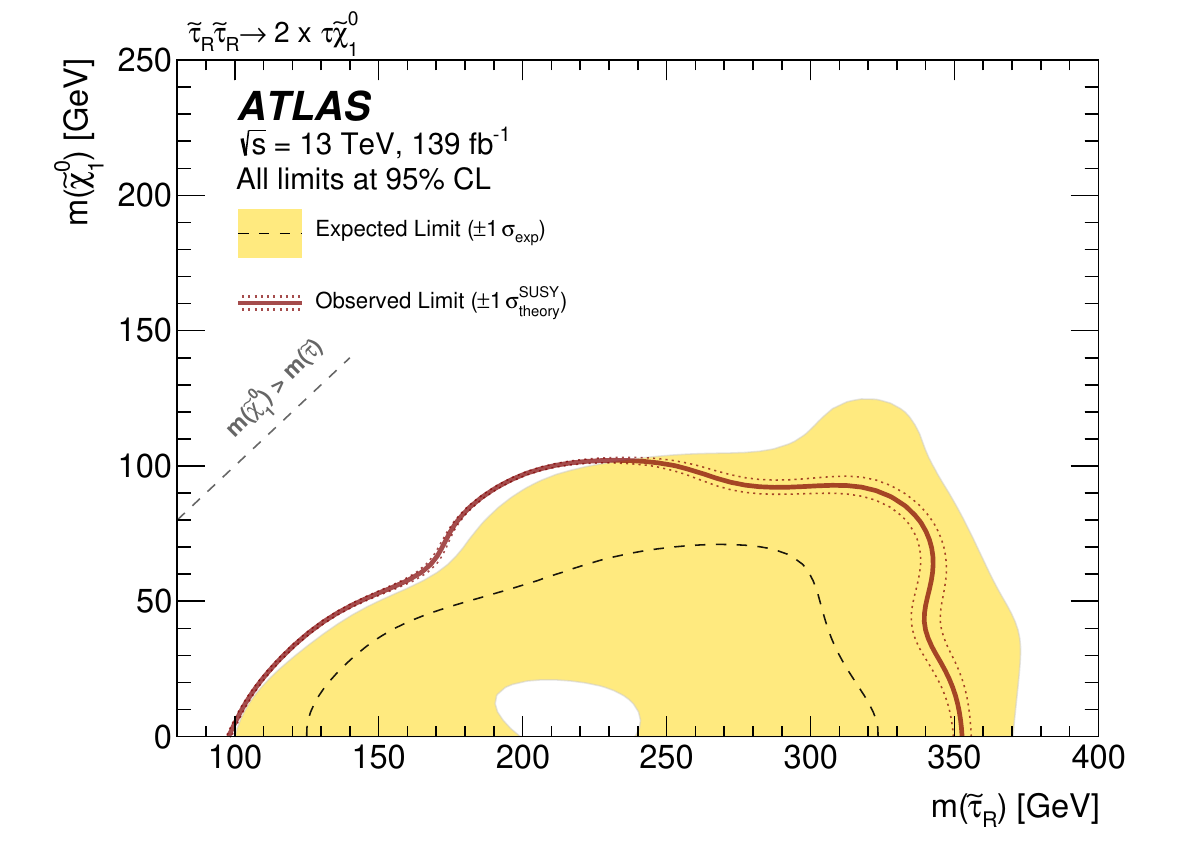}\label{fig:ExclStauR}}
\caption{The 95\% CL exclusion contours for simplified models of (a) $\stau_{L,R}\stau_{L,R}$ production, (b) $\stau_{L}\stau_{L}$ production, and (c) $\stau_{R}\stau_{R}$ production.
The solid (dashed) lines show the observed (expected) exclusion contours. The band around the expected limit shows the $\pm1\sigma$ variations, including all uncertainties except theoretical uncertainties in the signal cross-section.
The expected exclusion contour from the previous ATLAS result in Ref.~\cite{SUSY-2018-04} is shown as a gray-filled contour in (a) and (b).
The observed contour from LEP in Ref.~\cite{LEPSUSYWG:01-03.1} is shown as a brown-filled contour in (a).
\label{fig:alimit-stauDS}
}
\end{figure}

The model-dependent fits using the results from the signal and control regions are used to place exclusion limits at 95\% CL on Direct stau production. SR-BDT4 has a low predicted yield, so 10,000 pseudoexperiments are used to calculate the exclusion limits.
Since the four main SR-BDT overlap, the SR with the lowest expected CLs value for each signal scenario is used to set the limit, statistically combining the two bins each within SR-BDT1, SR-BDT2, and SR-BDT3 as two distinct regions.
The exclusion limits for mass-degenerate $\stau_{L,R}$ production are shown in Figure~\ref{fig:ExclStauLR}, where stau masses up to $500\,\GeV$ are excluded for massless $\ninoone$. The limits are improved relative to previous results, particularly towards smaller $\stau$-$\ninoone$ mass splittings, as well as at higher $\stau$ masses. The observed limits are stronger than the expected limit due to the small observed deficit. The feature in the observed limit around stau masses of $250-300\,\GeV$ is due to a transition from one SR-BDT to another being used to set the limit (a similar effect is also seen for Figures~\ref{fig:ExclStauL}-\ref{fig:ExclStauR}).
 
The exclusion limits for $\stau_L$ and $\stau_R$ production separately are shown in Figure~\ref{fig:ExclStauL} and Figure~\ref{fig:ExclStauR}, respectively. Similar improvements are seen in the sensitivity to $\stau_L$ production as in the mass-degenerate case, with stau masses excluded up to $425\,\GeV$. Sensitivity to $\stau_R$ production is obtained for the first time at ATLAS, with masses excluded up to $350\,\GeV$.
 
\subsection{Intermediate stau analysis}
\begin{figure}[h!]
\centering
\subfloat[][]{\includegraphics[width=0.49\textwidth]{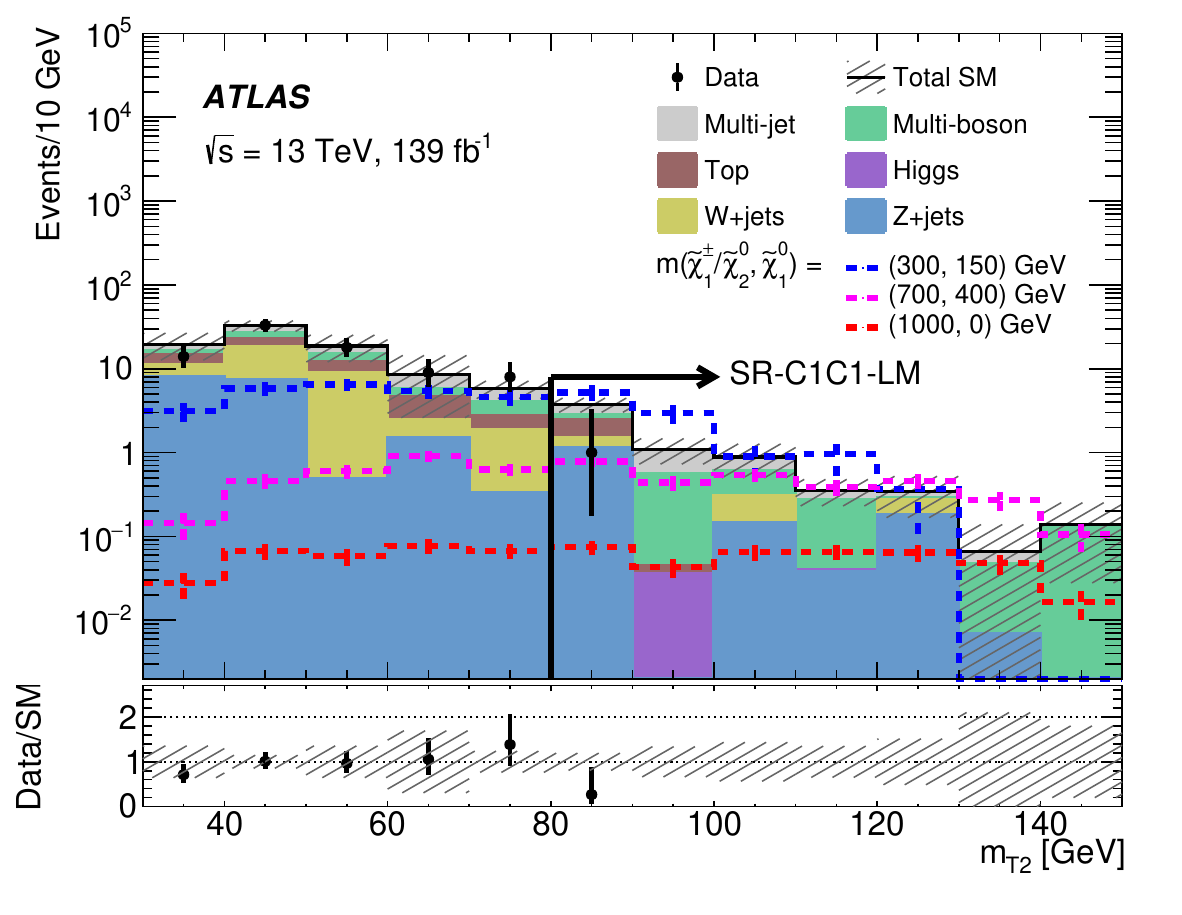}}   
\subfloat[][]{\includegraphics[width=0.49\textwidth]{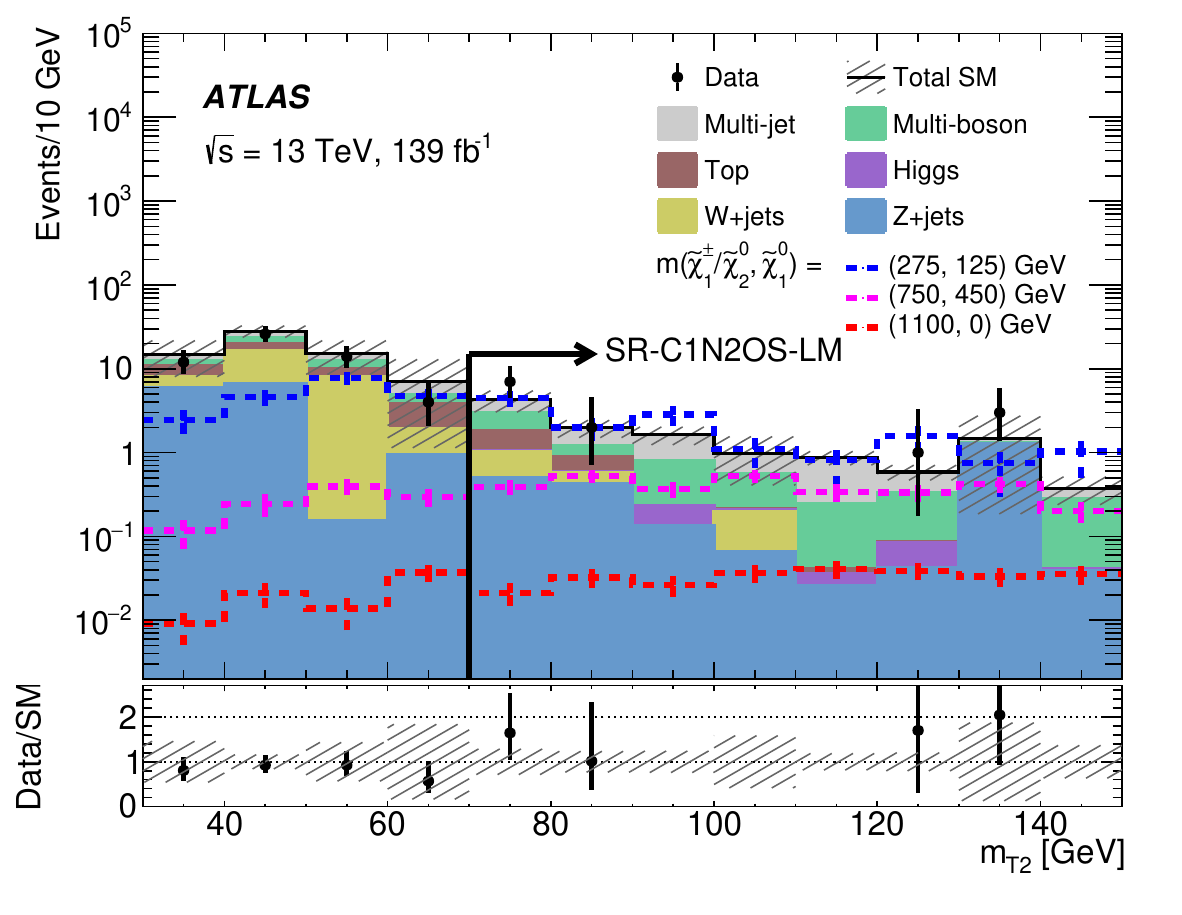}}\\ 
\subfloat[][]{\includegraphics[width=0.49\textwidth]{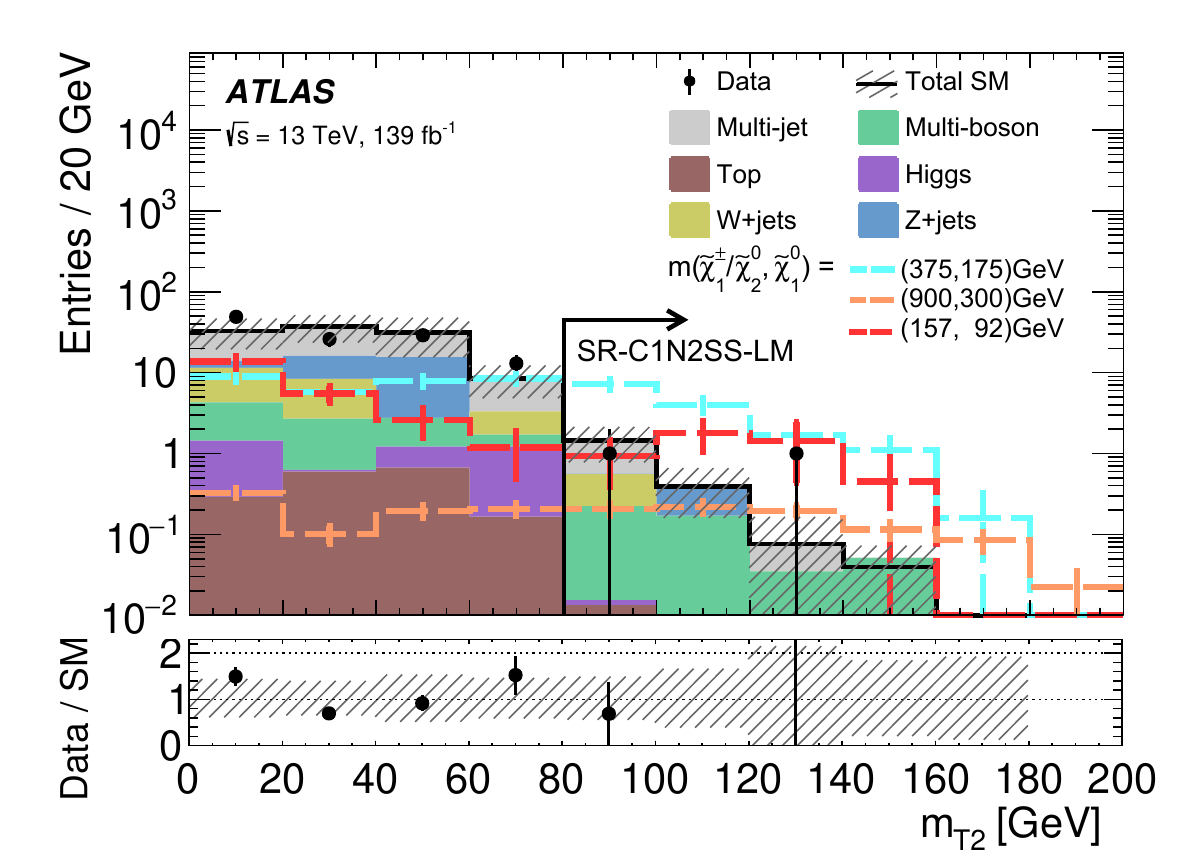}} 
\caption{ The kinematic distributions for the Intermediate stau channel after the background-only fit, showing the $\mttwo$ distribution in (a) SR-C1C1-LM, (b) SR-C1N2OS-LM, and (c) SR-C1N2SS-LM before the selection on $\mttwo$ is made.
The black arrow depicts the selection for the signal region.
An example SUSY scenario is overlaid for illustration.
The hashed bands represent the combined statistical and systematic uncertainties of the total SM background.
The lower panels show the ratio of data to the total SM background estimate.
}
\label{fig:SRBDTplots_IntStau}
\end{figure}

The observed number of events in each SR and the expected contributions from SM processes
are given in Table~\ref{tab:results1}. The contributions of multi-jet,
$W$+jets and top quark events are scaled with the normalization factors
obtained from the background-only fit.
The multi-jet normalization with respect to the prediction from the
ABCD method in all SRs is
compatible with unity and has an uncertainty of \percent{29--40} (\percent{4}), due to
the small number of observed events in the multi-jet CR-A in C1C1 and
C1N2OS (C1N2SS) scenarios.
The $W$+jets normalization factor is measured to be 0.98 $\pm$ 0.12 (1.04 $\pm$ 0.09) in C1C1 and
C1N2OS (C1N2SS) scenarios and the top quark background normalization factor is found to be 0.71 $\pm$ 0.11 in the C1N2SS scenario.
The normalization factors are summarised in Table \ref{tab:fit:NF}.
The $\mttwo$ distributions of events in the signal regions are shown in Figure~\ref{fig:SRBDTplots_IntStau}.
In all SRs, observations and background predictions are
found to be compatible within uncertainties.
The one-sided $p_0$-values, and the observed and expected \percent{95} CL upper limits on the visible non-SM cross-section ($\sigma_{\mathrm{vis}}^{95}$) are shown in Table~\ref{tab:results1}.

\begin{table}[h!]
\centering
\caption{
Observed and expected numbers of events for the background-only fit in the signal regions targeting chargino/neutralino production and decay via intermediate staus.
Expected event yields for a few SUSY reference points are also shown.
The uncertainties correspond to the sum in quadrature of statistical  and systematic uncertainties.
The correlation of systematic uncertainties among control regions and
among background processes is taken into account.
The one-sided $p_0$-values, and the observed and expected \percent{95} CL
upper limits on the visible non-SM cross-section ($\sigma_{\mathrm{vis}}^{95}$) from the model-independent fit are given.
}
\footnotesize
\renewcommand\arraystretch{1.3}
\begin{tabular}{l||c|c||c|c}
\toprule
SM process  & SR-C1C1-LM       & SR-C1C1-HM   & SR-C1N2OS-LM & SR-C1N2OS-HM   \\
\midrule
\midrule
Multi-boson  &  1.6 $\pm$  0.6   &  2.2 $\pm$  1.6    &  3.2 $\pm$ 1.2   &  2.4 $\pm$  1.6  \\
$W$+jets      &  0.4 $\pm$  0.4    &  $0.29_{-0.29}^{+0.35}$  &  $0.6_{-0.6}^{+2.2}$   &  $0.29_{-0.29}^{+0.35}$ \\ [+0.5mm]
Top quark     &  1.0 $\pm$  0.5    &   0.36 $\pm$  0.13 &  $1.1_{-1.1}^{+1.2}$   &0.36 $\pm$  0.14  \\
$Z$+jets      &   $1.4_{-1.4}^{+1.5}$ &  0.78 $\pm$  0.34 &2.5 $\pm$ 1.7    &   0.9 $\pm$  0.4  \\
Higgs           &   0.27 $\pm$ 0.06 &  $0.01_{-0.01}^{+0.13}$&  0.40 $\pm$ 0.22      &  0.73 $\pm$ 0.23\\
Multi-jet      &  1.5 $\pm$ 0.5      &  0.37 $\pm$ 0.21  &  4.5 $\pm$ 1.0      &  0.31 $\pm$ 0.17 \\
\midrule
SM total       &  6.2 $\pm$  2.0 &  4.0 $\pm$  1.8     &  12.2 $\pm$  4.8 &  5.0 $\pm$  2.0  \\
\midrule
Observed     &   1   &    4   &   14   &    4 \\
\midrule
$m(\chinoonepm,\ninoone) = (700, 400)\,\GeV$ &   $3.0 \pm 0.6$ & $7.8 \pm 1.6$ &  $4.7 \pm 1.0$ & $14.1 \pm 2.8$ \\
$m(\chinoonepm/\ninotwo,\ninoone) = (1100, 0)\,\GeV$ &   $0.20 \pm 0.05$  & $3.1 \pm 0.6$ &  $0.39 \pm 0.11$  & $4.6 \pm 1.0$  \\
\midrule
$p_0$                                  &0.5   & 0.5          &0.4    & 0.5     \\
Expected $\sigma_\text{vis}^{95}$ [fb]  & $0.04$  &    $0.05$ & $0.10$  &    $0.05$ \\ [+1mm]
 
Observed $\sigma_\text{vis}^{95}$ [fb]  & $0.02$  &    $0.05$ & $0.10$  &    $0.05$  \\
\bottomrule
\end{tabular}
 
\smallskip
 
\begin{tabular}{l||c|c}
\toprule
SM process  & SR-C1N2SS-LM & SR-C1N2SS-HM   \\
\midrule
\midrule
Multi-boson  &   0.47 $\pm$  0.20  &  0.8 $\pm$  0.4 \\
$W$+jets      &    0.33 $\pm$  0.25  &  0.10 $\pm$  0.05\\
Top quark     &   $0.01_{-0.01}^{+0.02}$&  0.59 $\pm$ 0.20 \\
$Z$+jets      &   0.20 $\pm$  0.15    &  $0.6_{-0.6}^{+0.8}$\\
Higgs           &   $<0.01$&  0.02 $\pm$ 0.01\\
Multi-jet      &   0.9 $\pm$ 0.5 &  0.00 $\pm$ 0.00 \\
\midrule
SM total       &   2.0 $\pm$  0.7     &  2.1 $\pm$  1.1\\
\midrule
Observed     &   2   &    3 \\
\midrule
$m(\chinoonepm/\ninotwo,\ninoone) = (157, 92)\,\GeV$ & $4.6 \pm 1.3$  & $0.00 \pm 0.00$  \\
$m(\chinoonepm/\ninotwo,\ninoone) = (900, 300)\,\GeV$ & $0.84 \pm 0.07$  & $6.23 \pm 0.21$  \\
\midrule
$p_0$                          &0.4    & 0.3       \\
Expected $\sigma_\text{vis}^{95}$ [fb]  & $0.03$  &    $0.04$ \\ [+1mm]
 
Observed $\sigma_\text{vis}^{95}$ [fb] & $0.03$  &    $0.04$ \\
\bottomrule
\end{tabular}
 
\label{tab:results1}
\end{table}

\begin{figure}[h!]
\centering
\subfloat[][]{\includegraphics[width=0.49\textwidth]{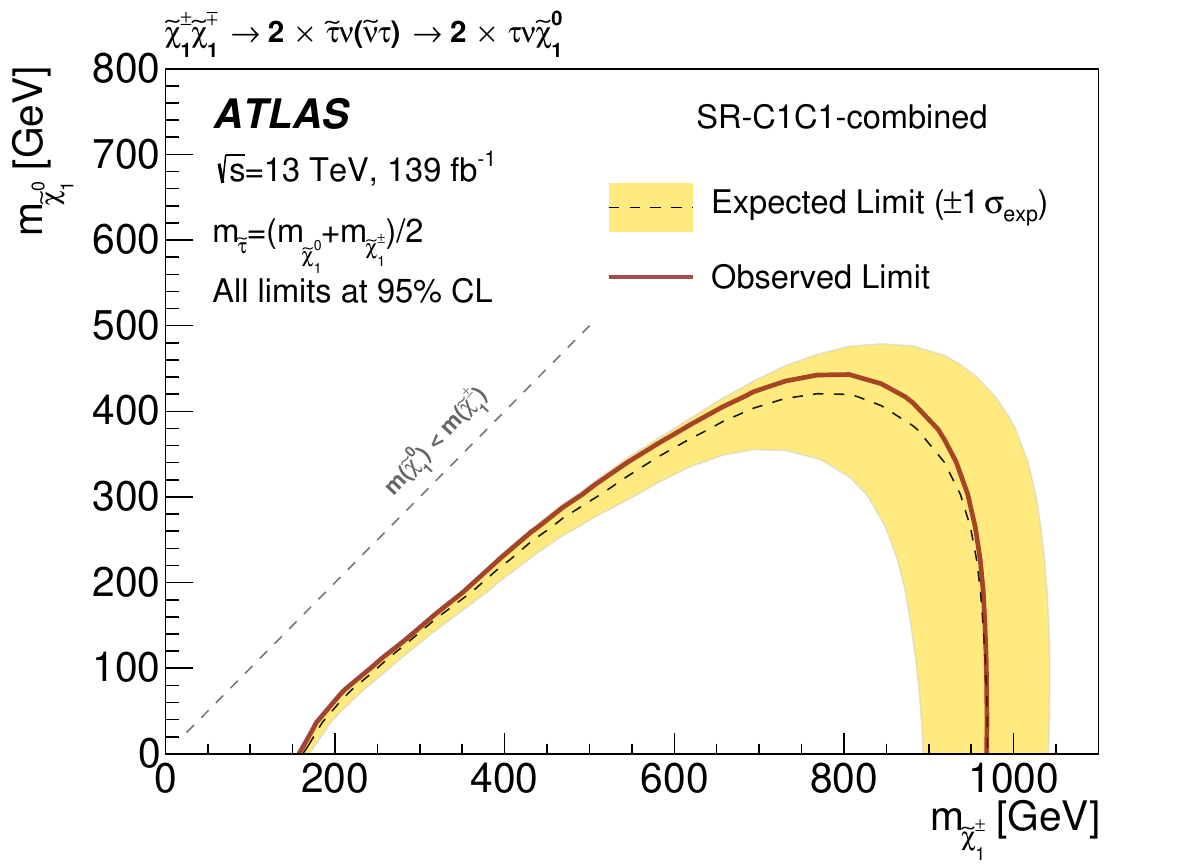}} \\ 
\subfloat[][]{\includegraphics[width=0.49\textwidth]{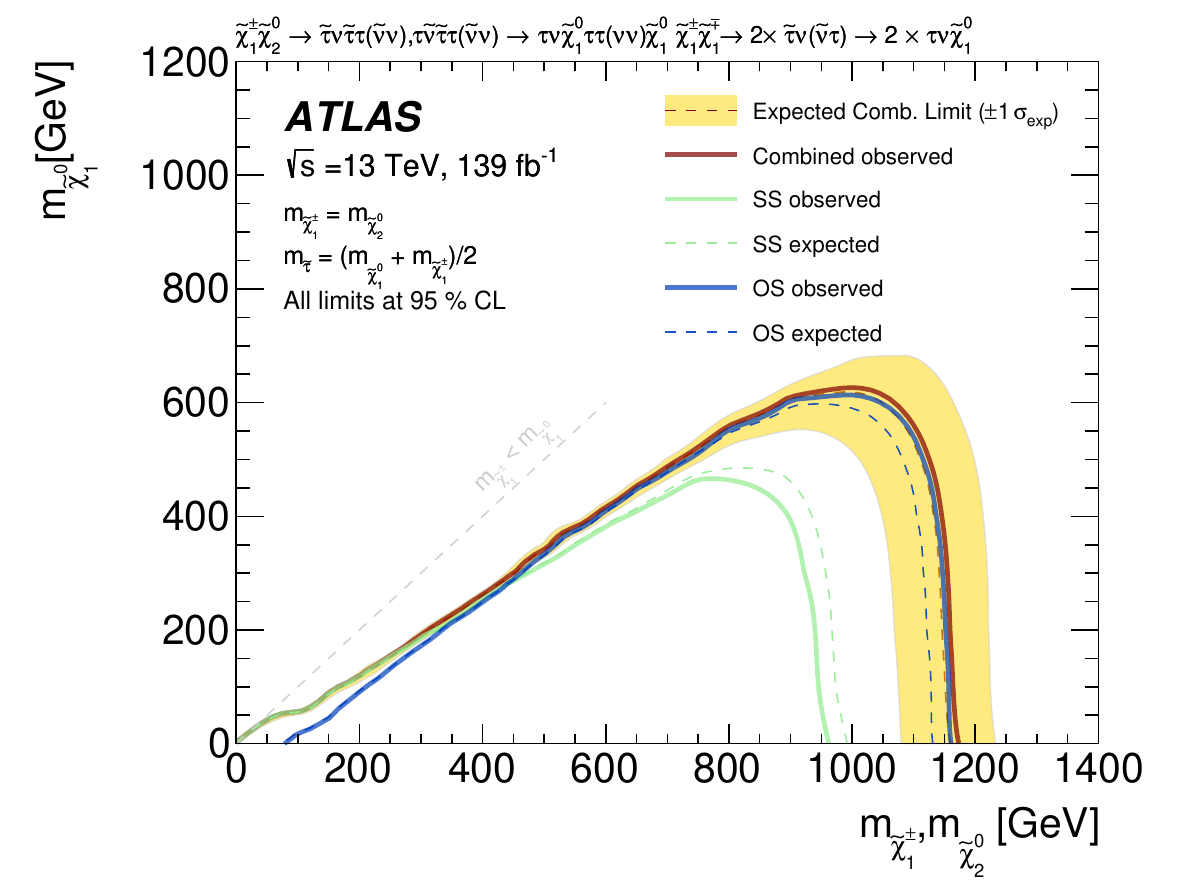}}  
\subfloat[][]{\includegraphics[width=0.49\textwidth]{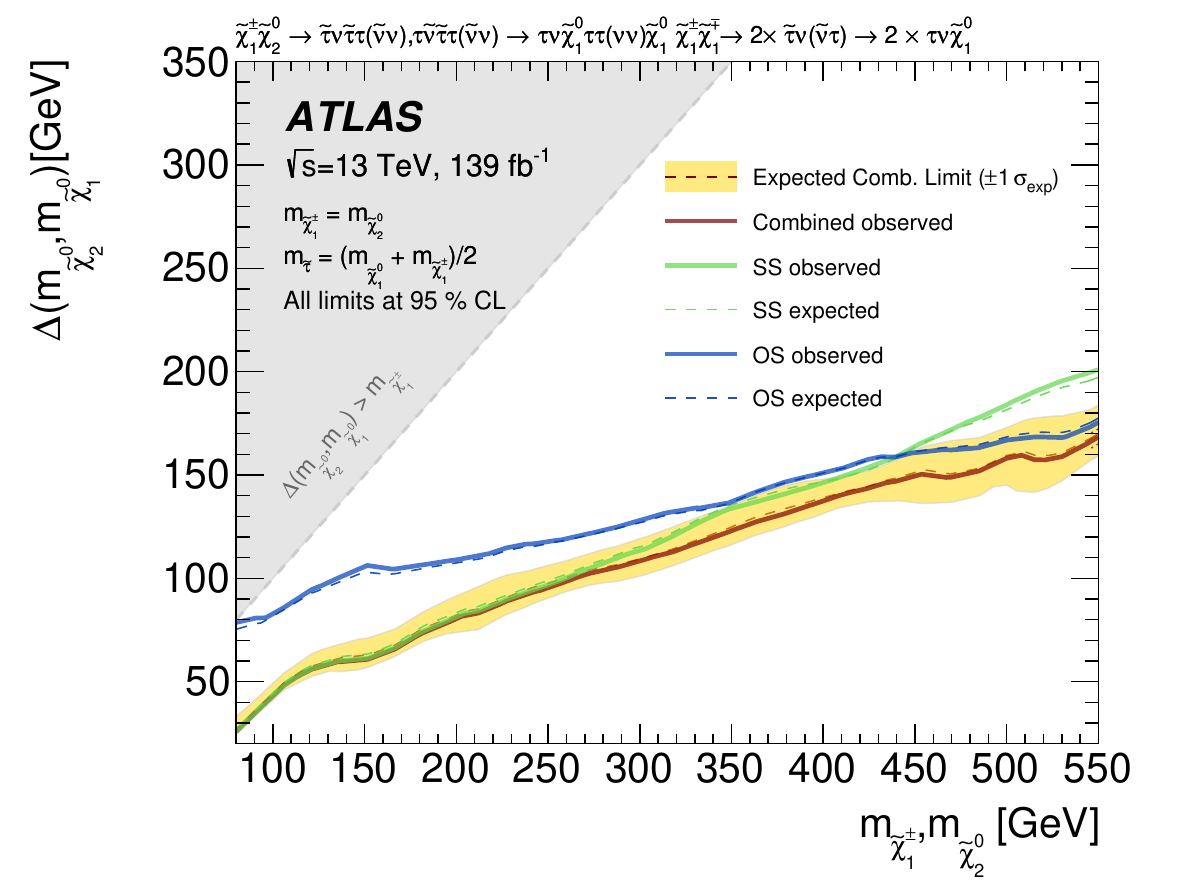}} 
\caption{ The \percent{95} CL exclusion contours for simplified
models of (a) \ConepConem\ production, and of (b-c) \ConepConem\ and \ConepmNtwo\ production.
The solid (dashed) lines show the observed (expected)
exclusion contours. The band around the expected limit shows the
$\pm1\sigma$ variations, including all
uncertainties except theoretical uncertainties in the signal cross-section.
The green curves are from the contribution of C1N2SS scenario, while
the blue curves are from the contribution of C1C1 and C1N2OS scenarios, and
the red curves are the combination of the channels.
The gray solid area in (c) shows the forbidden area where $m(\ninoone) > m(\ninotwo)$.
}
\label{fig:alimit-stau}
\end{figure}

In the absence of a significant excess over the expected SM background, the observed and expected numbers of events in the signal regions are used to
place exclusion limits at \percent{95} CL using the
model-dependent fit. SR-C1C1-LM and SR-C1C1-HM are statistically combined to derive limits on \ConepConem\ production, and
SR-C1N2OS-LM, SR-C1N2OS-HM, SR-C1N2SS-LM and SR-C1N2SS-HM are combined to derive limits for the production of \ConepConem\ and \ConepmNtwo.
The exclusion limits for simplified models
are shown in Figure~\ref{fig:alimit-stau}. Only
\ConepConem\ production is assumed for Figure~\ref{fig:alimit-stau} (a),
whereas both production processes are considered simultaneously for
the Figure~\ref{fig:alimit-stau} (b) and (c).
The C1N2SS channel contributes significantly to the combination in the lower mass regions where this channel does not contain significant SM backgrounds.
 
Chargino masses up to $970\,\GeV$ are excluded for decays into a massless neutralino in the direct production of chargino pairs.
For production of chargino pairs of mass-degenerate charginos and next-to-lightest neutralinos,
chargino masses up to $1160\,\GeV$ are excluded for a massless  neutralino. Both the limits apply to scenarios where
the neutralinos and charginos decay solely via intermediate staus and $\tau$-sneutrinos.
These limits significantly extend previous
results~\cite{SUSY-2016-23, CMS-SUS-19-012} in the high $\chinoonepm$/$\ninotwo$ mass
region. The improvement at compressed and low $\chinoonepm$/$\ninotwo$
masses is mainly driven by the C1N2SS analysis.
 
\subsection{Intermediate $Wh$ analysis}
\begin{figure}[h!]
\centering
\subfloat[][]{\includegraphics[width=0.49\textwidth]{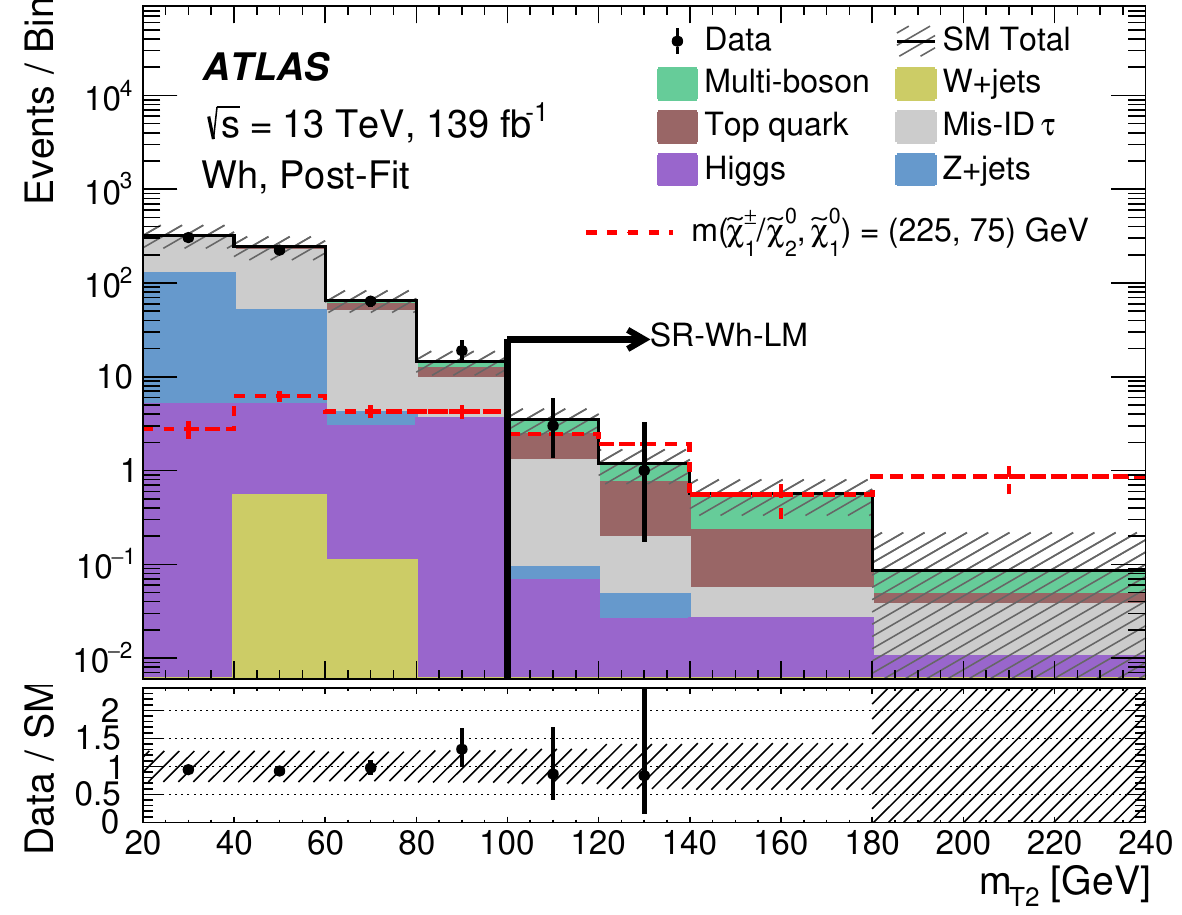}} 
\subfloat[][]{\includegraphics[width=0.49\textwidth]{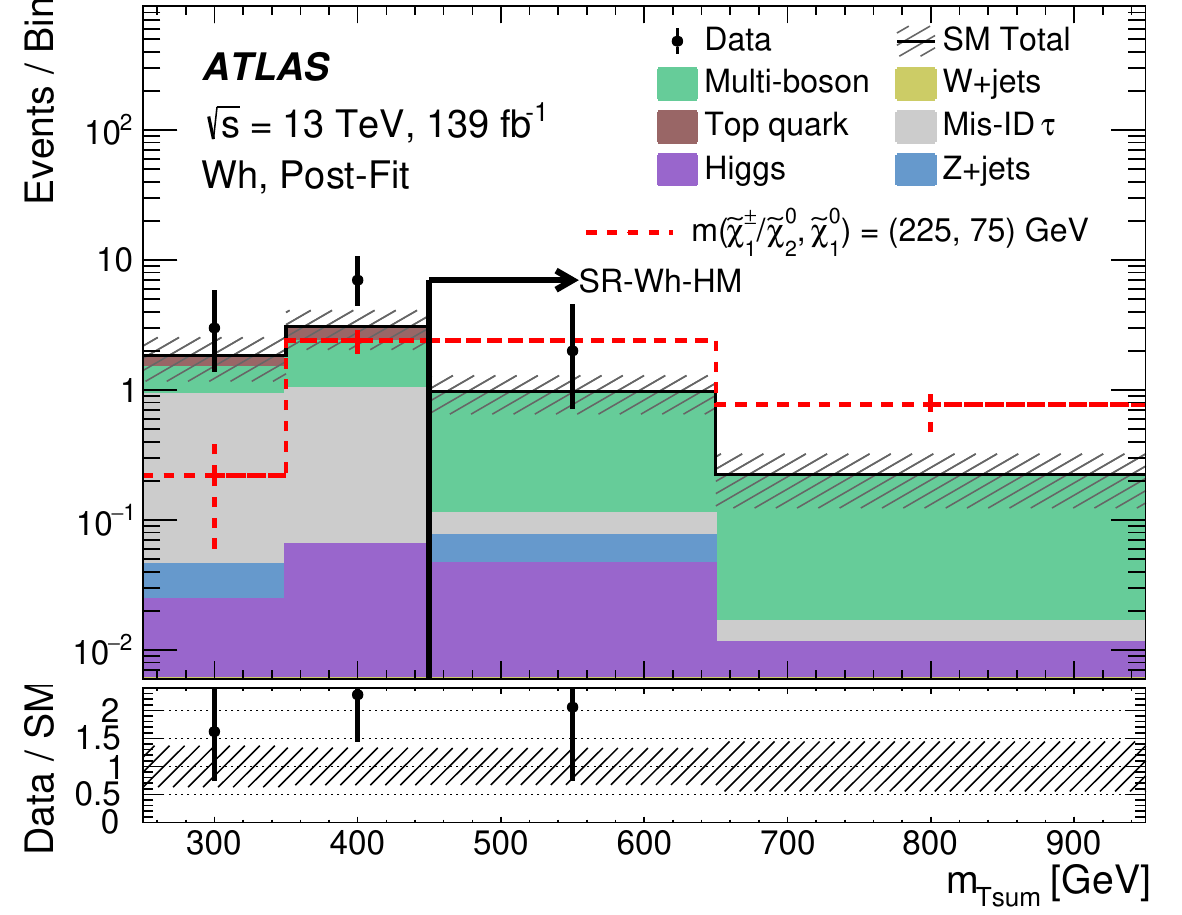}} 
\caption{The kinematic distributions for the Intermediate $Wh$ channel after the background-only fit, showing (a) $\mttwo$ in SR-Wh-LM and (b) $m_{{\mathrm T}\mathrm{sum}}$ in SR-Wh-HM, before the selection on that kinematic is made.
The black arrow depicts the selection for the signal region.
An example SUSY scenario is overlaid for illustration.
The hashed bands represent the combined statistical and systematic uncertainties of the total SM background.
The lower panels show the ratio of data to the total SM background estimate.
}
\label{fig:SRBDTplots_Wh}
\end{figure}

The observed number of events in the two SRs and the expected contributions from SM processes
are given in Table~\ref{tab:results2}. The contribution of top quark background events is scaled with the normalization factor
obtained from the background-only fit.
The top quark background normalization factor is fitted to be 1.00 $\pm$ 0.14.
Kinematic distributions of events in the signal regions are shown in Figure~\ref{fig:SRBDTplots_Wh}.
In all SRs, the observed number of events from data and the background predictions are found to be compatible within uncertainties.
The one-sided $p_0$-values, the observed and expected \percent{95} CL upper limits on the visible non-SM cross-section ($\sigma_{\mathrm{vis}}^{95}$) are shown in Table~\ref{tab:results2}.

\begin{table}[h!]
\centering
\caption{
Observed and expected numbers of events for the background-only fit in the signal regions targeting chargino-neutralino production and decay via $Wh$.
Expected event yields for a few SUSY reference points are also shown.
The uncertainties correspond to the sum in quadrature of statistical  and systematic uncertainties.
The correlation of systematic uncertainties among control regions and among background processes is fully taken into account.
The one-sided $p_0$-values, and the observed and expected \percent{95} CL upper limits on the visible non-SM cross-section ($\sigma_{\mathrm{vis}}^{95}$) from the model-independent fit are given.
}
\footnotesize
\renewcommand\arraystretch{1.3}
\begin{tabular}{l|c|c}
\toprule
SM process  & SR-Wh-LM       & SR-Wh-HM     \\
\midrule
\midrule
Multi-boson  &  1.85 $\pm$  0.5\phantom{0}   &  1.1 $\pm$  0.4    \\
Misidentified processes      &  1.4 $\pm$ 0.6      &  0.06 $\pm$ 0.03  \\
Top quark     &  1.9 $\pm$  0.6    &    $0.04_{-0.04\phantom{0}}^{+0.06\phantom{0}}$ \\
$Z$+jets      &   0.05 $\pm$ 0.02 &  0.03 $\pm$  0.01 \\
Higgs           &  $0.13_{-0.13\phantom{0}}^{+0.99\phantom{0}}$ &   0.06 $\pm$ 0.02 \\ [+1mm]
\midrule
SM total       &  5.3 $\pm$  1.4 &  1.3 $\pm$ 0.4   \\
\midrule
Observed     &   4   &    2   \\
\midrule
$m(\chinoonepm/\ninotwo,\ninoone) = (225, 75)\,\GeV$&    $5.8 \pm 1.5$  & $3.3 \pm 0.9$ \\
\midrule
$p_0$                                  &0.5\phantom{0}    & 0.3\phantom{0}            \\
Expected $\sigma_\text{vis}^{95}$ [fb]  & $0.05$ &     $0.03$ \\ [+1mm]
 
Observed $\sigma_\text{vis}^{95}$ [fb]  & $0.04$  &    $0.03$ \\
\bottomrule
\end{tabular}
\label{tab:results2}
\end{table}

\begin{figure}[h!]
\centering
\includegraphics[width=0.6\textwidth]{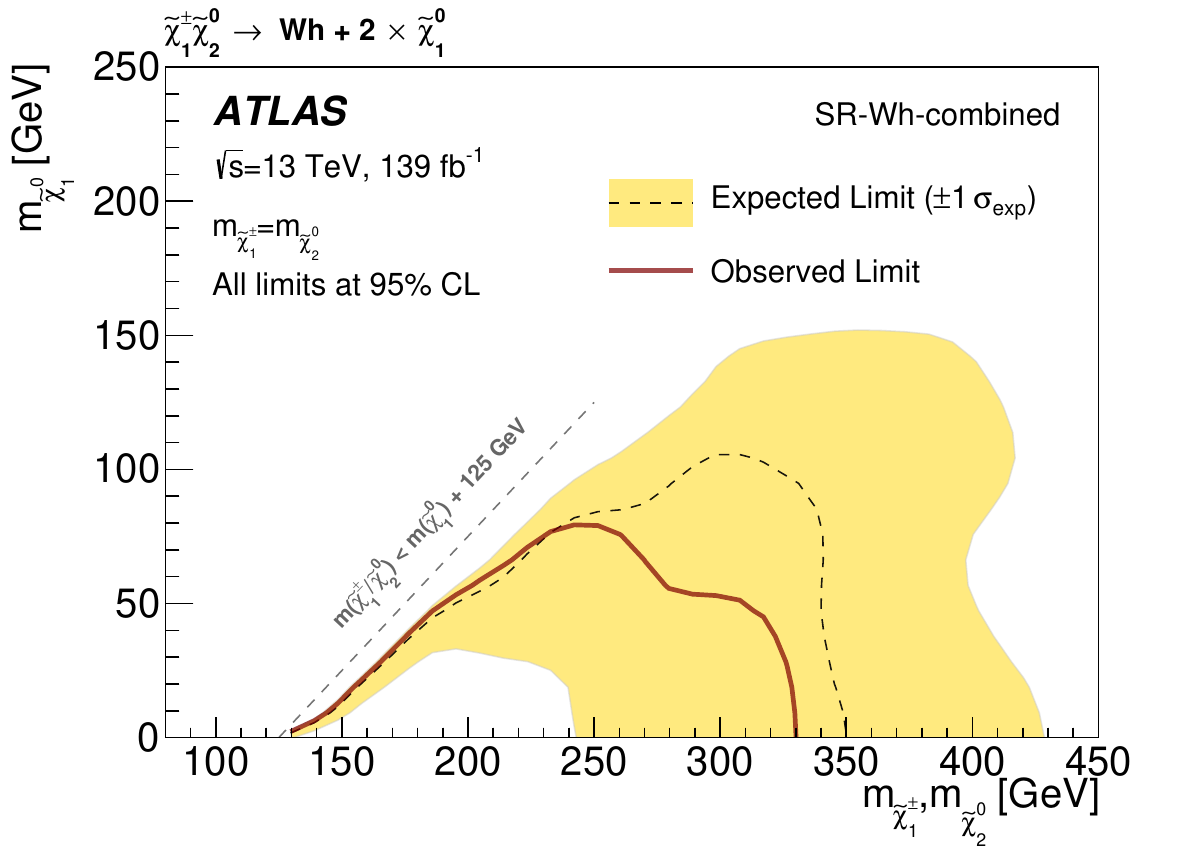} 
\caption{The \percent{95} CL exclusion contours for simplified models of \ConepmNtwo production decaying via an intermediate $Wh$.
The solid (dashed) lines show the observed (expected) exclusion contours. The band around the expected limit shows the $\pm1\sigma$ variations, including all uncertainties except theoretical uncertainties in the signal cross-section.
The best expected limits for SR-Wh-LM and SR-Wh-HM are used.
}
\label{fig:alimit-wh}
\end{figure}

Since no significant excess over the expected SM background is observed, the observed and expected number of events in the SRs are used to place exclusion limits at \percent{95} CL using the model-dependent fit.
The SR with the best expected limit from SR-Wh-LM and SR-Wh-HM is used to derive limits on \ConepmNtwo production decaying via an intermediate $Wh$, and are shown in Figure~\ref{fig:alimit-wh}.
The slightly weaker observed limits than expected for $\chinoonepm/\ninotwo$ masses above $250\,\GeV$ are driven by the results in SR-Wh-HM, as seen in Table~\ref{tab:results2}.
Chargino and next-to-lightest neutralino masses up to $330\,\GeV$ are excluded for a massless lightest neutralino.
 
\FloatBarrier
\section{Summary}
\label{sec:conclusion}
 
Searches for direct $\tau$-slepton (stau) production and direct wino pair production decaying via an intermediate stau or $Wh$ with at least two hadronically decaying $\tau$-leptons in the final state are presented. The searches use 139 fb$^{-1}$ of integrated luminosity of $pp$ collisions at $\sqrt{s} = 13\,\TeV$ collected by the ATLAS detector from 2015 to 2018.
Agreement between data and SM expectation is observed in all signal regions and the results are used to set limits on the visible cross-section for processes beyond the Standard Model.
Exclusion limits at 95\% CL are also placed on simplified models of direct stau production, excluding mass-degenerate $\stau_{L,R}$ up to $500\,\GeV$, $\stau_{L}$ up to $425\,\GeV$, and $\stau_{R}$ up to $350\,\GeV$. The sensitivity to $\stau_{R}\stau_{R}$ production obtained here is the first time for this process reported by ATLAS, achieved with a Boosted Decision Tree for improved signal-background separation.
 
In the scenario of direct production of wino-like chargino pairs decaying into the lightest neutralino via an intermediate on-shell stau, exclusion limits are placed for chargino masses up to $970\,\GeV$ for a massless lightest neutralino.
In the case of the associated production of pairs of charginos and of mass-degenerate charginos and next-to-lightest neutralinos,
masses up to $1160\,\GeV$ are excluded for a massless lightest neutralino.
These limits improve upon previous results by 340--400$\,\GeV$, mainly due to an increased amount of integrated luminosity and
improvements in the recurrent neural network $\tau$-lepton identification.
The sensitivity for more compressed mass scenarios is also improved by the addition of the same-sign $\tau$-lepton channel.
For pairs of degenerate charginos and next-to-lightest neutralinos decaying via intermediate $W$ and $h$ bosons, chargino and next-to-lightest neutralino masses up to $330\,\GeV$ are excluded for a massless lightest neutralino.

\section*{Acknowledgements}

% The next lines are included from the .//acknowledgements/Acknowledgements.tex input file

We thank CERN for the very successful operation of the LHC and its injectors, as well as the support staff at
CERN and at our institutions worldwide without whom ATLAS could not be operated efficiently.
 
The crucial computing support from all WLCG partners is acknowledged gratefully, in particular from CERN, the ATLAS Tier-1 facilities at TRIUMF/SFU (Canada), NDGF (Denmark, Norway, Sweden), CC-IN2P3 (France), KIT/GridKA (Germany), INFN-CNAF (Italy), NL-T1 (Netherlands), PIC (Spain), RAL (UK) and BNL (USA), the Tier-2 facilities worldwide and large non-WLCG resource providers. Major contributors of computing resources are listed in Ref.~\cite{ATL-SOFT-PUB-2023-001}.
 
We gratefully acknowledge the support of ANPCyT, Argentina; YerPhI, Armenia; ARC, Australia; BMWFW and FWF, Austria; ANAS, Azerbaijan; CNPq and FAPESP, Brazil; NSERC, NRC and CFI, Canada; CERN; ANID, Chile; CAS, MOST and NSFC, China; Minciencias, Colombia; MEYS CR, Czech Republic; DNRF and DNSRC, Denmark; IN2P3-CNRS and CEA-DRF/IRFU, France; SRNSFG, Georgia; BMBF, HGF and MPG, Germany; GSRI, Greece; RGC and Hong Kong SAR, China; ISF and Benoziyo Center, Israel; INFN, Italy; MEXT and JSPS, Japan; CNRST, Morocco; NWO, Netherlands; RCN, Norway; MEiN, Poland; FCT, Portugal; MNE/IFA, Romania; MESTD, Serbia; MSSR, Slovakia; ARRS and MIZ\v{S}, Slovenia; DSI/NRF, South Africa; MICINN, Spain; SRC and Wallenberg Foundation, Sweden; SERI, SNSF and Cantons of Bern and Geneva, Switzerland; MOST, Taipei; TENMAK, T\"urkiye; STFC, United Kingdom; DOE and NSF, United States of America.
 
Individual groups and members have received support from BCKDF, CANARIE, CRC and DRAC, Canada; COST, ERC, ERDF, Horizon 2020, ICSC-NextGenerationEU and Marie Sk{\l}odowska-Curie Actions, European Union; Investissements d'Avenir Labex, Investissements d'Avenir Idex and ANR, France; DFG and AvH Foundation, Germany; Herakleitos, Thales and Aristeia programmes co-financed by EU-ESF and the Greek NSRF, Greece; BSF-NSF and MINERVA, Israel; Norwegian Financial Mechanism 2014-2021, Norway; NCN and NAWA, Poland; La Caixa Banking Foundation, CERCA Programme Generalitat de Catalunya and PROMETEO and GenT Programmes Generalitat Valenciana, Spain; G\"{o}ran Gustafssons Stiftelse, Sweden; The Royal Society and Leverhulme Trust, United Kingdom.
 
In addition, individual members wish to acknowledge support from Chile: Agencia Nacional de Investigación y Desarrollo (FONDECYT 1190886, FONDECYT 1210400, FONDECYT 1230987); China: National Natural Science Foundation of China (NSFC - 12175119, NSFC 12275265); European Union: European Research Council (ERC - 948254), Horizon 2020 Framework Programme (MUCCA - CHIST-ERA-19-XAI-00), Italian Center for High Performance Computing, Big Data and Quantum Computing (ICSC, NextGenerationEU), Marie Sklodowska-Curie Actions (EU H2020 MSC IF GRANT NO 101033496); France: Agence Nationale de la Recherche (ANR-20-CE31-0013, ANR-21-CE31-0013, ANR-21-CE31-0022), Investissements d'Avenir Idex (ANR-11-LABX-0012), Investissements d'Avenir Labex (ANR-11-LABX-0012); Germany: Baden-Württemberg Stiftung (BW Stiftung-Postdoc Eliteprogramme), Deutsche Forschungsgemeinschaft (DFG - CR 312/5-1); Italy: Istituto Nazionale di Fisica Nucleare (FELLINI G.A. n. 754496, ICSC, NextGenerationEU); Japan: Japan Society for the Promotion of Science (JSPS KAKENHI 22H01227, JSPS KAKENHI JP21H05085, JSPS KAKENHI JP22H04944); Netherlands: Netherlands Organisation for Scientific Research (NWO Veni 2020 - VI.Veni.202.179); Norway: Research Council of Norway (RCN-314472); Poland: Polish National Agency for Academic Exchange (PPN/PPO/2020/1/00002/U/00001), Polish National Science Centre (NCN 2021/42/E/ST2/00350, NCN UMO-2019/34/E/ST2/00393, UMO-2020/37/B/ST2/01043, UMO-2021/40/C/ST2/00187); Slovenia: Slovenian Research Agency (ARIS grant J1-3010); Spain: BBVA Foundation (LEO22-1-603), Generalitat Valenciana (Artemisa, FEDER, IDIFEDER/2018/048), La Caixa Banking Foundation (LCF/BQ/PI20/11760025), Ministry of Science and Innovation (RYC2019-028510-I, RYC2020-030254-I), PROMETEO and GenT Programmes Generalitat Valenciana (CIDEGENT/2019/023, CIDEGENT/2019/027); Sweden: Swedish Research Council (VR 2022-03845, VR 2022-04683), Knut and Alice Wallenberg Foundation (KAW 2017.0100, KAW 2018.0157, KAW 2019.0447); Switzerland: Swiss National Science Foundation (SNSF - PCEFP2\_194658); United Kingdom: Leverhulme Trust (Leverhulme Trust RPG-2020-004); United States of America: Neubauer Family Foundation.

% End of text imported from the .//acknowledgements/Acknowledgements.tex input file

\printbibliography

@article{Vasquez:2011,
      author         = "Albornoz V\'asquez, Daniel and B\'elanger, Genevieve and B\oe{}hm, Celine",
      title          = "{Revisiting light neutralino scenarios in the MSSM}",
      journal        = "Phys.~Rev. D",
      volume         = "84",
      year           = "2011",
      pages          = "095015",
      eprint         = "1108.1338",
      archivePrefix  = "arXiv",
      primaryClass   = "hep-ph",
      doi            = "10.1103/PhysRevD.84.095015"
}

@Article{Lester:1999tx,
  Title                    = {{Measuring masses of semi-invisibly decaying particle pairs
 produced at hadron colliders}},
  Author                   = {Lester, C. G. and Summers, D. J.},
  Journal                  = {Phys. Lett. B},
  Year                     = {1999},
  Pages                    = {99},
  Volume                   = {463},

  Archiveprefix            = {arXiv},
  Doi                      = {10.1016/S0370-2693(99)00945-4},
  Eprint                   = {hep-ph/9906349},
  Slaccitation             = {%%CITATION = HEP-PH/9906349;%%}
}

@Article{Barr:2003rg,
  Title                    = {A variable for measuring masses at hadron colliders when missing energy is expected; $m_{\mathrm{T2}}$: the truth behind the glamour},
  Author                   = {Barr, Alan and Lester, Christopher and Stephens, P.},
  Journal                  = {J. Phys. G},
  Year                     = {2003},
  Pages                    = {2343},
  Volume                   = {29},

  Archiveprefix            = {arXiv},
  Doi                      = {10.1088/0954-3899/29/10/304},
  Eprint                   = {hep-ph/0304226},
  Slaccitation             = {%%CITATION = HEP-PH/0304226;%%}
}

@Article{Tovey:2008,
  Title                    = {{On measuring the masses of pair-produced semi-invisibly decaying particles at hadron colliders}},
  Author                   = {Tovey, D.R.},
  Journal                  = {JHEP},
  Year                     = {2008},
  Pages                    = {034},
  Volume                   = {04},

  Archiveprefix            = {arXiv},
  Eprint                   = {0802.2879},
  Doi                      = {10.1088/1126-6708/2008/04/034},
  Slaccitation             = {}
}

@article{CLs:2002,
      author         = "Read, A. L.",
      title          = "{Presentation of search results: the CL$_s$ technique}",
      journal        = "J. Phys. G",
      volume         = "28",
      pages          = "2693",
      year           = "2002",
      doi            = "10.1088/0954-3899/28/10/313",
}

@Article{statforumlimitsl,
    Author = {Cowan, Glen and Cranmer, Kyle and Gross, Eilam and Vitells, Ofer},
    Eprint = {1007.1727},
    Archiveprefix = {arXiv},
    Primaryclass = {physics.data-an},
    Journal = {Eur. Phys. J. C},
    Pages = {1554},
    doi = "10.1140/epjc/s10052-011-1554-0",
    Title = {{Asymptotic formulae for likelihood-based tests of new physics}},
    Volume = {71},
    Year = {2011},
    Addendum = "[Erratum: \href{http://dx.doi.org/10.1140/epjc/s10052-013-2501-z}{Eur. Phys. J. C {\textbf{73}} (2013) 2501}]",
      note           = "{{[Erratum: \href{http://dx.doi.org/10.1140/epjc/s10052-013-2501-z}{Eur. Phys. J. C {\textbf{73}} (2013) 2501}]}}",
}

@article{Catani:2009sm,
      author         = "Catani, Stefano and Cieri, Leandro and Ferrera, Giancarlo
                        and de Florian, Daniel and Grazzini, Massimiliano",
      title          = "{Vector Boson Production at Hadron Colliders: A Fully
                        Exclusive QCD Calculation at Next-to-Next-to-Leading Order}",
      journal        = "Phys. Rev. Lett.",
      volume         = "103",
      year           = "2009",
      pages          = "082001",
      doi            = "10.1103/PhysRevLett.103.082001",
      eprint         = "0903.2120",
      archivePrefix  = "arXiv",
      primaryClass   = "hep-ph",
      SLACcitation   = "%%CITATION = ARXIV:0903.2120;%%"
}

@article{Lazopoulos:2008de,
      author         = "Lazopoulos, Achilleas and McElmurry, Thomas and Melnikov,
                        Kirill and Petriello, Frank",
      title          = "{Next-to-leading order QCD corrections to $t \bar{t} Z$
                        production at the LHC}",
      journal        = "Phys. Lett. B",
      volume         = "666",
      year           = "2008",
      pages          = "62",
      doi            = "10.1016/j.physletb.2008.06.073",
      eprint         = "0804.2220",
      archivePrefix  = "arXiv",
      primaryClass   = "hep-ph",
      reportNumber   = "UH-511-1125-08",
      SLACcitation   = "%%CITATION = ARXIV:0804.2220;%%"
}

@article{Campbell:2012dh,
      author         = "Campbell, John M. and Ellis, R. Keith",
      title          = "{$t \bar{t} W^{\pm}$ production and decay at NLO}",
      journal        = "JHEP",
      volume         = "07",
      year           = "2012",
      pages          = "052",
      doi            = "10.1007/JHEP07(2012)052",
      eprint         = "1204.5678",
      archivePrefix  = "arXiv",
      primaryClass   = "hep-ph",
      reportNumber   = "FERMILAB-PUB-12-109-T",
      SLACcitation   = "%%CITATION = ARXIV:1204.5678;%%"
}

@article{CKKW,
      author         = "L{\"o}nnblad, L. and Prestel, S.",
      title          = "{Matching tree-level matrix elements with interleaved showers}",
      journal        = "JHEP",
      volume         = "03",
      year           = "2012",
      pages          = "019",
      doi            = "10.1007/JHEP03(2012)019",
      eprint         = "1109.4829",
      archivePrefix  = "arXiv",
      primaryClass   = "hep-ph",
}

@Article{EvtGen,
    author         = "{D. J. Lange}",
    title          = "{The EvtGen particle decay simulation package}",
    journal        = "Nucl. Instrum. Meth. A",
    volume         = "462",
    year           = "2001",
    pages          = "152",
    doi            = "10.1016/S0168-9002(01)00089-4",
    primaryClass   = "hep-ex",
}

@article{Cacciari:2011ma,
        author  = "Cacciari, Matteo and Salam, Gavin P. and Soyez, Gregory",
	title = "{FastJet user manual}",
	journal = "Eur. Phys. J. C",
	volume = "72",
	year = "2012",
	pages = "1896",
	doi = "10.1140/epjc/s10052-012-1896-2",
	eprint = "1111.6097",
	archivePrefix = "arXiv",
	reportNumber = "CERN-PH-TH-2011-297",
	Primaryclass = {hep-ph},
	SLACcitation = "%%CITATION = ARXIV:1111.6097;%%"
}

@article{jetarea,
      author         = "{M. Cacciari and G. P. Salam}",
      title          = "{Pileup subtraction using jet areas}",
      journal        = "Phys. Lett. B",
      volume         = "659",
      year           = "2008",
      pages          = "119",
      doi = "10.1016/j.physletb.2007.09.077",
      archivePrefix  = "arXiv",
      eprint         = "0707.1378"
}

@article{Aaboud:2016mmw,
      author         = "{ATLAS Collaboration}",
      title          = "{Measurement of the Inelastic Proton-Proton Cross Section
                        at $\sqrt{s}$ = 13 TeV with the ATLAS Detector at the LHC}",
      journal        = "Phys. Rev. Lett.",
      volume         = "117",
      year           = "2016",
      number         = "18",
      pages          = "182002",
      doi            = "10.1103/PhysRevLett.117.182002",
      eprint         = "1606.02625",
      archivePrefix  = "arXiv",
      primaryClass   = "hep-ex",
      reportNumber   = "CERN-EP-2016-140",
      SLACcitation   = "%%CITATION = ARXIV:1606.02625;%%"
}

@misc{LEPSUSYWG:01-03.1,
     author         = "{The LEP SUSY Working Group and the ALEPH, DELPHI, L3 and OPAL experiments}",
     %year          = "",
     howpublished        = "{notes LEPSUSYWG/01-03.1, 04-01.1}",
     url            = "http://lepsusy.web.cern.ch/lepsusy/Welcome.html",
}

@article{LUCID2,
  author = "{G. Avoni et al.}",
  title = "{The new LUCID-2 detector for luminosity measurement and monitoring in ATLAS}",
  journal = "{JINST}",
  year = "2018",
  volume = "13",
  pages = "P07017",
    doi            = "10.1088/1748-0221/13/07/P07017"


}

@article{PDF4LHC,
author = "{J. Butterworth et al.}",
title = "{PDF4LHC recommendations for LHC Run II}",
journal = "{J. Phys. G}",
year = "2016",
volume = "43",
pages = "023001",
    doi            = "10.1088/0954-3899/43/2/023001",
    eprint         = "1510.03865",
    archivePrefix  = "arXiv",
    primaryClass   = "hep-ex",
}

@article{LightGBM,
author = {Ke, Guolin and Meng, Qi and Finley, Thomas and Wang, Taifeng and Chen, Wei and Ma, Weidong and Ye, Qiwei and Liu, Tie-Yan},
title = {LightGBM: A Highly Efficient Gradient Boosting Decision Tree},
year = {2017},
isbn = {9781510860964},
publisher = {Curran Associates Inc.},
address = {Red Hook, NY, USA},
booktitle = {Proceedings of the 31st International Conference on Neural Information Processing Systems},
pages = {3149–3157},
numpages = {9},
location = {Long Beach, California, USA},
series = {NIPS’17}
}

@article{Borschensky:2014cia,
      author         = "Borschensky, Christoph and Kr{\"a}mer, Michael and Kulesza,
                        Anna and Mangano, Michelangelo and Padhi, Sanjay and
                        Plehn, Tilman and Portell, Xavier",
      title          = "{Squark and gluino production cross sections in pp
                        collisions at \(\sqrt{s} = 13, 14, 33\) and \(100\,\text{TeV}\)}",
      journal        = "Eur. Phys. J. C",
      volume         = "74",
      year           = "2014",
      pages          = "3174",
      doi            = "10.1140/epjc/s10052-014-3174-y",
      eprint         = "1407.5066",
      archivePrefix  = "arXiv",
      primaryClass   = "hep-ph",
      reportNumber   = "MS-TP-14-25, CERN-PH-TH-2014-137, TTK-14-13",
      SLACcitation   = "%%CITATION = ARXIV:1407.5066;%%"
}

@article{Golfand:1971iw,
      author         = {Golfand, Y.A.  and Likhtman, E.P.},
      title          = "{Extension of the Algebra of Poincare Group Generators
                        and Violation of P Invariance}",
      journal        = "JETP Lett.",
      volume         = "13",
      year           = "1971",
      pages          = "323",
      note           = "[Pisma Zh. Eksp. Teor. Fiz. \textbf{13} (1971) 452]",
      SLACcitation   = "%%CITATION = JTPLA,13,323;%%"
}

@article{Volkov:1973ix,
      author         = "Volkov, D.V. and Akulov, V.P.",
      title          = "{Is the neutrino a goldstone particle?}",
      journal        = "Phys. Lett. B",
      volume         = "46",
      year           = "1973",
      pages          = "109",
      doi            = "10.1016/0370-2693(73)90490-5",
      SLACcitation   = "%%CITATION = PHLTA,B46,109;%%"
}

@article{Wess:1974tw,
      author         = "Wess, J. and Zumino, B.",
      title          = "{Supergauge transformations in four dimensions}",
      journal        = "Nucl. Phys. B",
      volume         = "70",
      year           = "1974",
      pages          = "39",
      doi            = "10.1016/0550-3213(74)90355-1",
      SLACcitation   = "%%CITATION = NUPHA,B70,39;%%"
}

@article{Salam:1974ig,
      author         = "Salam, Abdus and Strathdee, J.",
      title          = "{Super-symmetry and non-Abelian gauges}",
      journal        = "Phys. Lett. B",
      volume         = "51",
      year           = "1974",
      pages          = "353",
      doi            = "10.1016/0370-2693(74)90226-3",
      reportNumber   = "IC/74/36",
      SLACcitation   = "%%CITATION = PHLTA,B51,353;%%"
}

@article{Wess:1974jb,
      author         = "Wess, J. and Zumino, B.",
      title          = "{Supergauge invariant extension of quantum electrodynamics}",
      journal        = "Nucl. Phys. B",
      volume         = "78",
      year           = "1974",
      pages          = "1",
      doi            = "10.1016/0550-3213(74)90112-6",
      reportNumber   = "CERN-TH-1857",
      SLACcitation   = "%%CITATION = NUPHA,B78,1;%%"
}

@article{Ferrara:1974pu,
      author         = "Ferrara, S. and Zumino, B.",
      title          = "{Supergauge invariant Yang-Mills theories}",
      journal        = "Nucl. Phys. B",
      volume         = "79",
      year           = "1974",
      pages          = "413",
      doi            = "10.1016/0550-3213(74)90559-8",
      reportNumber   = "CERN-TH-1866",
      SLACcitation   = "%%CITATION = NUPHA,B79,413;%%"
}

@article{Martin:1997ns,
      author         = "Martin, Stephen P.",
      title          = "{A Supersymmetry Primer}",
      journal        = "Adv. Ser. Direct. High Energy Phys.",
      volume         = "18",
      pages          = "1",
      year           = "1998",
      doi            = "10.1142/9789812839657_0001",
      eprint         = "hep-ph/9709356",
      archivePrefix  = "arXiv",
      reportNumber   = "FERMILAB-PUB-97-425-T",
      SLACcitation   = "%%CITATION = HEP-PH/9709356;%%"
}

@article{Farrar:1978xj,
      author         = "Farrar, Glennys R. and Fayet, Pierre",
      title          = "{Phenomenology of the production, decay, and detection of
                        new hadronic states associated with supersymmetry}",
      journal        = "Phys. Lett. B",
      volume         = "76",
      year           = "1978",
      pages          = "575",
      doi            = "10.1016/0370-2693(78)90858-4",
      reportNumber   = "CALT-68-648",
      SLACcitation   = "%%CITATION = PHLTA,B76,575;%%"
}

@article{Goldberg:1983nd,
      author         = "Goldberg, H.",
      title          = "{Constraint on the Photino Mass from Cosmology}",
      journal        = "Phys. Rev. Lett.",
      volume         = "50",
      year           = "1983",
      pages          = "1419",
      doi            = "10.1103/PhysRevLett.50.1419",
      related        = "Goldberg:1983nd-err",
      relatedstring  = "Erratum:",
      reportNumber   = "NUB-2592",
      SLACcitation   = "%%CITATION = PRLTA,50,1419;%%"
}

@article{Ellis:1983ew,
      author         = "Ellis, John and Hagelin, J.S. and Nanopoulos, Dimitri V. and Olive, Keith A. and Srednicki, M.",
      title          = "{Supersymmetric relics from the big bang}",
      booktitle      = "{IN *BATAVIA 1984, PROCEEDINGS, INNER SPACE/OUTER SPACE*,
                        458-459., In *Srednicki, M.A. (ed.): Particle physics and
                        cosmology* 223-246}",
      journal        = "Nucl. Phys. B",
      volume         = "238",
      year           = "1984",
      pages          = "453",
      doi            = "10.1016/0550-3213(84)90461-9",
      reportNumber   = "SLAC-PUB-3171",
      SLACcitation   = "%%CITATION = NUPHA,B238,453;%%"
}

@article{Fuks:2012qx,
     author         = "Fuks, Benjamin and Klasen, Michael and Lamprea, David R. and Rothering, Marcel",
     title          = "{Gaugino production in proton-proton collisions at a center-of-mass energy of \(8\,\text{TeV}\)}",
     journal        = "JHEP",
     volume         = "10",
     pages          = "081",
     doi            = "10.1007/JHEP10(2012)081",
     year           = "2012",
     eprint         = "1207.2159",
     archivePrefix  = "arXiv",
     primaryClass   = "hep-ph",
     reportNumber   = "IPHC-PHENO-12-07, MS-TP-12-05",
     SLACcitation   = "%%CITATION = ARXIV:1207.2159;%%",
}

@article{Fuks:2013vua,
     author         = "Fuks, Benjamin and Klasen, Michael and Lamprea, David R. and Rothering, Marcel",
     title          = "{Precision predictions for electroweak superpartner production at hadron colliders with \textsc{resummino}}",
     journal        = "Eur. Phys. J. C",
     volume         = "73",
     pages          = "2480",
     doi            = "10.1140/epjc/s10052-013-2480-0",
     year           = "2013",
     eprint         = "1304.0790",
     archivePrefix  = "arXiv",
     primaryClass   = "hep-ph",
     reportNumber   = "CERN-PH-TH-2013-064, IPHC-PHENO-13-02, MS-TP-13-06",
     SLACcitation   = "%%CITATION = ARXIV:1304.0790;%%",
}

@article{Baak:2014wma,
      author         = "Baak, M. and Besjes, G.J. and C{\^o}t{\'e}, D. and Koutsman, A. and Lorenz, J. and Short, D.",
      title          = "{HistFitter software framework for statistical data analysis}",
      journal        = "Eur. Phys. J. C",
      volume         = "75",
      year           = "2015",
      pages          = "153",
      doi            = "10.1140/epjc/s10052-015-3327-7",
      eprint         = "1410.1280",
      archivePrefix  = "arXiv",
      primaryClass   = "hep-ex",
      SLACcitation   = "%%CITATION = ARXIV:1410.1280;%%",
}

@Article{Cacciari:2008gp,
     author    = "Cacciari, Matteo and Salam, Gavin P. and Soyez, Gregory",
     title     = "{The anti-\(k_{t}\) jet clustering algorithm}",
     journal   = "JHEP",
     volume    = "04",
     year      = "2008",
     pages     = "063",
     eprint    = "0802.1189",
     archivePrefix = "arXiv",
     primaryClass  =  "hep-ph",
     doi       = "10.1088/1126-6708/2008/04/063",
     SLACcitation  = "%%CITATION = 0802.1189;%%"
}

@article{Ball:2012cx,
      author         = "{NNPDF Collaboration} and Ball, Richard D. and others",
      title          = "{Parton distributions with LHC data}",
      journal        = "Nucl. Phys. B",
      volume         = "867",
      year           = "2013",
      pages          = "244",
      doi            = "10.1016/j.nuclphysb.2012.10.003",
      eprint         = "1207.1303",
      archivePrefix  = "arXiv",
      primaryClass   = "hep-ph",
      reportNumber   = "EDINBURGH-2012-08, IFUM-FT-997, FR-PHENO-2012-014,
                        RWTH-TTK-12-25, CERN-PH-TH-2012-037, SFB-CPP-12-47\,
                        --CERN-PH-TH-2012-037",
      SLACcitation   = "%%CITATION = ARXIV:1207.1303;%%"
}

@article{Ball:2014uwa,
      author         = "{NNPDF Collaboration} and Ball, Richard D. and others",
      title          = "{Parton distributions for the LHC run II}",
      journal        = "JHEP",
      volume         = "04",
      year           = "2015",
      pages          = "040",
      doi            = "10.1007/JHEP04(2015)040",
      eprint         = "1410.8849",
      archivePrefix  = "arXiv",
      primaryClass   = "hep-ph",
      reportNumber   = "EDINBURGH-2014-15, IFUM-1034-FT, CERN-PH-TH-2013-253,
                        OUTP-14-11P, CAVENDISH-HEP-14-11",
      SLACcitation   = "%%CITATION = ARXIV:1410.8849;%%"
}

@article{Czakon:2012pz,
      author         = "Czakon, Michal and Mitov, Alexander",
      title          = "{NNLO corrections to top pair production at hadron
                        colliders: the quark-gluon reaction}",
      journal        = "JHEP",
      volume         = "01",
      pages          = "080",
      doi            = "10.1007/JHEP01(2013)080",
      year           = "2013",
      eprint         = "1210.6832",
      archivePrefix  = "arXiv",
      primaryClass   = "hep-ph",
      SLACcitation   = "%%CITATION = ARXIV:1210.6832;%%",
}

@Article{Sjostrand:2007gs,
	author    = "Sj{\"o}strand, T. and Mrenna, S. and Skands, P.",
	title     = "{A brief introduction to PYTHIA 8.1}",
	journal   = "Comput. Phys. Commun.",
	volume    = "178",
	year      = "2008",
	pages     = "852-867",
	eprint    = "0710.3820",
	archivePrefix = "arXiv",
	primaryClass  =  "hep-ph",
	doi       = "10.1016/j.cpc.2008.01.036",
	SLACcitation  = "%%CITATION = 0710.3820;%%"
}

@Article{Alwall:2014hca,
      author         = "Alwall, J. and Frederix, R. and Frixione, S. and Hirschi,
                        V. and Maltoni, F. and Mattelaer, O. and Shao, H. -S. and
                        Stelzer, T. and Torrielli, P. and Zaro, M.",
      title          = "{The automated computation of tree-level and
                        next-to-leading order differential cross sections, and
                        their matching to parton shower simulations}",
      journal        = "JHEP",
      volume         = "07",
      year           = "2014",
      pages          = "079",
      doi            = "10.1007/JHEP07(2014)079",
      eprint         = "1405.0301",
      archivePrefix  = "arXiv",
      primaryClass   = "hep-ph",
      reportNumber   = "CERN-PH-TH-2014-064, CP3-14-18, LPN14-066, MCNET-14-09,
                        ZU-TH-14-14",
      SLACcitation   = "%%CITATION = ARXIV:1405.0301;%%"
}

@Article{Nason:2004rx,
      author         = "Nason, Paolo",
      title          = "{A new method for combining NLO QCD with shower Monte Carlo algorithms}",
      journal        = "JHEP",
      volume         = "11",
      pages          = "040",
      doi            = "10.1088/1126-6708/2004/11/040",
      year           = "2004",
      eprint         = "hep-ph/0409146",
      archivePrefix  = "arXiv",
}

@Article{Frixione:2007vw,
      author         = "Frixione, Stefano and Nason, Paolo and Oleari, Carlo",
      title          = "{Matching NLO QCD computations with parton shower
                        simulations: the POWHEG method}",
      journal        = "JHEP",
      volume         = "11",
      pages          = "070",
      doi            = "10.1088/1126-6708/2007/11/070",
      year           = "2007",
      eprint         = "0709.2092",
      archivePrefix  = "arXiv",
      primaryClass   = "hep-ph",
}

@Article{Alioli:2010xd,
      author         = "Alioli, Simone and Nason, Paolo and Oleari, Carlo and Re,
                        Emanuele",
      title          = "{A general framework for implementing NLO calculations in
                        shower Monte Carlo programs: the POWHEG BOX}",
      journal        = "JHEP",
      volume         = "06",
      pages          = "043",
      doi            = "10.1007/JHEP06(2010)043",
      year           = "2010",
      eprint         = "1002.2581",
      archivePrefix  = "arXiv",
      primaryClass   = "hep-ph",
}

@article{Aliev:2010zk,
      author         = "Aliev, M. and Lacker, H. and Langenfeld, U. and Moch, S.
                        and Uwer, P. and Wiedermann, M.",
      title          = "{HATHOR -- HAdronic Top and Heavy quarks crOss section
                        calculatoR}",
      journal        = "Comput. Phys. Commun.",
      volume         = "182",
      year           = "2011",
      pages          = "1034-1046",
      doi            = "10.1016/j.cpc.2010.12.040",
      eprint         = "1007.1327",
      archivePrefix  = "arXiv",
      primaryClass   = "hep-ph",
      reportNumber   = "DESY-10-091, HU-EP-10-33, SFB-CPP-10-60",
      SLACcitation   = "%%CITATION = ARXIV:1007.1327;%%"
}

@article{Kant:2014oha,
      author         = "Kant, P. and Kind, O. M. and Kintscher, T. and Lohse, T.
                        and Martini, T. and Mölbitz, S. and Rieck, P. and Uwer, P.",
      title          = "{HatHor for single top-quark production: Updated
                        predictions and uncertainty estimates for single top-quark
                        production in hadronic collisions}",
      journal        = "Comput. Phys. Commun.",
      volume         = "191",
      year           = "2015",
      pages          = "74-89",
      doi            = "10.1016/j.cpc.2015.02.001",
      eprint         = "1406.4403",
      archivePrefix  = "arXiv",
      primaryClass   = "hep-ph",
      reportNumber   = "HU-EP-14-22",
      SLACcitation   = "%%CITATION = ARXIV:1406.4403;%%"
}

@article{Beneke:2011mq,
      author         = "Beneke, M. and Falgari, P. and Klein, S. and Schwinn, C.",
      title          = "{Hadronic top-quark pair production with NNLL threshold
                        resummation}",
      journal        = "Nucl. Phys. B",
      volume         = "855",
      year           = "2012",
      pages          = "695-741",
      doi            = "10.1016/j.nuclphysb.2011.10.021",
      eprint         = "1109.1536",
      archivePrefix  = "arXiv",
      primaryClass   = "hep-ph",
      reportNumber   = "TTK-11-38, ITP-UU-11-26, SPIN-11-19, FR-PHENO-2011-015,
                        SFB-CPP-11-49",
      SLACcitation   = "%%CITATION = ARXIV:1109.1536;%%"
}

@article{Cacciari:2011hy,
      author         = "Cacciari, Matteo and Czakon, Michal and Mangano,
                        Michelangelo and Mitov, Alexander and Nason, Paolo",
      title          = "{Top-pair production at hadron colliders with
                        next-to-next-to-leading logarithmic soft-gluon
                        resummation}",
      journal        = "Phys. Lett. B",
      volume         = "710",
      pages          = "612-622",
      doi            = "10.1016/j.physletb.2012.03.013",
      year           = "2012",
      eprint         = "1111.5869",
      primaryClass   = "hep-ph",
      archivePrefix  = "arXiv",
      reportNumber   = "CERN-PH-TH-2011-277, TTK-11-54",
      SLACcitation   = "%%CITATION = ARXIV:1111.5869;%%",
}

@article{Czakon:2012zr,
      author         = "Czakon, Michal and Mitov, Alexander",
      title          = "{NNLO corrections to top-pair production at hadron
                        colliders: the all-fermionic scattering channels}",
      journal        = "JHEP",
      volume         = "12",
      pages          = "054",
      doi            = "10.1007/JHEP12(2012)054",
      year           = "2012",
      eprint         = "1207.0236",
      archivePrefix  = "arXiv",
      primaryClass   = "hep-ph",
      SLACcitation   = "%%CITATION = ARXIV:1207.0236;%%",
}

@article{Czakon:2013goa,
      author         = "Czakon, Michal and Fiedler, Paul and Mitov, Alexander",
      title          = "{Total Top-Quark Pair-Production Cross Section at
                        Hadron Colliders Through \(O(\alpha_S^4)\)}",
      journal        = "Phys. Rev. Lett.",
      volume         = "110",
      pages          = "252004",
      doi            = "10.1103/PhysRevLett.110.252004",
      year           = "2013",
      eprint         = "1303.6254",
      archivePrefix  = "arXiv",
      primaryClass   = "hep-ph",
      reportNumber   = "CERN-PH-TH-2013-056, TTK-13-08",
      SLACcitation   = "%%CITATION = ARXIV:1303.6254;%%",
}

@article{Baernreuther:2012ws,
      author         = "B{\"a}rnreuther, Peter and Czakon, Michal and Mitov,
                        Alexander",
      title          = "{Percent-Level-Precision Physics at the
                  Tevatron: Next-to-Next-to-Leading Order QCD
                  Corrections to \(q \bar{q} \to t \bar{t} + X\)}",
      journal        = "Phys. Rev. Lett.",
      volume         = "109",
      pages          = "132001",
      doi            = "10.1103/PhysRevLett.109.132001",
      year           = "2012",
      eprint         = "1204.5201",
      archivePrefix  = "arXiv",
      primaryClass   = "hep-ph",
      SLACcitation   = "%%CITATION = ARXIV:1204.5201;%%",
}

@article{Frixione:2007nw,
      author         = "Frixione, Stefano and Ridolfi, Giovanni and Nason, Paolo",
      title          = "{A positive-weight next-to-leading-order Monte Carlo for
                        heavy flavour hadroproduction}",
      journal        = "JHEP",
      volume         = "09",
      pages          = "126",
      doi            = "10.1088/1126-6708/2007/09/126",
      year           = "2007",
      eprint         = "0707.3088",
      archivePrefix  = "arXiv",
      primaryClass   = "hep-ph",
      SLACcitation   = "%%CITATION = ARXIV:0707.3088;%%",
}

@article{deFlorian:2016spz,
      author         = "de Florian, D. and others",
      title          = "{Handbook of LHC Higgs Cross Sections: 4. Deciphering the
                        Nature of the Higgs Sector}",
      collaboration  = "LHC Higgs Cross Section Working Group",
      doi            = "10.23731/CYRM-2017-002",
      year           = "2016",
      eprint         = "1610.07922",
      archivePrefix  = "arXiv",
      primaryClass   = "hep-ph",
      reportNumber   = "FERMILAB-FN-1025-T, CERN-2017-002-M",
      SLACcitation   = "%%CITATION = ARXIV:1610.07922;%%"
}

@Article{Gleisberg:2008ta,
      author         = "Gleisberg, T. and H{\"o}che, S. and Krauss, F. and Sch{\"o}nherr, M. and Schumann, S. and Siegert, F. and Winter J.",
      title          = "{Event generation with SHERPA 1.1}",
      journal        = "JHEP",
      volume         = "02",
      pages          = "007",
      doi            = "10.1088/1126-6708/2009/02/007",
      year           = "2009",
      eprint         = "0811.4622",
      archivePrefix  = "arXiv",
      primaryClass   = "hep-ph",
}

@Article{Hoeche:2009rj,
      author         = "H{\"o}che, Stefan and Krauss, Frank and Schumann, Steffen and Siegert, Frank",
      title          = "{QCD matrix elements and truncated showers}",
      journal        = "JHEP",
      volume         = "05",
      pages          = "053",
      doi            = "10.1088/1126-6708/2009/05/053",
      year           = "2009",
      eprint         = "0903.1219",
      archivePrefix  = "arXiv",
      primaryClass   = "hep-ph",
}

@Article{Gleisberg:2008fv,
      author         = "Gleisberg, Tanju and H{\"o}che, Stefan",
      title          = "{Comix, a new matrix element generator}",
      journal        = "JHEP",
      volume         = "12",
      pages          = "039",
      doi            = "10.1088/1126-6708/2008/12/039",
      year           = "2008",
      eprint         = "0808.3674",
      archivePrefix  = "arXiv",
      primaryClass   = "hep-ph",
}

@Article{Schumann:2007mg,
      author         = "Schumann, Steffen and Krauss, Frank",
      title          = "{A parton shower algorithm based on Catani--Seymour dipole factorisation}",
      journal        = "JHEP",
      volume         = "03",
      pages          = "038",
      doi            = "10.1088/1126-6708/2008/03/038",
      year           = "2008",
      eprint         = "0709.1027",
      archivePrefix  = "arXiv",
      primaryClass   = "hep-ph",
}

@article{Hoeche:2012yf,
      author         = "H{\"o}che, Stefan and Krauss, Frank and Sch{\"o}nherr, Marek and
                        Siegert, Frank",
      title          = "{QCD matrix elements + parton showers. The NLO case}",
      journal        = "JHEP",
      volume         = "04",
      year           = "2013",
      pages          = "027",
      doi            = "10.1007/JHEP04(2013)027",
      eprint         = "1207.5030",
      archivePrefix  = "arXiv",
      primaryClass   = "hep-ph",
      reportNumber   = "SLAC-PUB-15191, IPPP-12-52, DCPT-12-104, LPN12-081,
                        FR-PHENO-2012-017, MCNET-12-09, --FR-PHENO-2012-017",
      SLACcitation   = "%%CITATION = ARXIV:1207.5030;%%"
}

@article{Catani:2001cc,
      author         = "Catani, S. and Krauss, F. and Webber, B. R. and Kuhn, R.",
      title          = "{QCD Matrix Elements + Parton Showers}",
      journal        = "JHEP",
      volume         = "11",
      year           = "2001",
      pages          = "063",
      doi            = "10.1088/1126-6708/2001/11/063",
      eprint         = "hep-ph/0109231",
      archivePrefix  = "arXiv",
      reportNumber   = "CERN-TH-2000-367, CAVENDISH-HEP-00-03",
      SLACcitation   = "%%CITATION = HEP-PH/0109231;%%"
}

@article{Cascioli:2011va,
      author         = "Cascioli, Fabio and Maierh{\"o}fer, Philipp and Pozzorini, Stefano",
      title          = "{Scattering Amplitudes with Open Loops}",
      journal        = "Phys. Rev. Lett.",
      volume         = "108",
      year           = "2012",
      pages          = "111601",
      doi            = "10.1103/PhysRevLett.108.111601",
      eprint         = "1111.5206",
      archivePrefix  = "arXiv",
      primaryClass   = "hep-ph",
      reportNumber   = "ZU-TH-23-11, LPN11-66",
      SLACcitation   = "%%CITATION = ARXIV:1111.5206;%%"
}

@article{Denner:2016kdg,
      author         = "Denner, Ansgar and Dittmaier, Stefan and Hofer, Lars",
      title          = "{\textsc{Collier}: A fortran-based complex one-loop library in
                        extended regularizations}",
      journal        = "Comput. Phys. Commun.",
      volume         = "212",
      year           = "2017",
      pages          = "220-238",
      doi            = "10.1016/j.cpc.2016.10.013",
      eprint         = "1604.06792",
      archivePrefix  = "arXiv",
      primaryClass   = "hep-ph",
      reportNumber   = "FR-PHENO-2016-003, ICCUB-16-016",
      SLACcitation   = "%%CITATION = ARXIV:1604.06792;%%"
}

@article{Czakon:2011xx,
  author         = "Czakon, Michal and Mitov, Alexander",
  title          = "{Top++: A program for the calculation of the top-pair cross-section at hadron colliders}",
  journal        = "Comput. Phys. Commun.",
  volume         = "185",
  year           = "2014",
  pages          = "2930",
  doi            = "10.1016/j.cpc.2014.06.021",
  eprint         = "1112.5675",
  archivePrefix  = "arXiv",
  primaryClass   = "hep-ph",
  reportNumber   = "CERN-PH-TH-2011-303, TTK-11-58",
}

@article{Agostinelli:2002hh,
      author         = "Agostinelli, S. and others",
      title          = "{\textsc{Geant4} -- a simulation toolkit}",
      journal        = "Nucl. Instrum. Meth. A",
      volume         = "506",
      year           = "2003",
      pages          = "250",
      doi            = "10.1016/S0168-9002(03)01368-8",
      reportNumber   = "SLAC-PUB-9350, FERMILAB-PUB-03-339",
      SLACcitation   = "%%CITATION = NUIMA,A506,250;%%"
}

@Article{PERF-2007-01,
    author         = "{ATLAS Collaboration}",
    title          = "{The ATLAS Experiment at the CERN Large Hadron Collider}",
    journal        = "JINST",
    volume         = "3",
    year           = "2008",
    pages          = "S08003",
    doi            = "10.1088/1748-0221/3/08/S08003",
    primaryClass   = "hep-ex",
}

@Article{SOFT-2010-01,
    author         = "{ATLAS Collaboration}",
    title          = "{The ATLAS Simulation Infrastructure}",
    journal        = "Eur. Phys. J. C",
    volume         = "70",
    year           = "2010",
    pages          = "823",
    doi            = "10.1140/epjc/s10052-010-1429-9",
    eprint         = "1005.4568",
    archivePrefix  = "arXiv",
    primaryClass   = "physics.ins-det",
}

@Article{PERF-2011-03,
    author         = "{ATLAS Collaboration}",
    title          = "{Jet energy measurement with the ATLAS detector in proton--proton collisions at \(\sqrt{s} = 7\,\text{TeV}\)}",
    journal        = "Eur. Phys. J. C",
    volume         = "73",
    year           = "2013",
    pages          = "2304",
    doi            = "10.1140/epjc/s10052-013-2304-2",
    reportNumber   = "CERN-PH-EP-2011-191",
    eprint         = "1112.6426",
    archivePrefix  = "arXiv",
    primaryClass   = "hep-ex",
}

@Article{PERF-2013-06,
    author         = "{ATLAS Collaboration}",
    title          = "{Identification and energy calibration of hadronically decaying tau leptons with the ATLAS experiment in \(pp\) collisions at \(\sqrt{s} = 8\,\text{TeV}\)}",
    journal        = "Eur. Phys. J. C",
    volume         = "75",
    year           = "2015",
    pages          = "303",
    doi            = "10.1140/epjc/s10052-015-3500-z",
    reportNumber   = "CERN-PH-EP-2014-227",
    eprint         = "1412.7086",
    archivePrefix  = "arXiv",
    primaryClass   = "hep-ex",
}

@Article{SUSY-2013-14,
    author         = "{ATLAS Collaboration}",
    title          = "{Search for the direct production of charginos, neutralinos and staus in final states with at least two hadronically decaying taus and missing transverse momentum in \(pp\) collisions at \(\sqrt{s} = 8\,\text{TeV}\) with the ATLAS detector}",
    journal        = "JHEP",
    volume         = "10",
    year           = "2014",
    pages          = "096",
    doi            = "10.1007/JHEP10(2014)096",
    reportNumber   = "CERN-PH-EP-2014-121",
    eprint         = "1407.0350",
    archivePrefix  = "arXiv",
    primaryClass   = "hep-ex",
}

@Article{PERF-2014-03,
    author         = "{ATLAS Collaboration}",
    title          = "{Performance of pile-up mitigation techniques for jets in \(pp\) collisions at \(\sqrt{s} = 8\,\text{TeV}\) using the ATLAS detector}",
    journal        = "Eur. Phys. J. C",
    volume         = "76",
    year           = "2016",
    pages          = "581",
    doi            = "10.1140/epjc/s10052-016-4395-z",
    reportNumber   = "CERN-PH-EP-2015-206",
    eprint         = "1510.03823",
    archivePrefix  = "arXiv",
    primaryClass   = "hep-ex",
}

@Article{SUSY-2014-05,
    author         = "{ATLAS Collaboration}",
    title          = "{Search for the electroweak production of supersymmetric particles in \(\sqrt{s} = 8\,\text{TeV}\) \(pp\) collisions with the ATLAS detector}",
    journal        = "Phys. Rev. D",
    volume         = "93",
    year           = "2016",
    pages          = "052002",
    doi            = "10.1103/PhysRevD.93.052002",
    reportNumber   = "CERN-PH-EP-2015-218",
    eprint         = "1509.07152",
    archivePrefix  = "arXiv",
    primaryClass   = "hep-ex",
}

@Article{PERF-2015-09,
    author         = "{ATLAS Collaboration}",
    title          = "{Jet reconstruction and performance using particle flow with the ATLAS Detector}",
    journal        = "Eur. Phys. J. C",
    volume         = "77",
    year           = "2017",
    pages          = "466",
    doi            = "10.1140/epjc/s10052-017-5031-2",
    reportNumber   = "CERN-EP-2017-024",
    eprint         = "1703.10485",
    archivePrefix  = "arXiv",
    primaryClass   = "hep-ex",
}

@Article{SUSY-2016-23,
    author         = "{ATLAS Collaboration}",
    title          = "{Search for the direct production of charginos and neutralinos in final states with tau leptons in \(\sqrt{s} = 13\,\text{TeV}\) \(pp\) collisions with the ATLAS detector}",
    journal        = "Eur. Phys. J. C",
    volume         = "78",
    year           = "2018",
    pages          = "154",
    doi            = "10.1140/epjc/s10052-018-5583-9",
    reportNumber   = "CERN-EP-2017-173",
    eprint         = "1708.07875",
    archivePrefix  = "arXiv",
    primaryClass   = "hep-ex",
}

@Article{DAPR-2018-01,
    author         = "{ATLAS Collaboration}",
    title          = "{ATLAS data quality operations and performance for 2015--2018 data-taking}",
    journal        = "JINST",
    volume         = "15",
    year           = "2020",
    pages          = "P04003",
    doi            = "10.1088/1748-0221/15/04/P04003",
    reportNumber   = "CERN-EP-2019-207",
    eprint         = "1911.04632",
    archivePrefix  = "arXiv",
    primaryClass   = "physics.ins-det",
}

@Article{EGAM-2018-01,
    author         = "{ATLAS Collaboration}",
    title          = "{Electron and photon performance measurements with the ATLAS detector using the 2015--2017 LHC proton--proton collision data}",
    journal        = "JINST",
    volume         = "14",
    year           = "2019",
    pages          = "P12006",
    doi            = "10.1088/1748-0221/14/12/P12006",
    reportNumber   = "CERN-EP-2019-145",
    eprint         = "1908.00005",
    archivePrefix  = "arXiv",
    primaryClass   = "hep-ex",
}

@Article{FTAG-2018-01,
    author         = "{ATLAS Collaboration}",
    title          = "{ATLAS \(b\)-jet identification performance and efficiency measurement with \(t\bar{t}\) events in \(pp\) collisions at \(\sqrt{s} = 13\,\text{TeV}\)}",
    journal        = "Eur. Phys. J. C",
    volume         = "79",
    year           = "2019",
    pages          = "970",
    doi            = "10.1140/epjc/s10052-019-7450-8",
    reportNumber   = "CERN-EP-2019-132",
    eprint         = "1907.05120",
    archivePrefix  = "arXiv",
    primaryClass   = "hep-ex",
}

@Article{MUON-2018-03,
    author         = "{ATLAS Collaboration}",
    title          = "{Muon reconstruction and identification efficiency in ATLAS using the full Run~2 \(pp\) collision data set at \(\sqrt{s} = 13\,\text{TeV}\)}",
    journal        = "Eur. Phys. J. C",
    volume         = "81",
    year           = "2021",
    pages          = "578",
    doi            = "10.1140/epjc/s10052-021-09233-2",
    reportNumber   = "CERN-EP-2020-199",
    eprint         = "2012.00578",
    archivePrefix  = "arXiv",
    primaryClass   = "hep-ex",
}

@Article{SUSY-2018-04,
    author         = "{ATLAS Collaboration}",
    title          = "{Search for direct stau production in events with two hadronic \(\tau\)-leptons in \(\sqrt{s} = 13\,\text{TeV}\) \(pp\) collisions with the ATLAS detector}",
    journal        = "Phys. Rev. D",
    volume         = "101",
    year           = "2020",
    pages          = "032009",
    doi            = "10.1103/PhysRevD.101.032009",
    reportNumber   = "CERN-EP-2019-191",
    eprint         = "1911.06660",
    archivePrefix  = "arXiv",
    primaryClass   = "hep-ex",
}

@Article{TRIG-2018-01,
    author         = "{ATLAS Collaboration}",
    title          = "{Performance of the ATLAS muon triggers in Run~2}",
    journal        = "JINST",
    volume         = "15",
    year           = "2020",
    pages          = "P09015",
    doi            = "10.1088/1748-0221/15/09/p09015",
    reportNumber   = "CERN-EP-2020-031",
    eprint         = "2004.13447",
    archivePrefix  = "arXiv",
    primaryClass   = "physics.ins-det",
}

@Article{FTAG-2019-07,
    author         = "{ATLAS Collaboration}",
    title          = "{ATLAS flavour-tagging algorithms for the LHC Run~2 \(pp\) collision dataset}",
    journal        = "Eur. Phys. J. C",
    volume         = "83",
    year           = "2023",
    pages          = "681",
    doi            = "10.1140/epjc/s10052-023-11699-1",
    reportNumber   = "CERN-EP-2022-226",
    eprint         = "2211.16345",
    archivePrefix  = "arXiv",
    primaryClass   = "physics.data-an",
}

@Article{TRIG-2019-04,
    author         = "{ATLAS Collaboration}",
    title          = "{Operation of the ATLAS trigger system in Run~2}",
    journal        = "JINST",
    volume         = "15",
    year           = "2020",
    pages          = "P10004",
    doi            = "10.1088/1748-0221/15/10/P10004",
    reportNumber   = "CERN-EP-2020-109",
    eprint         = "2007.12539",
    archivePrefix  = "arXiv",
    primaryClass   = "hep-ex",
}

@Booklet{ATL-SOFT-PUB-2021-001,
    author         = "{ATLAS Collaboration}",
    title          = "{The ATLAS Collaboration Software and Firmware}",
    howpublished   = "{ATL-SOFT-PUB-2021-001}",
    url            = "https://cds.cern.ch/record/2767187",
    year           = "2021",
}

@Booklet{ATL-SOFT-PUB-2023-001,
    author         = "{ATLAS Collaboration}",
    title          = "{ATLAS Computing Acknowledgements}",
    howpublished   = "{ATL-SOFT-PUB-2023-001}",
    url            = "https://cds.cern.ch/record/2869272",
    year           = "2023",
}

@Article{CMS-SUS-13-006,
    author         = "{CMS Collaboration}",
    title          = "{Searches for electroweak production of charginos, neutralinos, and sleptons decaying to leptons and \(W\), \(Z\), and Higgs bosons in \(pp\) collisions at \(8\,\text{TeV}\)}",
    journal        = "Eur. Phys. J. C",
    volume         = "74",
    year           = "2014",
    pages          = "3036",
    doi            = "10.1140/epjc/s10052-014-3036-7",
    reportNumber   = "CERN-PH-EP-2014-098",
    eprint         = "1405.7570",
    archivePrefix  = "arXiv",
    primaryClass   = "hep-ex",
}

@Article{CMS-SUS-19-012,
    author         = "{CMS Collaboration}",
    title          = "{Search for electroweak production of charginos and neutralinos in proton--proton collisions at \(\sqrt{s} = 13\,\text{TeV}\)}",
    journal        = "JHEP",
    volume         = "04",
    year           = "2022",
    pages          = "147",
    doi            = "10.1007/JHEP04(2022)147",
    reportNumber   = "CERN-EP-2021-097",
    eprint         = "2106.14246",
    archivePrefix  = "arXiv",
    primaryClass   = "hep-ex",
}

@Article{CMS-SUS-21-001,
    author         = "{CMS Collaboration}",
    title          = "{Search for direct pair production of supersymmetric partners of \(\tau\) leptons in the final state with two hadronically decaying \(\tau\) leptons and missing transverse momentum in proton--proton collisions at \(\sqrt{s} = 13\,\text{TeV}\)}",
    year           = "2023",
    reportNumber   = "CERN-EP-2022-032",
    eprint         = "2207.02254",
    archivePrefix  = "arXiv",
    primaryClass   = "hep-ex",
}

@Booklet{ATLAS-CONF-2015-029,
    author         = "{ATLAS Collaboration}",
    title          = "{Selection of jets produced in \(13~\text{TeV}\) proton--proton collisions with the ATLAS detector}",
    howpublished   = "{ATLAS-CONF-2015-029}",
    url            = "https://cds.cern.ch/record/2037702",
    year           = "2015",
}

@Booklet{ATLAS-CONF-2017-029,
    author         = "{ATLAS Collaboration}",
    title          = "{Measurement of the tau lepton reconstruction and identification performance in the ATLAS experiment using \(pp\) collisions at \(\sqrt{s} = 13~\text{TeV}\)}",
    howpublished   = "{ATLAS-CONF-2017-029}",
    url            = "https://cds.cern.ch/record/2261772",
    year           = "2017",
}

@Booklet{ATLAS-CONF-2018-023,
    author         = "{ATLAS Collaboration}",
    title          = "{\(E_{\text{T}}^{\text{miss}}\) performance in the ATLAS detector using 2015--2016 LHC \(pp\) collisions}",
    howpublished   = "{ATLAS-CONF-2018-023}",
    url            = "https://cds.cern.ch/record/2625233",
    year           = "2018",
}

@Booklet{ATLAS-CONF-2019-021,
    author         = "{ATLAS Collaboration}",
    title          = "{Luminosity determination in \(pp\) collisions at \(\sqrt{s} = 13\,\text{TeV}\) using the ATLAS detector at the LHC}",
    howpublished   = "{ATLAS-CONF-2019-021}",
    url            = "https://cds.cern.ch/record/2677054",
    year           = "2019",
}

@Booklet{ATL-PHYS-PUB-2014-021,
    author         = "{ATLAS Collaboration}",
    title          = "{ATLAS Pythia~8 tunes to \(7~\text{TeV}\) data}",
    howpublished   = "{ATL-PHYS-PUB-2014-021}",
    url            = "https://cds.cern.ch/record/1966419",
    year           = "2014",
}

@Booklet{ATL-SOFT-PUB-2014-001,
    author         = "{ATLAS Collaboration}",
    title          = "{Performance of the Fast ATLAS Tracking Simulation (FATRAS) and the ATLAS Fast Calorimeter Simulation (FastCaloSim) with single particles}",
    howpublished   = "{ATL-SOFT-PUB-2014-001}",
    url            = "https://cds.cern.ch/record/1669341",
    year           = "2014",
}

@Booklet{ATL-PHYS-PUB-2015-015,
    author         = "{ATLAS Collaboration}",
    title          = "{Jet Calibration and Systematic Uncertainties for Jets Reconstructed in the ATLAS Detector at \(\sqrt{s} = 13~\text{TeV}\)}",
    howpublished   = "{ATL-PHYS-PUB-2015-015}",
    url            = "https://cds.cern.ch/record/2037613",
    year           = "2015",
}

@Booklet{ATL-PHYS-PUB-2015-023,
    author         = "{ATLAS Collaboration}",
    title          = "{Expected performance of missing transverse momentum reconstruction for the ATLAS detector at \(\sqrt{s} = 13~\text{TeV}\)}",
    howpublished   = "{ATL-PHYS-PUB-2015-023}",
    url            = "https://cds.cern.ch/record/2037700",
    year           = "2015",
}

@Booklet{ATL-PHYS-PUB-2015-026,
    author         = "{ATLAS Collaboration}",
    title          = "{Vertex Reconstruction Performance of the ATLAS Detector at \(\sqrt{s} = 13~\text{TeV}\)}",
    howpublished   = "{ATL-PHYS-PUB-2015-026}",
    url            = "https://cds.cern.ch/record/2037717",
    year           = "2015",
}

@Booklet{ATL-PHYS-PUB-2015-027,
    author         = "{ATLAS Collaboration}",
    title          = "{Performance of missing transverse momentum reconstruction with the ATLAS detector in the first proton--proton collisions at \(\sqrt{s} = 13~\text{TeV}\)}",
    howpublished   = "{ATL-PHYS-PUB-2015-027}",
    url            = "https://cds.cern.ch/record/2037904",
    year           = "2015",
}

@Booklet{ATL-PHYS-PUB-2015-045,
    author         = "{ATLAS Collaboration}",
    title          = "{Reconstruction, Energy Calibration, and Identification of Hadronically Decaying Tau Leptons in the ATLAS Experiment for Run-2 of the LHC}",
    howpublished   = "{ATL-PHYS-PUB-2015-045}",
    url            = "https://cds.cern.ch/record/2064383",
    year           = "2015",
}

@Booklet{ATL-PHYS-PUB-2016-017,
    author         = "{ATLAS Collaboration}",
    title          = "{The Pythia~8 A3 tune description of ATLAS minimum bias and inelastic measurements incorporating the Donnachie--Landshoff diffractive model}",
    howpublished   = "{ATL-PHYS-PUB-2016-017}",
    url            = "https://cds.cern.ch/record/2206965",
    year           = "2016",
}

@Booklet{ATL-PHYS-PUB-2019-033,
    author         = "{ATLAS Collaboration}",
    title          = "{Identification of hadronic tau lepton decays using neural networks in the ATLAS experiment}",
    howpublished   = "{ATL-PHYS-PUB-2019-033}",
    url            = "https://cds.cern.ch/record/2688062",
    year           = "2019",
}
\clearpage
% ATLAS Collaboration author list
% Reference date of SUSY-2023-03 is 2023-05-02
% Author list last updated on date 09-APR-24
% Data extracted on 09-Apr-2024 for paper reference SUSY-2023-03
% at 10:57am
 
\begin{flushleft}
\hypersetup{urlcolor=black}
{\Large The ATLAS Collaboration}

\bigskip

\AtlasOrcid[0000-0002-6665-4934]{G.~Aad}$^\textrm{\scriptsize 102}$,
\AtlasOrcid[0000-0002-5888-2734]{B.~Abbott}$^\textrm{\scriptsize 120}$,
\AtlasOrcid[0000-0002-1002-1652]{K.~Abeling}$^\textrm{\scriptsize 55}$,
\AtlasOrcid[0000-0001-5763-2760]{N.J.~Abicht}$^\textrm{\scriptsize 49}$,
\AtlasOrcid[0000-0002-8496-9294]{S.H.~Abidi}$^\textrm{\scriptsize 29}$,
\AtlasOrcid[0000-0002-9987-2292]{A.~Aboulhorma}$^\textrm{\scriptsize 35e}$,
\AtlasOrcid[0000-0001-5329-6640]{H.~Abramowicz}$^\textrm{\scriptsize 151}$,
\AtlasOrcid[0000-0002-1599-2896]{H.~Abreu}$^\textrm{\scriptsize 150}$,
\AtlasOrcid[0000-0003-0403-3697]{Y.~Abulaiti}$^\textrm{\scriptsize 117}$,
\AtlasOrcid[0000-0002-8588-9157]{B.S.~Acharya}$^\textrm{\scriptsize 69a,69b,m}$
% ...
\bigskip
\\

% ...
$^{29}$Physics Department, Brookhaven National Laboratory, Upton NY; United States of America.\\
% ...
$^{35}$$^{(a)}$Facult\'e des Sciences Ain Chock, R\'eseau Universitaire de Physique des Hautes Energies - Universit\'e Hassan II, Casablanca;$^{(b)}$Facult\'{e} des Sciences, Universit\'{e} Ibn-Tofail, K\'{e}nitra;$^{(c)}$Facult\'e des Sciences Semlalia, Universit\'e Cadi Ayyad, LPHEA-Marrakech;$^{(d)}$LPMR, Facult\'e des Sciences, Universit\'e Mohamed Premier, Oujda;$^{(e)}$Facult\'e des sciences, Universit\'e Mohammed V, Rabat;$^{(f)}$Institute of Applied Physics, Mohammed VI Polytechnic University, Ben Guerir; Morocco.\\
% ...
$^{49}$Fakult\"{a}t Physik , Technische Universit{\"a}t Dortmund, Dortmund; Germany.\\
% ...
$^{55}$II. Physikalisches Institut, Georg-August-Universit\"{a}t G\"ottingen, G\"ottingen; Germany.\\
% ...
$^{69}$$^{(a)}$INFN Gruppo Collegato di Udine, Sezione di Trieste, Udine;$^{(b)}$ICTP, Trieste;$^{(c)}$Dipartimento Politecnico di Ingegneria e Architettura, Universit\`a di Udine, Udine; Italy.\\
% ...
$^{102}$CPPM, Aix-Marseille Universit\'e, CNRS/IN2P3, Marseille; France.\\
% ...
$^{117}$Department of Physics, New York University, New York NY; United States of America.\\
% ...
$^{120}$Homer L. Dodge Department of Physics and Astronomy, University of Oklahoma, Norman OK; United States of America.\\
% ...
$^{150}$Department of Physics, Technion, Israel Institute of Technology, Haifa; Israel.\\
$^{151}$Raymond and Beverly Sackler School of Physics and Astronomy, Tel Aviv University, Tel Aviv; Israel.\\
% ...
$^{a}$ Also Affiliated with an institute covered by a cooperation agreement with CERN.\\
$^{b}$ Also at An-Najah National University, Nablus; Palestine.\\
$^{c}$ Also at Borough of Manhattan Community College, City University of New York, New York NY; United States of America.\\
$^{d}$ Also at Center for High Energy Physics, Peking University; China.\\
$^{e}$ Also at Center for Interdisciplinary Research and Innovation (CIRI-AUTH), Thessaloniki; Greece.\\
$^{f}$ Also at Centro Studi e Ricerche Enrico Fermi; Italy.\\
% ...
$^{m}$ Also at Department of Physics, King's College London, London; United Kingdom.\\
% ...

\end{flushleft}

% Created with Glance <Atlas.Glance@cern.ch>

\end{document}